\documentclass[11pt,a4paper]{article}                           

\usepackage{amsmath}
\usepackage{amssymb}
\usepackage[usenames]{color} 
\usepackage{multirow}
\usepackage{graphicx}
\usepackage{epstopdf}
\usepackage{array}
\usepackage{hhline}
\usepackage{cite}
\usepackage{microtype,colortbl}
\usepackage[bf]{caption}
\usepackage{color}
\usepackage[usenames,dvipsnames]{xcolor}
\usepackage{pstool}
\usepackage{tikz}
\usepackage{ytableau}
\usepackage{latexsym,stmaryrd}
\usepackage[bookmarks=false]{hyperref}
 \usepackage{lscape}
 \usepackage{subfig}

\usepackage{ifpdf}
\def\hybrid{\topmargin -20pt    \oddsidemargin 0pt
        \headheight 0pt \headsep 0pt
        \textwidth 6.25in       
        \textheight 9 in       
        \marginparwidth .875in
        \parskip 5pt plus 1pt 
          \jot = 1.5ex
   }
\hybrid
\numberwithin{equation}{section}
\numberwithin{table}{section}

\newcommand{\beq}{\begin{equation}}  \newcommand{\eeq}{\end{equation}}
\newcommand{\bal}{\begin{aligned}}   \newcommand{\eal}{\end{aligned}}
\newcommand{\bea}{\begin{eqnarray}}  \newcommand{\eea}{\end{eqnarray}}

\newcommand{\bmat}{\left(\begin{array}}
\newcommand{\emat}{\end{array}\right)}

\newcommand{\bbC}{\mathbb{C}}
\newcommand{\bbR}{\mathbb{R}}

\newcommand{\cO}{\mathcal{O}}

\newcommand{\cE}{\mathcal{E}}

\newcommand{\cK}{\mathcal{K}}
\newcommand{\cN}{\mathcal{N}}

\newcommand{\cA}{\mathcal{A}}

\newcommand{\cF}{\mathcal{F}}

\newcommand{\cR}{\mathcal{R}}

\newcommand{\cV}{\mathcal{V}}

\newcommand{\cM}{\mathcal M}
\newcommand{\cQ}{\mathcal Q}

\renewcommand{\Im}{\mathrm{Im}\,}
\renewcommand{\Re}{\mathrm{Re}\,}

\newcommand{\I}{\text{Im}}
\newcommand{\R}{\text{Re}}

\usepackage{cancel}

\newcommand{\be}{\begin{equation}}
\newcommand{\ee}{\end{equation}}

\newcommand*\Bell{\ensuremath{\boldsymbol\ell}}

\definecolor{Gray}{gray}{0.95}

\begin{document}
\baselineskip=14pt
\parskip 5pt plus 1pt 

\vspace*{2cm}
\begin{center}
{\LARGE\bfseries Weak Gravity Bounds in}\\[.3cm]
{\LARGE\bfseries  Asymptotic String Compactifications}\\[5mm]

\vspace{1cm}
{\bf Brice Bastian}\footnote{b.bastian@uu.nl},
{\bf Thomas W.~Grimm}\footnote{t.w.grimm@uu.nl},
{\bf Damian van de Heisteeg}\footnote{d.t.e.vandeheisteeg@uu.nl},

{\small
\vspace*{.5cm}
Institute for Theoretical Physics, Utrecht University\\ Princetonplein 5, 3584 CC Utrecht, The Netherlands\\[3mm]
}
\end{center}
\vspace{1cm}
\begin{abstract}\noindent
We study the charge-to-mass ratios of BPS states in four-dimensional $\mathcal{N}=2$ supergravities arising from Calabi-Yau threefold compactifications of Type IIB string theory. We present a formula for the asymptotic charge-to-mass ratio valid for all limits in complex structure moduli 
space. This is achieved by using the sl(2)-structure that emerges in any such limit as described by asymptotic Hodge theory. The asymptotic charge-to-mass formula applies for sl(2)-elementary states that couple to the graviphoton asymptotically. Using this formula, we determine the radii of the ellipsoid that forms the extremality region of electric BPS black holes, which provides us with a general asymptotic bound on the charge-to-mass ratio for these theories. Finally, we comment on how these bounds for the Weak Gravity Conjecture relate to their counterparts in the asymptotic de Sitter Conjecture and Swampland Distance Conjecture.
\end{abstract}

\newpage

\tableofcontents
\setcounter{footnote}{0}

\newpage
\section{Introduction}
The aim of the swampland programme is to find criteria that distinguish low energy effective theories that 
can be embedded into a UV complete theory of quantum gravity from those which cannot. The former class of theories has been dubbed the Landscape whereas the latter is known as the Swampland. In the last decades several conditions have been formulated that 
should be satisfied in order for a theory to belong to the Landscape, such as the Swampland Distance and Weak Gravity Conjectures \cite{Ooguri:2006in,ArkaniHamed:2006dz}. Since our understanding of quantum gravity is still rather incomplete
these criteria were often not derived from microscopic principles, but rather reflect a large collection of empirical evidence 
gathered from effective theories that are known to be UV consistent. Hence, we generally refer to this set of criteria as the 
Swampland conjectures. In recent years much effort has been dedicated to formulate, refine, test, and interconnect these conjectures with the aim 
that a clear picture emerges which general structures must be present in any valid theory of quantum gravity. 
A comprehensive introduction and review of the programme is given in \cite{Palti:2019pca}.

One fruitful way to test Swampland conjectures, is to try to verify whether these conjectures hold true in compactifications of string theory, and investigate the constraints obeyed by all the resulting effective theories. Eventually this might also lead to a deeper understanding of the underlying 
structures that ensure the validity of these conjectures. Clearly, the success of this program crucially depends on probing a very large or 
possibly general set of string compactifications, rather than studying specific examples. This is an active field of research, for example for the Swampland Distance Conjecture various approaches have been suggested in \cite{Grimm:2018ohb,Heidenreich:2018kpg,Blumenhagen:2018nts,Lee:2018urn,Grimm:2018cpv,Corvilain:2018lgw,Font:2019cxq,Baume:2019sry}. One approach to reach such generality was 
put forward in \cite{Grimm:2018ohb,Grimm:2018cpv,Corvilain:2018lgw,Font:2019cxq,Grimm:2019wtx,Grimm:2019ixq,Gendler:2020dfp,Lanza:2020qmt,Grimm:2020cda}, where it was argued that in supersymmetric compactifications the asymptotic region in which the geometric
compactification space degenerates are universally described by the principles of asymptotic Hodge theory. This powerful mathematical 
framework allows one to perform a general analysis and does not rely on specific examples. One key result of  asymptotic Hodge theory is 
that when taking any $n$-parameter limit towards the boundary of moduli space an sl$(2)^n$-symmetry emerges. This symmetry algebra can 
be used to classify limits and group the states  of the effective theory into representations. It was argued that this emergent symmetry structure can be viewed as the underlying reason that some of the Swampland conjectures are satisfied.\footnote{More recently, it was also suggested 
in \cite{Cecotti:2020rjq,Cecotti:2020uek} that a certain structural principle of Hodge theory should be translated to a Swampland criterium. } 
In this work we will show that the techniques from asymptotic Hodge theory can also be used to derive rather general numerical bounds in the 
Swampland conjectures.

The main focus of this work will be the Weak Gravity Conjecture \cite{ArkaniHamed:2006dz}, which states that a quantum theory of gravity containing at least one $U(1)$ gauge field should have a superextremal particle, i.e.~that its charge-to-mass ratio is larger than or equal to the black hole extremality bound. When multiple gauge fields are present one has to study the extremality region of electrically charged black holes in more detail. Namely, instead of requiring the existence of a single superextremal particle, the Weak Gravity Conjecture has to be satisfied for every direction in the charge lattice. In \cite{Cheung:2014vva} this observation was formalized into the statement that there should exist a set of electrically charged particles whose charge-to-mass vectors span a convex hull that contains the black hole extremality region, and even stronger versions of this condition were proposed with the sublattice \cite{Heidenreich:2015nta,Heidenreich:2016aqi,Montero:2016tif} and tower Weak Gravity Conjecture \cite{Andriolo:2018lvp}. The constraints put by the Weak Gravity Conjecture motivated many detailed studies of asymptotic string compactifications \cite{Palti:2017elp,Grimm:2018ohb,Lee:2018urn,Lee:2018spm,Lee:2019tst,Font:2019cxq,Marchesano:2019ifh,Lee:2019xtm,Grimm:2019wtx,Demirtas:2019lfi,Lee:2019wij,Baume:2019sry,Enriquez-Rojo:2020pqm,Gendler:2020dfp,Lanza:2020qmt,Heidenreich:2020ptx,Klaewer:2020lfg}\footnote{The Weak Gravity Conjecture can also be used to constrain the field ranges for axions. Implications of the convex hull condition and its stronger versions were investigated in \cite{Rudelius:2015xta,Montero:2015ofa,Brown:2015iha,Bachlechner:2015qja,Junghans:2015hba,Hebecker:2015zss}.}. In this work we study Calabi-Yau threefold compactifications of Type IIB string theory. The compactification yields four-dimensional $\mathcal{N}=2$ supergravity theories that are, due to their string theory origin, in the Landscape and therefore should satisfy the Swampland conjectures. The gauge fields then arise from expanding the R-R four-form potential of Type IIB in terms of harmonic three-forms on the Calabi-Yau manifold, and in turn we expect the Weak Gravity Conjecture to be satisfied by wrapped D3-brane states charged under these gauge fields. It is then natural to pose the question whether one can in fact identify these D3-brane states, and moreover if charge lattice sites populated by BPS states suffice or if non-BPS states are also necessary in order to satisfy the convex hull condition. In this work we set ourselves a more modest goal, and we merely aim to make the bounds put by the extremality region of electrically charged BPS black holes as precise as possible.

As a first step in approaching Weak Gravity Conjecture bounds we single out a special set of candidate BPS states that are elementary with respect to the aforementioned asymptotic sl$(2)^n$-structure, i.e.~they sit in a single eigenspace under the sl$(2)^n$-decomposition. If in addition these particles couple to the graviphoton asymptotically, we find that their asymptotic charge-to-mass ratio is constant and for any limit in complex structure moduli space given by 
\begin{equation} \label{intro-Q/M}
 \lim_{\gamma \to \infty} \bigg( \frac{Q}{M} \bigg)^{-2} \bigg|_{q_{\rm G}}= 2^{1-d_n}  \prod_{i=1}^n {\Delta d_i\choose{(\Delta d_i - \Delta \ell_i)/2}} \times \begin{cases}
1 \text{ for $d_n = 3$}\, ,\\
\frac{1}{2} \text{ for $d_n \neq 3$}\, . 
\end{cases}\, 
\end{equation}
In this formula the $d_i$ and $\ell_i$ correspond to discrete data characterizing the candidate BPS state and the type of limit, with $\Delta d_i = d_i - d_{i-1}$ and similarly for $\Delta \ell_i$. The notation $q_{\rm G}$ stands for sl(2)-elementary charges that couple asymptotically to the graviphoton. Of course, all these notions and the formula itself will be thoroughly explained in the main text. Our results significantly extend the recent formula of \cite{Gendler:2020dfp}, which was derived using asymptotic Hodge theory for a large class of infinite distance limits. While the two formulas look rather different we explain that they agree in most cases, with some particular exceptions where our formula contains additional terms.

\begin{figure}[h!]
\vspace*{-.2cm}
\centering
\hspace{-1.5cm} \subfloat[$\gamma_1=2/\sqrt{3},\gamma_2=2$]{\label{fig:ellipsoid1}\hspace{1cm} \includegraphics[height=2cm]{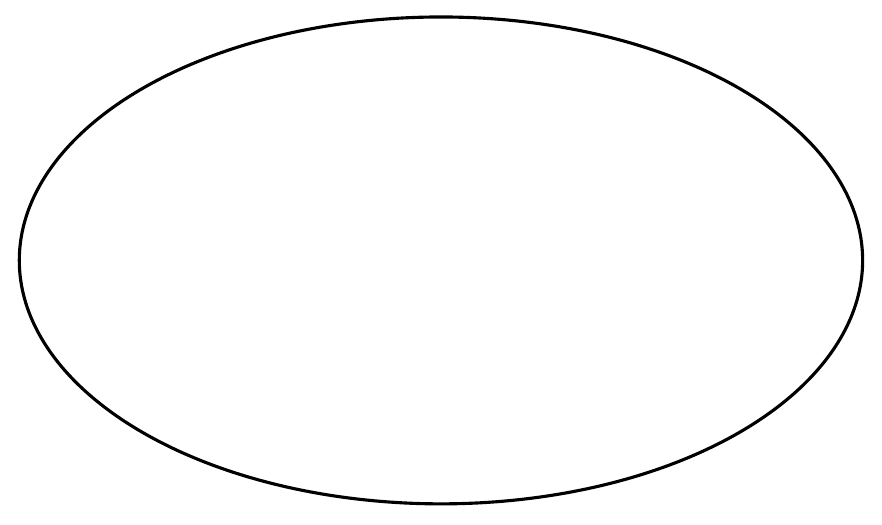}  \hspace{1cm}}
\hspace{0.2cm}
\subfloat[$\gamma^{-2}_1=\gamma^{-2}_2=1/2$]{\label{fig:ellipsoid2}\hspace{.3cm}\includegraphics[height=3cm]{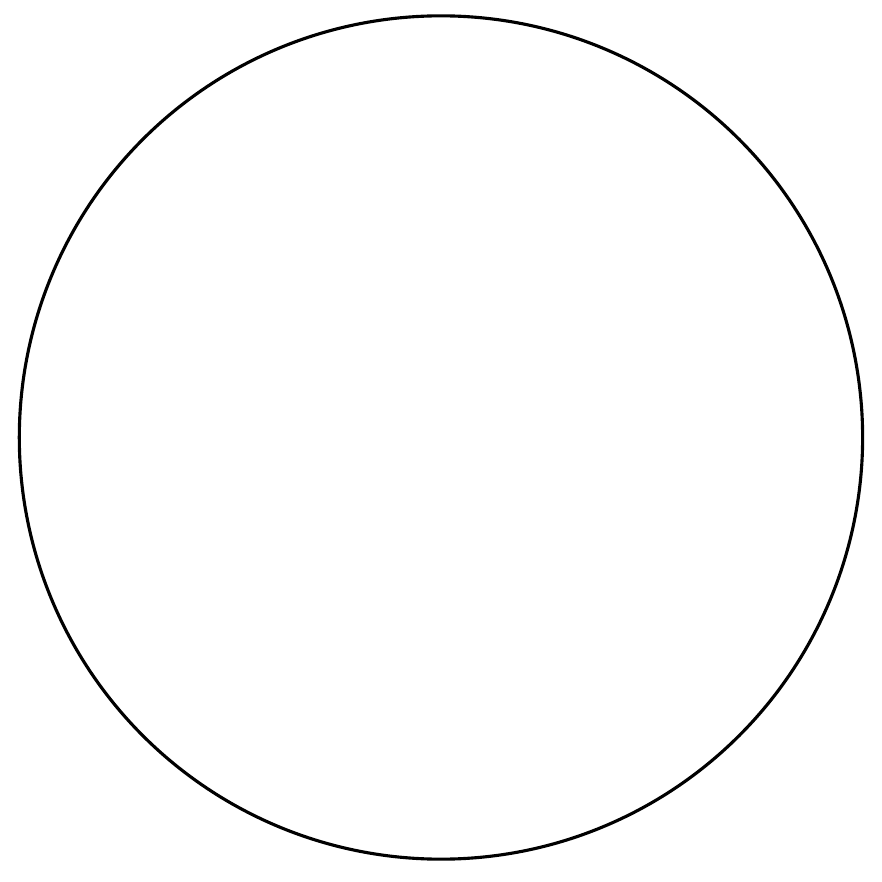}}
\hspace{1cm}
\subfloat[$\gamma^{2}_1=1,\gamma^{-2}_2=0$]{\label{fig:ellipsoid3}\hspace{0.2cm}\includegraphics[height=1.9cm]{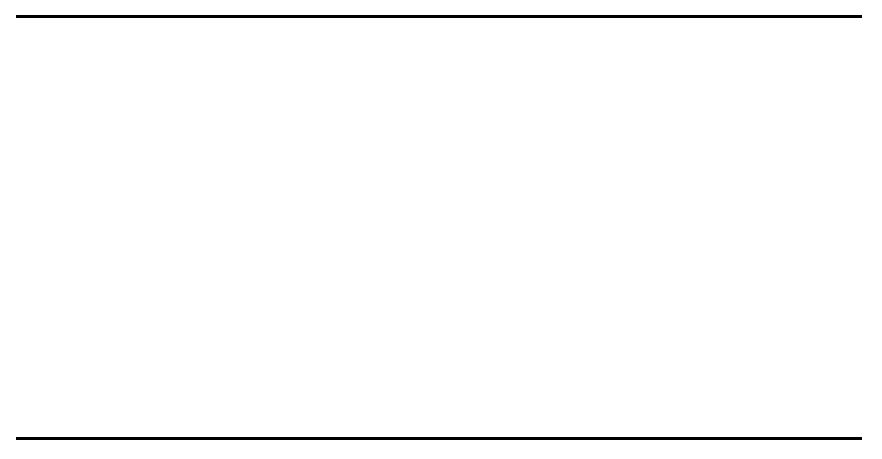}}
\caption{The three possible asymptotic shapes of the ellipsoid that forms the extremality region of electric BPS black holes. Figures \ref{fig:ellipsoid1} and \ref{fig:ellipsoid2} occur for asymptotic regions at infinite distance, whereas figure \ref{fig:ellipsoid3} occurs for finite distance limits. }\label{fig:ellipsoids}
\end{figure}

The formula \eqref{intro-Q/M} will be essential in establishing actual bounds on the charge-to-mass spectrum of electric BPS states in four-dimensional $\mathcal{N}=2$ supergravities in the asymptotic regime. In order to do this we elaborate on a result of \cite{Gendler:2020dfp} that the charge-to-mass vectors of electric BPS states lie on an ellipsoid with two non-degenerate directions $\gamma_1,\gamma_2$. We will compute the asymptotic values of these radii using the above general formula for all limits in complex structure moduli space, both at finite and infinite distance. We find that there are only three possible sets of values for these radii corresponding to the three ellipsoids depicted in figure \ref{fig:ellipsoids}. We note that for finite distance singularities $\gamma_2^{-2}=0$, so another direction of the ellipsoid degenerates, and only a single non-degenerate direction remains. Besides specifying a structure for the charge-to-mass spectrum, the smallest radius of this ellipsoid also serves as a lower bound on the charge-to-mass ratio for electric BPS states. Although we only computed the limit values of these radii, the framework of asymptotic Hodge theory also dictates how corrections enter when we move away from the boundary. To be precise, these corrections are polynomially suppressed in the scalar fields that are taken to be large. While we do not manage to control the coefficients that appear with these polynomially suppressed corrections, this does allow us to designate so-called strict asymptotic regimes where these corrections can be taken to be small. We thus claim that the lowest value of the radii also provides a lower bound on electric charge-to-mass ratios in these strict asymptotic regime.  Inserting the numerical values for the radii, we find that for infinite distance singularities the asymptotic charge-to-mass ratio of electric BPS states is bounded from below by $2/\sqrt{3}$.

There is evidence that the various Swampland conjectures are not independent but rather seem to form an intricate web. In particular relating their $\cO(1)$ coefficients is crucial for the phenomenological constraints imposed by these conjectures. Having obtained a numerical bound for the charge-to-mass ratio, it is natural to investigate how this relates to other Swampland $\cO(1)$ coefficients. In that sense, we are able to determine asymptotic values for the relevant order-one coefficients appearing in the asymptotic de Sitter \cite{Obied:2018sgi,Ooguri:2018wrx,Garg:2018reu} and Swampland Distance Conjectures \cite{Ooguri:2006in,Klaewer:2016kiy}. A relation with the de Sitter conjecture is established by making particular flux choices and rewriting the flux potential such that the order-one coefficient of the de Sitter conjecture can be evaluated by using the above general formula \eqref{intro-Q/M}. We find bounds that are known from the literature (see \cite{Andriot:2020lea,Lanza:2020qmt} and references therein) depending on which contributions to the flux potential are taken into account, i.e.~we separate the contributions coming from the axio-dilaton, complex structure and tree-level K\"ahler moduli. In particular, we are able to saturate the recently proposed Trans-Planckian Censorship Conjecture bound \cite{Bedroya:2019snp} $c \geq \sqrt{2/3}$, when we only consider complex structure moduli for infinite distance singularities. Furthermore, we use the relation of the Weak Gravity Conjecture with the Swampland Distance Conjecture suggested in \cite{Palti:2017elp,Grimm:2018ohb,Lee:2018spm, Gendler:2020dfp}. This allows us to match the order-one coefficients as outlined in \cite{Lee:2018spm, Gendler:2020dfp}. We find agreement with results from the literature \cite{Grimm:2018ohb, Andriot:2020lea,Gendler:2020dfp} and in particular with the lowest value of $\lambda = 1/\sqrt{6}$, which matches with the recently proposed relation $\lambda = c/2$.

The paper is structured as follows. In section \ref{sec:Review} we review the basics of 4d $\mathcal{N}=2$ supergravity theories, and we look at two examples where we evaluate charge-to-mass ratios to gain some intuition for the question at hand. In section \ref{sec:asympHodgetheory} we introduce techniques from asymptotic Hodge theory. These techniques are put to use in section \ref{sec:generalanalysis} to perform a general analysis of the charge-to-mass spectrum at asymptotic regions in complex structure moduli space. Finally, in section \ref{sec:remarks} we discuss connections between the Weak Gravity Conjecture and the de Sitter and Distance Conjectures, and how bounds obtained for the former relate to those of the latter two.

\section{Charge-to-mass ratios and limits in moduli space}
\label{sec:Review}
The goal of this section is twofold. First, we review the basics of four-dimensional (4d) $\mathcal{N}=2$ supergravities arising from Type IIB string compactifications on Calabi-Yau threefolds together with the relevant notions associated with BPS states of these theories. This allows us to fix notation and set the stage for the main calculation of interest in this paper. Second, we outline a remarkable observation made by the authors of \cite{Gendler:2020dfp}, namely that the so-called charge-to-mass vectors (defined in \ref{CMVect}) of BPS states of these theories lie on a degenerate ellipsoid with exactly two finite radii. Furthermore, we provide two examples where these radii can be computed explicitly as the prepotential is known. Studying these examples gives a feeling for the task at hand and serves as a precursor to section \ref{sec:generalanalysis}, where we perform a general analysis that holds for any type of singularity in complex structure moduli space and does not rely on a prepotential formulation of the underlying supergravity theory. 

\subsection{Review of 4d $\mathcal{N}=2$ supergravities from Calabi-Yau compactifications} \label{N=2basics}

To set the stage for our analysis, let us introduce the relevant background needed in 4d $\mathcal{N}=2$ supergravities. We consider Type IIB string theory compactified on a Calabi-Yau threefold $Y_3$, and focus on the $h^{2,1}$ scalar fields $t^i$ and $h^{2,1}+1$ vectors with field strengths $F^I_{\mu \nu}$ that arise in this compactification. The hypermultiplet sector can be ignored for the purposes of our work. Let us begin by writing down the bosonic action, which reads
\begin{equation}\label{eq:action}
S^{(4)}= \int_{\mathbb{M}^{3,1}} \bigg( \frac{1}{2} R *_{4}\!1- K_{i\bar \jmath}\ d t^i \wedge *_{4} d \bar{t}^j + \frac{1}{4}\, \text{Im} \cN_{IJ} F^I \wedge *_{4} 
F^{J} +  \frac{1}{4} \text{Re} \cN_{IJ} F^{I} \wedge F^{J} \bigg)\, ,
\end{equation}
where $*_{4}$ is the 4d Hodge star. The K\"ahler metric $K_{i\bar \jmath}$ and gauge kinetic functions $\cN_{IJ}$ depend on the scalars $t^i,\bar t^i$. 
It is well known that these supergravity theories enjoy electro-magnetic duality for which $Sp(2(h_{2,1}+1),\mathbb{Z})$ is the relevant symmetry group. Depending on the symplectic frame, there exists a single holomorphic function, the prepotential $\mathcal{F}(t)$, in terms of which the K\"ahler metric and the gauge kinetic functions can be computed. It can be shown that in every symplectic orbit there exists a prepotential formulation \cite{Craps:1997gp}.  

Geometrically, the vector multiplet scalar fields $t^i$ correspond to the complex structure moduli of the Calabi-Yau threefold $Y_3$. On the latter there exists a unique holomorphic $(3,0)$-form $\Omega$ whose integrals over a suitable symplectic basis of three-cycles $A_I, B^I \in H_3(Y_3,\mathbb{Z})$ give rise to the period vector 
\begin{align}
\mathbf{\Pi}=\begin{pmatrix}
\int_{A^I} \Omega \\
\int_{B^I} \Omega 
\end{pmatrix} = \begin{pmatrix}
X^I \\\mathcal{F}_I
\end{pmatrix} \label{eq:PeriodVector} \ ,
\end{align}
where $X^I$ and $\mathcal{F_I}$ are holomorphic functions of the complex structure moduli $t^i$. Using the dual cohomology elements of our three-cycle basis denoted by $\alpha_I, \beta^I$, we can define a natural symplectic pairing matrix $\eta$.
Let us first define the wedge product pairing between two three forms $\alpha,\alpha' \in H^3(Y_3,\bbR)$ as
\beq \label{inner-produced}
    \langle \alpha,\alpha'\rangle = \int_{Y_3} \alpha \wedge \alpha'\ ,
\eeq
and note that it is skew-symmetric. 
If we choose this basis appropriately, we can bring $\eta$ to the simple form
\begin{align}
\eta^{\ I}_J =  \langle \alpha_I ,\beta^J  \rangle = \delta^I_J \ , \qquad \eta_{IJ}= \langle  \alpha_I , \alpha_J \rangle = 0 \ ,\qquad \eta^{IJ} = \langle \beta^I , \beta^J \rangle = 0 \,. \label{SympParing}
\end{align}
The K\"ahler metric can be obtained via $K_{i \bar \jmath}=\partial_{t^i} \partial_{\bar t^j} K$
with the K\"ahler potential $K$ given by
\begin{equation}\label{eq:kahlerpotential}
K =- \log i  \, \bar{\mathbf{\Pi}}^T \eta \mathbf{\Pi} = - \log i(\bar{X}^I \cF_I - X^I \bar{\cF}_I)\, .
\end{equation}
Another important quantity that will appear throughout this work is the K\"ahler covariant derivative of the period vector which is given by
\begin{equation}
D_i \mathbf{\Pi}=\partial_i \mathbf{\Pi}+ (\partial_i K) \mathbf{\Pi}= \begin{pmatrix}
D_i X^I \\ D_i \mathcal{F}_I
\end{pmatrix}
\end{equation}
In terms of the latter we can define the gauge kinetic functions that appear in the supergravity action (\ref{eq:action}) in terms of the following matrix relation
\begin{equation}\label{eq:gaugefunctions}
\cN_{IJ} = \begin{pmatrix}
 \mathcal{F}_I & D_{\bar{\imath}} \bar{\mathcal{F}}_I
\end{pmatrix} \begin{pmatrix}
  X^J & D_{\bar{\imath}} \bar{X}^J
\end{pmatrix}^{-1} \,,
\end{equation}
where the invertibility of the second matrix is guaranteed by the positivity of the kinetic terms for the scalars and the field strengths in (\ref{eq:action}).

We finish this lightning review by giving some relevant formulas for BPS states in 4d $\mathcal{N}=2$ supergravities. As we consider Type IIB string theory compactified on a Calabi-Yau threefold $Y_3$, the BPS states we are interested in arise from D3-branes wrapping a particular class of three-cycles of $Y_3$. We denote the dual three-form class by $q$, which can be specified by quantized charges $q_I,p^I$
in an integral basis $\alpha_I,\beta^I$ as $q= q^I \alpha_I + p_I \beta^I$. 
The mass of a given BPS state then follows by definition from its central charge $M=|Z|$, which is given by
\begin{equation}
\label{eq:centralcharge}
Z= e^{K/2} \langle q, \Omega \rangle  = e^{K/2} \mathbf{q}\eta \mathbf{\Pi}  \,   ,
\end{equation}
where $\mathbf{q}=(p^I,q_I)$. 
The physical charge of a BPS state is given by (see for instance \cite{Palti:2017elp})
\begin{equation}\label{eq:charge}
Q^2 = -\frac{1}{2}\, \mathbf{q}^T  \cM \mathbf{q} \, , 
\end{equation}
where we have defined 
\beq \label{def-cM}
     \cM = \begin{pmatrix}
\Im \cN + \Re\cN (\I \cN)^{-1} \R \cN & -\Re\cN (\I \cN)^{-1}\\
-\Re\cN (\I \cN)^{-1} &  (\I \cN)^{-1}
\end{pmatrix}\, .
\eeq
There is a useful  identity in $\cN=2$ supergravity theories that relates this physical charge to the central charge via \cite{Ceresole_1996}
\begin{equation}
\label{eq:N=2identity}
Q^2 = |Z|^2 + K^{i \bar{\jmath}} D_i Z D_{\bar{\jmath}} \bar{Z}\, ,
\end{equation}
where the K\"ahler covariant derivative acts on $Z$ as $D_i Z=\partial_i Z+ \frac{1}{2} (\partial_i K) Z$.  
Note that the identity \eqref{eq:N=2identity} has a simple interpretation when considered 
in the compactification setting, since it arises from a change of the real basis $\alpha_I,\beta^I$ into a basis 
of $(3,0)$- and $(2,1)$-forms given by $\Omega$ and $D_i \Omega$. The matrix $-\cM$ defined in \eqref{def-cM}  
is the Hodge star on $Y_3$ evaluated in the basis $\alpha_I,\beta^I$. We introduce the notation 
\beq \label{Hodge-product}
      \langle \alpha | \alpha'\rangle = \int_{Y_3} \bar \alpha \wedge * \alpha'\ ,\qquad  \|  \alpha\|^2= \int_{Y_3} \bar \alpha \wedge * \alpha\ ,
\eeq
for the Hodge product and the Hodge norm of three-forms $\alpha,\alpha' \in H^3(Y_3,\bbC)$. 
We then have
\beq \label{eq:normtogkfunctions}
   \langle \alpha_I | \alpha_J \rangle= - \cM_{IJ}\ , \qquad \langle \beta^I | \beta^J \rangle= - \cM^{IJ}\ , \qquad 
    \langle \alpha_I | \beta^J \rangle= - \cM_{I}^{\ J}\ . 
\eeq
Furthermore, we note that the physical charge $Q^2$ is thus related to the Hodge norm of the three-form $q$ via
\begin{equation}\label{eq:chargehodgenorm}
Q^2 = \frac{1}{2} \|q \|^2\, .
\end{equation}
We can use these facts to realize that the right-hand side of \eqref{eq:N=2identity} arises from evaluation of the Hodge star on $\Omega$ and $D_i \Omega$. Furthermore, the charge identity (\ref{eq:N=2identity}) can be recast by using the fact that $D_i(Z \bar{Z})=\partial_i (Z \bar{Z})$, i.e.~the squared norm of the central charge has zero K\"ahler weight. One obtains the form
\begin{align}\label{eq:N=2identityrewritten}
Q^2=|Z|^2 + 4  K^{i \bar{\jmath}} \partial_i |Z| \partial_{\bar{\jmath}} |Z|
\end{align}
which will be useful in later sections.

\subsection{Charge-to-mass spectrum of BPS states}
\label{ssec:examples}
Here we discuss the structure of the charge-to-mass spectrum of BPS states in 4d $\mathcal{N}=2$ supergravities. We consider primarily states with electric charge. We will review the argument of \cite{Gendler:2020dfp} that their charge-to-mass vectors lie on an ellipsoid with two non-degenerate directions, whose radii can be computed from the supergravity data. We conclude with a remark on the charge-to-mass spectrum of BPS states with generic charge.

First let us clarify what we mean by charge-to-mass vectors. In our setting we compute the physical charge of a BPS state via \eqref{eq:charge}. For electric states $\mathbf{q}=(0, q_I)$ this physical charge then follows from the right-bottom block of the matrix $\cM$. In order to determine the individual electric charges of this state, we have to decompose the matrix $\Im \cN_{IJ}$ that appears in this expression. By
introducing a symmetric matrix $G$ such that 
\beq
    - 2\, \Im \cN_{IJ}=G_{I}^{K} \delta_{KL} G^{L}_{J}\ ,
\eeq  
we then define the charge-to-mass vectors as
\begin{equation} 
    \mathfrak{z}_I = \frac{|Q|}{M} \hat Q_I\, , \label{CMVect}
\end{equation}
where $Q_I=(G^{-1})^{J}_{I} q_J$ and $\hat Q_I$ denotes the unit vector in this direction. 

The interest for these charge-to-mass vectors $ \mathfrak{z}_I$ stems from the electric Weak Gravity Conjecture. 
In the case of considering only a single gauge field,  
this conjecture states that there should always exist an electric state whose charge-to-mass ratio is larger than the black hole extremality bound. 
In a setting with multiple gauge fields, e.g.~for the general theory that we consider in \eqref{eq:action}, one has to consider the ratio between multiple electric charges and the mass of a state. In \cite{Cheung:2014vva} a generalized version of the Weak Gravity Conjecture with multiple gauge fields is proposed. Motivated by black hole remnant arguments the authors formulate a convex hull condition, which states that there should exist a set of particles such that the convex hull spanned by their charge-to-mass vectors contains the black hole extremality region. 

It is a non-trivial task to obtain the form of the extremality region in a general 4d $\mathcal{N}=2$ supergravity. A first 
step is to restrict to electric BPS black holes and determine the charge-to-mass spectrum of electric BPS states. In this case the defining equation for the shape of the charge-to-mass spectrum is the BPS condition $|Z|^2=M^2$, which can be written as
\begin{equation}
e^K \frac{(\mathbf{q}^T \eta \mathbf{\Pi} )(\mathbf{\bar \Pi}^T \eta \mathbf{q}) }{M^2}=1\, .
\end{equation}
Specializing to electric states $\mathbf{q}=(0,q_I)$ and using \eqref{CMVect}, we can express the quantized 
charge vectors $q_I$ in terms of the charge-to-mass vectors $\mathfrak{z}_I$ via $q_I=M G^J_I \mathfrak{z}_J$. The BPS condition then tells us that the charge-to-mass vectors obey
\begin{equation}\label{eq:spectrumequation}
e^K \big( \mathfrak{z}_J G_I^J X^I \big) \big(\bar{X}^K  G_K^L \mathfrak{z}_L \big) = 1\, .
\end{equation}
This condition can now be interpreted as a matrix equation for the charge-to-mass vectors. It can be written as
\begin{equation}
\mathfrak{z}_I \cA^{IJ} \mathfrak{z}_J = 1\, ,
\end{equation}
with the matrix $\mathbb{A}_E$ given by
\begin{equation}\label{eq:radiimatrix}
\cA^{IJ} =   e^K  G_K^I  \bigg(\Re X^K\, \Re X^L +\Im X^K\, \Im X^L \bigg) G_L^J \, .
\end{equation}
This tells us that the eigenvalues of this matrix specify the shape of the charge-to-mass spectrum. Looking at the form of $\cA$, we notice that this matrix has at most two non-zero eigenvalues, since there are only two independent vectors in its image. Denoting these eigenvalues by $\gamma_1^{-2}$ and $\gamma_2^{-2}$, and expanding the charge-to-mass vectors in terms of an eigenbasis for $\cA$
writing $ \mathfrak{\tilde z}_I$, we can write the BPS condition as
\begin{equation} \label{ellip1}
\gamma_1^{-2}  \mathfrak{\tilde z}_1^2 + \gamma_2^{-2} \mathfrak{ \tilde z}_2^2 =1\, ,
\end{equation}
where the components $\mathfrak{\tilde z}_i$ with $i\neq 1,2$ are unconstrained. Viewed as an equation constraining $\mathfrak{\tilde z}_I$ the condition \eqref{ellip1} describes an ellipsoid with two non-degenerate directions if $\gamma_1,\gamma_2 < \infty$. We can determine its radii $\gamma_{1,2}$ by computing the eigenvalues of the matrix $\cA$. This problem reduces to finding the eigenvalues of a $2\times 2$ matrix, by noting that only linear combinations of $ G_K^I  \Re X^K$ and $ G_K^I \Im X^K$ can be eigenvectors of $\cA$ with non-zero eigenvalues. Its eigenvalues are given by
\begin{equation}\label{eq:radiigeneral}
\gamma^{-2}_{1,2} = \frac{e^K}{R^2 I^2}  \bigg( (R^2+I^2)(R^2 I^2+P^2) \pm \sqrt{\Delta} \bigg) \, ,
\end{equation}
with
\begin{equation}
\Delta= 2R^2 I^2 \big(I^4+6R^2 I^2+R^4\big) P^2 +\big( R^4 I^4 + P^4 \big) (R^2- I^2)^2  \, ,
\end{equation}
and where we used short-hands
\begin{equation}
R^2=-  \Im \cN_{IJ} \Re X^I  \Re X^J \, ,\ \ \ I^2=- \Im \cN_{IJ}   \Im X^I  \Im X^J\, , \ \ \ P=  \Im \cN_{IJ} \Re X^I \Im X^J\, .
\end{equation}
In general, we have that $P \neq 0$. In order to set $P$ to zero one has to rescale\footnote{To be precise, this rescaling is given by $f=-(i/2) \arctan(2P/(I^2-R^2)) $. This can be verified by checking how this rescaling acts on $\Im \cN_{IJ} X^I X^J = I^2-R^2+2i P$, since it cancels out its complex phase and therefore sets the imaginary part to zero.} the period vector by $\mathbf{\Pi} \to e^f \mathbf{\Pi}$, which maps $X^I \to e^f X^I$. Note in particular that the defining equation for the shape of the charge-to-mass spectrum \eqref{eq:spectrumequation} is invariant under such rescalings, and therefore so are the formulas for the radii given in \eqref{eq:radiigeneral}. Let us point out that the function $f$ need not be holomorphic for our purposes, since the structure of the charge-to-mass spectrum does not involve derivatives. In particular the rescaling that sets $P$ to zero is generally not a holomorphic rescaling, and is therefore not a K\"ahler transformation. By using this rescaling to set $P=0$, the expressions for the radii \eqref{eq:radiigeneral} reduce to
\begin{equation}
\label{eq:radiiperiod}
\begin{aligned}
\gamma_1^{-2} &= -2e^K  \Im \cN_{IJ} \, \Re X^I \Re X^J \, ,   \\
\gamma_2^{-2} &= -2e^K  \Im \cN_{IJ} \, \Im X^I  \Im X^J \, .   \\
\end{aligned}
\end{equation}
This provides us with a simple way to compute the radii of the charge-to-mass spectrum from the supergravity data, i.e.~the K\"ahler potential $K$, the gauge kinetic functions $\Im \cN_{IJ}$ and the periods $X^I$. Note that the general $\cN=2$ special geometry identity $- 2 e^K  \Im \cN_{IJ} \,  X^I \bar X^J  = 1$ implies 
\beq \label{gamma_constr}
   \gamma_1^{-2}  +  \gamma_2^{-2} =1\ . 
\eeq
In other words the ellipsoid \eqref{ellip1} is not general but restricted by the $\cN=2$ condition \eqref{gamma_constr}.\footnote{Recently in \cite{Gonzalo:2018guu,Aalsma:2019ryi,Palti:2020qlc,Andriolo:2020lul,Loges:2020trf} supersymmetry and duality groups have been used as guiding principles in studying the Swampland.}

Let us close this subsection with two remarks. 
Firstly, the symplectic frame used to formulate this data is not necessarily the frame in which a prepotential formulation exists. Namely, we want to choose a symplectic frame in which we obtain a weakly-coupled description for the $U(1)$ gauge fields. In other words, we pick our electric charges based on the behavior of the physical charge \eqref{eq:charge}, since a small physical charge for electrically charged states indicates that the gauge kinetic functions $\Im \cN_{IJ}$ in the action \eqref{eq:action} are large. For now we assume that this choice of electric charges or symplectic frame has already been made for us. How to make this choice will be discussed in more detail in section \ref{ssec:bounds}, where we make use of an alternative manner to compute these radii that follows from \eqref{eq:radiiperiod}, namely via the charge-to-mass ratios of a particular set of electric states \eqref{eq:asymptoticradii}.

Secondly, we can also study the charge-to-mass spectrum of BPS states with generic charge. In order to define their charge-to-mass vectors we have to decompose the matrix $\cM$ that appears in the physical charge \eqref{eq:charge} via a symmetric matrix $G$ as $2\cM^{-1} = - Z Z^T$. The charge-to-mass vectors are then given by $\mathbf{z}=M Z^T \mathbf{q}$.\footnote{Note that we do not recover the electric charge-to-mass vectors $(0,\mathbf{z}_E)$ via $M Z^T (0,\mathbf{q}_E)$ due to the off-diagonal components of $\cM$. Namely, we find that $ Z^T (0,\mathbf{q}_E) \neq (0,  G^T \mathbf{q}_E)$ since application of $G^T$ on $(0,\mathbf{q}_E)$ generally results in a non-vanishing piece in the first component.} Following similar steps as in the analysis of electric states, the BPS condition can be rewritten as $\mathbf{z}^T \mathcal{Z} \mathbf{z} = 1$, 
where the matrix $\mathcal{Z}$ is given by
\begin{equation}\label{eq:radiimatrixgeneral}
\mathcal{Z} =   e^K \, Z \eta \Big(\Re \mathbf{\Pi}\  \Re \mathbf{\Pi}^T+\Im \mathbf{\Pi}\  \Im\mathbf{\Pi}^T \Big)\eta Z^T \, .
\end{equation}
Again we find a matrix with a two-dimensional image, in this case spanned by the vectors $Z\eta\Re \mathbf{\Pi}$ and $Z\eta\Im \mathbf{\Pi}$, so $\mathcal{Z}$ has only two non-vanishing eigenvalues. Let us denote these eigenvalues by $r_1^{-2}$ and $r_2^{-2}$ to avoid confusion with the eigenvalues $\gamma_1^{-2}$ and $\gamma_2^{-2}$ that were found for the electric charge-to-mass spectrum. We then obtain a similar relation for the charge-to-mass vectors by expanding in terms of an eigenbasis for $\mathcal{Z}$ as
\begin{equation}
r_1^{-2} \tilde z_1^2 + r_2^{-2} \tilde z_2^2 =1\, ,
\end{equation}
where the components $\tilde z_{\alpha}$ with $\alpha\neq 1,2$ are unconstrained. This means we are again dealing with an ellipsoid with two non-degenerate directions. We can determine the radii by computing the eigenvalues of the $2\times 2$ matrix by projecting onto the subspace spanned by $G \eta \Re \mathbf{\Pi}$ and $G \eta \Im \mathbf{\Pi}$. After some slightly involved computations, we find as radii $r_1 = r_2=1$. So we find that the ellipse forms a circle with unit radius at any point in moduli space. Note that this puts $Q/M \geq 1$ as lower bound on the charge-to-mass ratio of any BPS state, which is also expected from \eqref{eq:N=2identity}.

\subsection{Examples}
\label{ssec:examples2}
As promised, we now turn to two examples where we calculate the radii (\ref{eq:radiiperiod}) explicitly using the known prepotential formulations. 
We will highlight some of the ingredients that will play a central role in the more sophisticated general analysis of sections~\ref{sec:asympHodgetheory} and \ref{sec:generalanalysis}. Let us stress that the general approach is also essential to draw conclusions when a prepotential is hard to determine or unavailable. 

We start by recalling some relevant facts about the prepotential. The existence of the latter in a given duality frame relies on the condition that 
\begin{align}
\det \begin{pmatrix}
  X^I & D_{\imath} X^I
\end{pmatrix} \neq 0
\end{align}
In a duality frame where a prepotential $\mathcal{F}$ exists, the holomorphic functions $\mathcal{F}_I$ introduced in section \ref{ssec:examples} are given by the simple relation $\mathcal{F}_I= \partial_{X^I} \mathcal{F}$. In terms of the prepotential, the expression for the gauge kinetic function $\cN_{IJ}$ takes the form
\begin{equation}\label{eq:gaugefunctions}
\cN_{IJ} =\bar{\cF}_{IJ} + 2 i \frac{ \Im \cF_{IK}X^K \Im \cF_{JL}X^L}{\Im \cF_{MN}X^M X^N}\, , \\
\end{equation}
where $\cF_{IJ} = \partial_{X^I} \partial_{X^J} \cF$. As becomes already clear from this expression, the prepotential formulation, if available, simplifies the formulas one has to deal with and makes calculations more manageable.

\subsubsection{Example 1: conifold point}
\label{ssec:conifold}
Here we study the behavior of the charge-to-mass spectrum for an example of a finite distance singularity, namely the one-modulus conifold point. Such a singularity is realized in e.g.~the complex structure moduli space of the quintic \cite{Candelas:1990rm}. Other than for the obvious reason, which is the knowledge of the prepotential, we chose this example because in \cite{Gendler:2020dfp} only infinite distance singularities were treated. We show that the charge-to-mass spectrum of electric BPS states consists of two parallel lines separated from each other by a distance of 2. For the general asymptotic analysis of finite distance singularities, we refer to section \ref{sec:generalanalysis} and appendix \ref{app:Ichain}.  \\  \\
The conifold prepotential is given by \cite{Strominger_1995}
\begin{equation}\label{eq:conifoldprepotential}
\cF(X^0,X^1) = -i c_1 (X^0)^2 -i c_2 (X^1)^2 \log \frac{X^1}{X^0} \, ,
\end{equation}
with $c_1,c_2$ real positive constants. Then we obtain from \eqref{eq:PeriodVector} the period vector
\begin{equation}
\mathbf{\Pi}  = \begin{pmatrix}
1 \\
e^{2\pi i t} \\
-2 i c_1+ic_2 e^{4\pi i t}\\
 -\frac{c_2}{2\pi} t e^{2\pi i t} - i c_2 e^{2\pi i t}
\end{pmatrix}
\end{equation} 
where we set $X^0=1$ and $X^1=e^{2\pi i t}$. Under $t \to t+1  $ the period vector undergoes a monodromy transformation $\mathbf{\Pi}(t+1) = M \mathbf{\Pi}(t)$, with monodromy matrix
\begin{equation}\label{eq:conifoldmonodromy}
M = \begin{pmatrix}
1 & 0 & 0 & 0 \\
0 & 1 & 0 & 0 \\
0 & 0 & 1 & 0 \\
0 & -\frac{c_2}{2\pi} & 0 & 1
\end{pmatrix}.
\end{equation}
The associated log-monodromy matrix is given by $N=\log M = M-\mathbb{I}$.

From now on we will write $t=b+iv$ and set $b=0$ for simplicity, i.e.~the axions will not be relevant in what follows. By plugging the prepotential \eqref{eq:conifoldprepotential} into \eqref{eq:gaugefunctions} we find as gauge kinetic functions
\begin{equation}\label{eq:conifoldgaugecouplings}
\cN_{IJ} = i \begin{pmatrix}
 -\frac{2 c_{1}^2 e^{8 \pi  v}+c_{1} c_{2} e^{4 \pi  v} (4 \pi  v-5)+c_{2}^2 (2 \pi  v+1)}{c_{1} e^{8 \pi  v}-2 \pi  c_{2} e^{4 \pi  v} v} & \frac{c_{2} \left(4 c_{1} e^{4 \pi  v} (2 \pi  v-1)+c_{2}\right)}{c_{1} e^{6 \pi  v}-2 \pi  c_{2}
   e^{2 \pi  v} v} \\
 \frac{c_{2} \left(4 c_{1} e^{4 \pi  v} (2 \pi  v-1)+c_{2}\right)}{c_{1} e^{6 \pi  v}-2 \pi  c_{2} e^{2 \pi  v} v} & \frac{c_{2} \left(c_{1} e^{4 \pi  v} (4 \pi  v-3)+2 \pi  c_{2} v (4 \pi  v-1)+c_{2}\right)}{2 \pi  c_{2} v-c_{1} e^{4 \pi 
   v}} \\
\end{pmatrix}.
\end{equation}
Note that the $\Re \cN_{IJ} = 0$ because we set the axion to zero. The leading order part of this matrix is given by
\begin{equation}
\Im \cN_{IJ} = \begin{pmatrix}
-2 c_1 & 0 \\
0 & -4 \pi c_2  v
\end{pmatrix},
\end{equation}
where we ignored the off-diagonal components because they are exponentially suppressed for large $v$. Then we find that the leading order part of \eqref{eq:charge} becomes
\begin{equation}\label{eq:conifoldchargematrix}
\cM = \begin{pmatrix}
-2 c_1 & 0 & 0 & 0\\
0 & -4 \pi c_2  v & 0 & 0 \\
0 & 0 & -\frac{1}{2c_1} & 0 \\
0 & 0 & 0 & -\frac{1}{4\pi c_2 v}
\end{pmatrix}.
\end{equation}
In light of our later application of asymptotic Hodge theory, which is to be reviewed in section~\ref{sec:asympHodgetheory}, let us examine the form of this matrix in detail. We observe that $\cM$ takes a diagonal form, and that each diagonal component scales as a power-law in the modulus $v$. This is precisely the behavior that is predicted by asymptotic Hodge theory in \eqref{eq:growth}, and it is what makes this formalism so powerful. Namely, one is able to control the asymptotic behavior of couplings without any reference to a prepotential, which allows for general statements instead of being restricted to a particular example.

In our current choice of symplectic frame we take the electric charges to be of the form $\mathbf{q}_E=(0,0,q_0,q_1)$. Note that these are charges for which the physical charge becomes small (or finite) at the conifold point according to \eqref{eq:conifoldchargematrix}, which ensures a weakly-coupled description for the $U(1)$ gauge fields.\footnote{More generally one could pick $\mathbf{q}_E = (q_0 \sin \theta,0,q_0 \cos \theta, q_1)$ as charges for electric states. In the symplectic frame corresponding to this basis of electric charges one finds that $P=\Im \cN_{IJ} \Re X^I \Im X^J \neq 0$. This means that the expressions for the radii given in \eqref{eq:radiiperiod} no longer hold, but one should use \eqref{eq:radiigeneral} instead. However, instead of computing the radii via this formula there is another way to see that the radii do not depend on $\theta$. Namely, one can apply a K\"ahler transformation $\mathbf{\Pi} \to e^{i \theta} \mathbf{\Pi}$. This can be interpreted as a rotation of the charge vector back to $\mathbf{q}_E = (q_0 \sin \theta,0,q_0 \cos \theta, q_1) \to  (0,0,q_0, q_1)$, and it is precisely the transformation that sets $P=0$.}  Then the electric part of the period vector that couples to these charges is given by
\begin{equation}\label{eq:conifoldelectricperiods}
X^I = \big( 1,e^{- 2\pi v}  \big)\ .
\end{equation}
By inserting \eqref{eq:conifoldchargematrix} and \eqref{eq:conifoldelectricperiods} into the matrix \eqref{eq:radiimatrix} we obtain
\begin{equation}\label{eq:conifoldmatrix}
\cA = \frac{1}{2c_1}\begin{pmatrix}
2c_1 & 0  \\
0 & 4 \pi c_2 v e^{-4\pi v} \\
\end{pmatrix}\, .
\end{equation}
The eigenvalues of this matrix give the radii of the ellipsoid, and we find as asymptotic values
\begin{equation}
\gamma_1^{-2} =  1 \, , \qquad \gamma_2^{-2} = 0\, . 
\end{equation}
This structure of the charge-to-mass spectrum could also have been expected from the charge-to-mass ratios of the two states $\mathbf{q}_0 = (0,0,1,0)$ and $\mathbf{q}_1=(0,0,0,1)$. From the perspective of emergence, note that the state $\mathbf{q}_1$ is precisely the state that has to be integrated out to produce the conifold singularity \cite{Strominger_1995}. It does not couple to the polynomial part of the period vector in $t$ but to one of the exponentially suppressed terms, and this state therefore becomes massless at the singularity. Furthermore the monodromy matrix acts trivially on the charge vector, so the log-monodromy matrix annihilates $\mathbf{q}_1$ as $N \mathbf{q}_1 = 0$. This indicates that we are only dealing with a single state that has to be integrated out, instead of an entire tower that can be generated via monodromy transformations as was found for infinite distance limits in \cite{Grimm:2018ohb,Grimm:2018cpv}. We find that the leading order behavior of the charge-to-mass ratios of the states $\mathbf{q}_0,\mathbf{q}_1$ is given by
\begin{equation}
\bigg(\frac{Q}{M}\bigg)^2 \bigg|_{\mathbf{q}_0} = 1 \,, \qquad
\bigg(\frac{Q}{M}\bigg)^2 \bigg|_{\mathbf{q}_1} = \frac{2}{3 c_2}e^{4\pi v}\, . 
\end{equation}
These ratios match nicely with the structure observed for the charge-to-mass spectrum from the radii of the ellipse. On the one hand, we found that there is a state that attains the lowest value possible value for its charge-to-mass ratio. Namely, as can be seen from \eqref{eq:N=2identity}, the charge-to-mass ratio of BPS states in 4d $\mathcal{N}=2$ supergravities is always bounded from below by 1. On the other hand, we found a state for which its charge-to-mass ratio diverges at the conifold point. Together these results combine into a compelling picture: one radius diverges, and the ellipsoid degenerates into two lines separated from each other by a distance of 2. This shape can also be inferred from the radii, since the asymptotic value of the second radii is given by $\gamma_2^{-2}=0$, which means that this radius must diverge at the conifold point. It turns out that this behavior is characteristic for finite distance singularities, and we find in section \ref{sec:generalanalysis} that the ellipsoid always degenerates in this manner at finite distance points.

\subsubsection{Example 2: large complex structure point}
\label{ssec:LCS}
For our next example we turn to the large complex structure point, considering an arbitrary number of moduli. This choice of singularity allows us to study a large class of infinite distance limits all at once, since the prepotential always takes a cubic form at this point. Before we begin we should note that, depending on the form of the intersection numbers $\cK_{ijk}$ and the choice of path, there arise some subtleties in the choice of electric charges. To avoid distraction from the main purpose of the examples, we give here only the calculation for one of the two kinds of paths explicitly. The other path involves more technical details and will therefore not be considered here, but it is covered by the general analysis in section \ref{sec:generalanalysis}.

The prepotential at the large complex structure point can be conveniently written as
\begin{equation}\label{eq:LCSprepotential}
\cF(X^I) = -\frac{\cK_{ijk}X^i X^j X^k}{6X^0}\, ,
\end{equation} 
with $X^I = (X^0,X^i)$, and $\cK_{ijk}$ the intersection numbers of the mirror dual of the Calabi-Yau threefold $Y_3$. We can then write the period vector \eqref{eq:PeriodVector} as
\begin{equation}
\mathbf{\Pi} = \begin{pmatrix}
1 \\
t^i \\
\frac{1}{6}\cK_{klm}t^k t^l t^m \\
-\frac{1}{2} \cK_{ikl}t^k t^l
\end{pmatrix}, 
\end{equation}
where we used special coordinates $X^I=(1,t^i)$. In the following we write $t^i=b^i+iv^i$, and we set again the axions to zero for simplicity, i.e.~$b^i=0$. The K\"ahler potential \eqref{eq:kahlerpotential} can then be given in terms of these coordinates by
\begin{equation}
K = -\log \Big( \frac{4}{3}\cK_{ijk}v^i v^j v^k \Big) \, .
\end{equation}
Now we want to study electric BPS states at large complex structure. We can distinguish electric charges from magnetic charges by asking for what charges BPS states become light. Note that this approach deviates slightly from the prescription that will be used in section \ref{ssec:bounds}, where we look directly at how the physical charge \eqref{eq:charge} behaves asymptotically. Looking at the mass of states can become a problem when exponentially suppressed contributions to the period vector are important. These contributions can cause the mass associated with magnetic charges to vanish asymptotically, while their physical charge does diverge. For the large complex structure point this is not an issue and the two methods agree, but it can be an issue at e.g.~the conifold point, which is why we motivated our choice of electric charges via the physical charge in section \ref{ssec:conifold}. Taking a closer look at the mass of a BPS state \eqref{eq:centralcharge}, we find that
\begin{equation}
M^2 = \frac{3}{4\cK_{ijk}v^i v^j v^k}\big| q_0 + i q_i v^i + \frac{1}{6} p^0 i\cK_{ijk}v^i v^j v^k -\frac{1}{2} i \cK_{ikl}p^i v^j v^k \big|^2 \, ,
\end{equation}
where we wrote $\mathbf{q}=(q_0,q_i,p^0,p^i)$. The most natural choice of electric charges is given by $q_0,q_i$, and then $p^0,p^i$ form their dual magnetic charges. BPS states with these charges become light when asymptotically
\begin{equation}\label{eq:pathcondition}
\frac{v^i}{\sqrt{\cK_{ijk}v^i v^j v^k}} \ll 0\, .
\end{equation}
However, note that if for instance $\cK_{11i}=0$ for all $i$, then one can send the modulus $v^1$ to large complex structure at a rate much faster than all other moduli, say $v^1 \gg (v^i)^2$. In that case the charge $q_1$ is not electric, but one should consider $p^1$ as electric charge instead. One can then view \eqref{eq:pathcondition} as a constraint that specifies a certain sector of the moduli space around the large complex structure point. We only consider the charges $q_0,q_i$ to be electric in the following, and refer to the general analysis in section \ref{sec:generalanalysis} for other sectors around this singularity.

Having identified the electric charges, the electric periods that couple to these charges are simply the periods $X^I$. To compute the radii of the ellipsoid from \eqref{eq:radiiperiod}, we need to know the gauge kinetic functions. By plugging the cubic prepotential into \eqref{eq:gaugefunctions} we find that $\Re \cN_{IJ}=0$ and
\begin{equation}\label{eq:gaugecouplings}
\Im \cN_{IJ}  = -\frac{\cK}{6}\begin{pmatrix}
1 & 0 \\
0 & 4K_{ij}
\end{pmatrix}\,, \qquad K_{ij} = \partial_i \partial_{\bar{j}} K = -\frac{3}{2}\big(\frac{\cK_{ij}}{\cK}-\frac{3}{2} \frac{\cK_i \cK_j}{\cK^2} \big)\, ,
\end{equation}
where we wrote $\cK_{ij}=\cK_{ijk}v^k$, $\cK_{i}=\cK_{ijk}v^j v^k$ and $\cK=\cK_{ijk}v^i v^j v^k$.

We next evaluate the expressions for the radii \eqref{eq:radiiperiod} by 
writing $X^I=(1,iv^i)$ and using \eqref{eq:gaugecouplings} to find
\begin{equation}\label{eq:radii}
\gamma_1^{-2} = \frac{3}{2\cK} \frac{\cK}{6} =\frac{1}{4}\, ,\qquad \gamma_2^{-2} = \frac{3}{2\cK}  \frac{\cK}{2} = \frac{3}{4}\, .
\end{equation}
Let us also note that if one looks at paths that do not lie in the sector given by \eqref{eq:pathcondition}, then the radii are found to be $\gamma_1=\gamma_2=\sqrt{2}$ instead, and we elaborate further on this matter in section \ref{ssec:bounds}.

Even though large complex structure points form only a subset of all possible infinite distance singularities, we can already draw some lessons from our study of this singularity. First of all, note that the lower bound put by \eqref{eq:N=2identity} cannot be saturated, but that the charge-to-mass ratio of an electric BPS state is bounded from below by  $2/ \sqrt{3}$ or $\sqrt{2}$ instead of by 1 for these infinite distance limits.  Secondly, although the large complex structure point provides us with a large variety of infinite distance limits, we only obtain two different shapes that the ellipsoid can take. Quite remarkably, we will find that many of the observations made for the conifold point and large complex structure point apply generally, as we will see in section \ref{sec:generalanalysis}.

To conclude, we discuss how the large complex structure point provides us with infinite distance paths for which the analysis of \cite{Gendler:2020dfp} is not applicable. The reason for this was already stated by the authors of the latter work and explaining it requires us to say a few words about the mechanics behind asymptotic Hodge theory, which will be introduced in the next section. For each modulus $v^i$ that is scaled at a different rate compared to the others, one introduces an integer $d_i$. Assuming an ordered limit by $v^i \gg v^j$ for $i > j$, these integers need to satisfy $d_i \geq d_j$ if $i > j$. Furthermore, these integers are bounded from below by $d_i \geq 0$, and from above by the complex dimension of the Calabi-Yau manifold, so here $d_i \leq 3$. In the asymptotic analysis of \cite{Gendler:2020dfp} it was crucial that these integers satisfy $d_i \neq d_{i-1}$. However, this condition can clearly not be realized if one takes a limit with four different scalings of the moduli, and might not even be realized for a lower number of scalings depending on the values that the integers $d_i$ take. From this perspective, one can always find limits that are not covered by this analysis in moduli spaces with dimension $h^{2,1} \geq 4$. It is then interesting to point out that the above analysis of the large complex structure point did not require this assumption, and one is free to pick any relative scalings for the moduli. Even for the simple computation presented here that requires limits to obey \eqref{eq:pathcondition}, one can scale as many moduli at different rates as one wants, only how much these rates can differ is constrained. In the study of higher-dimensional moduli spaces there is thus a large class of limits still left unexplored, and this will be the subject of section \ref{sec:generalanalysis}.

\section{Techniques from asymptotic Hodge theory}
\label{sec:asympHodgetheory}
In the previous section we reviewed 4d $\mathcal{N}=2$ supergravities and their prepotential formulation and looked at two specific examples where the prepotential is known which allowed us to more or less directly calculate the charge to mass ratios of electric BPS states. Now we prepare for a more general approach that uses asymptotic Hodge theory and does not rely on the knowledge of a prepotential. Being a vast subject, we will only introduce the tools we need and refer the interested reader to \cite{Schmid,CKS} for the mathematical literature or \cite{Grimm:2018ohb,Grimm:2018cpv,Corvilain:2018lgw,Grimm:2019wtx,Grimm:2019ixq,Cecotti:2020rjq,Gendler:2020dfp,Lanza:2020qmt,Cecotti:2020uek,Grimm:2020cda} for recent applications in the Swampland programme. We begin by briefly reviewing the concept of Hodge structure and introduce the nilpotent orbit theorem which allows for a first approximation of the holomorphic three-form $\Omega$. Then we summarize how this formalism provides us with a classification of limits in complex structure moduli space. Last but not least, we describe how the nilpotent orbit can be further approximated by using the sl$(2)^n$-structure that emerges close to the boundary and how the presence of a pure Hodge structure at the boundary can be used to determine the numerical coefficient of the leading term in this expansion.

\subsection{Asymptotic expansion of periods as nilpotent orbits}
\label{ssec:nilpotentorbits}
In the following we describe tools to describe the general asymptotic behaviour 
of the $\cN=2$ supergravity data, discussed in section \ref{N=2basics}, for theories that arise from 
Calabi-Yau threefold compactifications. The first important ingredient will be the nilpotent orbit theorem which allows us to write down asymptotic expressions for the representatives of elements from $H^{3}(Y_3,\mathbb{C})$, in particular the holomorphic three-form $\Omega$. Before we turn to this asymptotic analysis, we rephrase the Hodge structure on the middle cohomology in a way that is more suitable for the discussion of the nilpotent orbit theorem later on.

The Hodge structure on $H^3(Y_3,\mathbb{Z})$ is usually described by the Hodge decomposition
\begin{align}
H^3(Y_3,\mathbb{C}) = \bigoplus_{k=0}^3 \,H^{k,3-k} \quad , \quad \overline{H^{p,q}} = H^{p,q}\ .
\end{align}
An equivalent definition can be given in terms of a finite decreasing filtration $F^p$, i.e.~the Hodge filtration
\begin{align}
0 \subset F^3 \subset F^2 \subset F^1 \subset F^0=H^{3}(Y_3,\mathbb{C}) \,,
\end{align}
such that
\begin{align}
F^p \cap \bar{F}^{4-p}=0 \quad \text{ and } \quad F^p \oplus \bar{F}^{4-p}=H^{3}(Y_3,\mathbb{C}) \,.
\end{align}
The relation with the Hodge decomposition is given by 
\begin{align}\label{eq:Hpqdef}
H^{p,q}= F^p  \cap \bar{F}^q \, , \qquad F^p = \bigoplus_{k=p}^{3}  H^{k,3-k}\ .
\end{align}
Due to the Calabi-Yau condition, there is (up to rescaling) a unique representative of $F^3$. Furthermore, we know that by taking derivatives with respect to the complex structure moduli we move down the filtration
\begin{align}
\partial_i F^p \subset F^{p-1} \quad i=1,\dots,h^{2,1} \label{eq:Transversality}\ .
\end{align}
Furthermore, we note that for a Calabi-Yau threefold the whole middle cohomology can be spanned by the derivatives of the unique representative of $F^3$, which in our case is provided by the holomorphic three-form $\Omega$. 

Let us now describe how to study the asymptotic regimes in complex structure moduli space  $\cM^{\text{cs}}(Y_3)$. Such an 
asymptotic regime is specified by a limit in $\cM^{\text{cs}}(Y_3)$, i.e.~is defined to be the region in the moduli space in which one or more moduli 
are pushed close to its boundary. A single parameter limit is described my moving onto a codimension-one boundary of $\cM^{\text{cs}}(Y_3)$, i.e.~a divisor, locally defined by $z_k=0$. 
Sending $n$ moduli to a limit then corresponds a higher codimension intersection of all the associated divisors, which after suitable relabeling of coordinates is locally described by $z^1=\ldots = z^n=0$.  In the following we will mostly work 
on the universal covering space of the asymptotic region and parametrize it in terms of coordinates
\begin{equation}
t^i \equiv b^i + i v^i= \frac{1}{2\pi i}\log z^i\, ,
\end{equation}
where we sometimes loosely refer to the $b^i$ as being axions and the $v^i$ as being saxions.\footnote{This notion can 
be justified for infinite distance limits by checking that $b^i \rightarrow b^i + c$ becomes an approximate 
shift symmetry.} 
The limit towards the boundary in the coordinates $t^i$ corresponds to 
\begin{equation}
t^i \to i \infty\, ,
\end{equation}
for $i=1,\ldots,n$. We note that we have already used the notation $t^i$ in section \ref{sec:Review} to denote the  
complex scalars in the vector multiplets. Let us stress, however, that we will slightly abuse notation, since the limit does not necessarily need to involve all $h^{2,1}$ moduli, but that the remaining moduli can be kept finite, which we shall henceforth denote by $\zeta^k$, with $k=n+1,\ldots, h^{2,1}$.

As we move through $\mathcal{M}^{\text{cs}}(Y_3)$ the Hodge structure varies, \emph{e.g.} the orientation of the complex line $F^3$ inside $H^3(Y_3,\mathbb{C})$ depends on the complex structure moduli and similarly for the other subspaces. A crucial point is that this variation provides some special data associated to each asymptotic regime. The first important information arises from  
the monodromy behavior of the elements $\omega^p \in F^p$. Circling the boundary divisor $z^i=0$ corresponds to sending $t^i \to t^i +1$ and it induces a monodromy transformation via\footnote{ The action of the monodromy matrix on a form is understood as an action on the basis of three-forms. The appearance of the inverse is conventional and implies that the action of monodromy on the period vector is defined as $\mathbf{\Pi}(t^i+1)= T^{-1}_i \mathbf{\Pi}(t^i)$. } 
\begin{align} \label{monodromy_v}
\omega^p(\ldots ,t^i+1,\ldots)=T_i\, \omega^p(\ldots ,t^i,\ldots ) \, ,
\end{align}
where the $T_i$ are unipotent monodromy matrices\footnote{It is known \cite{Landman} that in general these monodromy matrices are quasi-unipotent. Here we assume that the non-unipotent part has already been `removed' by appropriate coordinate redefinitions.} and sit in $Sp(2(h^{2,1}+1),\mathbb{R})$. Using the latter we define the log-monodromy matrices $N_i = \log T_i$ which are associated to each limiting coordinate $t^i$ and preserve the symplectic product, i.e.~$\langle  \, \cdot \, , N_i \, \cdot \, \rangle = - \langle N_i \, \cdot \, ,  \, \cdot \, \rangle$. Furthermore, they are nilpotent matrices of degree less or equal to four for Calabi-Yau threefolds. At this point, we have set the stage to introduce Schmid's nilpotent orbit theorem \cite{Schmid}, which states that given the filtration $F^p$ its associated nilpotent orbit
\begin{align}
F^p_{\rm nil}(t,\zeta)= e^{t^i N_i} F^p_0(\zeta) \label{eq:NilpOrbitGen}
\end{align}
still defines a proper Hodge structure on $H^3(Y_3,\mathbb{Z})$ provided that $\Im t^i \gg 0$, i.e.~we are close to the boundary. The elements of $F^p_0$ are holomorphic functions in $\zeta^k$ and do not depend on the $t^i$. In \cite{Schmid}, the author also gives a distance estimate on how well the nilpotent orbit approximates the complete filtration $F^p$. We want to remark that the elements of $F_0^p$ are far from arbitrary, as they are known to define  together with 
the $N_i$, a so called limiting mixed Hodge structure associated with the given asymptotic region \cite{CKS}. We will not go into further details about this last point, but we refer the interested reader to the referenced literature.

Having discussed the general setting, we now turn to our specific quantity of interest, namely the holomorphic three-form $\Omega$ which corresponds to a representative of $F^3$. Using the log-monodromy matrices $N_i$ associated with the boundary $t^i \to \infty$, we can write
\begin{equation}
\Omega(t,\zeta) = e^{t^i N_i} A(e^{2\pi i t},\zeta) \ . \label{eq:PeriodVectorExpansion}
\end{equation}
The three-form $A$ is holomorphic both in the coordinates $z^i = e^{2\pi i t^i}$ and the coordinates $\zeta^k$ not taken to a limit, so it can be expanded as
\begin{align}
A(e^{2\pi i t},\zeta)=a_0(\zeta) + \mathcal{O}(e^{2 \pi i t^i}) \ . \label{eq:AExpansion}
\end{align}
In general, the expression \eqref{eq:PeriodVectorExpansion} can be rather complicated to deal with and calculations quickly get out of hand. Luckily, the nilpotent orbit theorem introduced above makes things easier by identifying $a_{0}\in F^3_0$ and thus allowing us to write 
\begin{equation}
\Omega_{\text{nil}}(t,\zeta) = e^{t^i N_i} a_0(\zeta)\, , \label{eq:NilpotentOrbit}
\end{equation}
while guarantying us that the latter still describes a proper element of $F^3$ for $\Im t^i \gg 0$. In other words, the exponential terms in \eqref{eq:AExpansion} are not essential for describing a representative of $F^3$. However, it is important to note that at the same time some of the information about the other spaces $F^0,F^1,F^2$ has been dropped by taking this approximation. In that sense, it is no longer appropriate to rely on the relations \eqref{eq:Transversality} when using just $\Omega_{\text{nil}}$. From the definition \eqref{eq:NilpotentOrbit}, it is readily seen that 
\begin{align}
\partial_k (\Omega_{\text{nil}}) = e^{t^i N_i} N_k a_0(\zeta) \, , \label{eq:NilpDescend}
\end{align}
which is only non-trivial if $N_k a_{0} \neq 0$. Even if $N_k a_{0}$ is non-vanishing it is known that one only recovers all the information from $a_0$ in special cases, such as when all $h^{2,1}$ coordinates approach the large complex structure point \cite{Morrison1992}. In contrast, for finite distance singularities the expressions \eqref{eq:NilpDescend} would be zero for all $k$ so that all the non-trivial information would be lost. Hence, the derivative terms of $\Omega$ as representatives of $F^p$ should each be independently approximated by their nilpotent orbit \eqref{eq:NilpOrbitGen} to capture all the essential information. In the special geometry setting we often have derivatives of $\Omega$ that appear, so one can easily miss important terms by just plugging in the nilpotent orbit expressions $\Omega_{\text{nil}}$. An alternative to approximating the elements in $F^p$ independently is to determine how the essential information about the filtration $F^p$ is encoded in the higher order terms of \eqref{eq:AExpansion} and with that knowledge defining a `reduced' expression for $\Omega$ that consists only of these required terms. In this way, one can plug in this still much simpler version without missing out on essential information about the Hodge structure along the calculation. A step in this direction will be taken in a forthcoming project \cite{toappear}. However, for the relevant calculations in this project we will either bypass this complication or consider quantities that descend from $a_0$ in the sense of \eqref{eq:NilpDescend}. Therefore, we find that dealing with $\Omega_{\text{nil}}$ or, as explained in the next section, with an even further simplified version is enough.

\subsection{Classification of singularities in complex structure moduli space}
A useful feature of asymptotic Hodge theory is that it gives a systematic classification of the possible singularities that can occur, and also dictates how singularities can enhance when one moves towards the intersection of limiting divisors. Due to this feature, we can be sure that our analysis is completely general by covering simply all of the possible cases. While one could elaborate quite extensively on this subject, we just summarize how singularities are classified according to the properties of the relevant log-monodromy matrices and the symplectic pairing without providing a complete background. For detailed expositions on the singularity classification for Calabi-Yau threefolds, we refer the reader to \cite{Kerr2017}. 

\begin{table}[htb]\centering
\renewcommand{\arraystretch}{1.3}
\begin{tabular}{|c|c|c|}
\hline
Singularity type & Index range & Properties of $N$ and $\eta$                                                       \\ \hline 
$\text{I}_a$ & $0\leq a \leq h^{2,1}$        & \begin{minipage}[c]{0.5\textwidth}\centering \vspace{0.1cm}
$\text{rank}(N,N^2,N^3)=(a,0,0)$ \\
$\eta N$ has $a$ negative eigenvalues  \vspace{0.1cm}
\end{minipage}          \\ \hline
$\text{II}_b$   & $0\leq b \leq h^{2,1}-1$         &

\begin{minipage}[c]{0.5\textwidth}\centering \vspace{0.1cm}
 $\text{rank}(N,N^2,N^3)=(2+b,0,0)$ \\
$\eta N$ has $b$ negative and two positive eigenvalues \vspace{0.1cm}
\end{minipage}   \\ \hline
\rule[-.25cm]{.0cm}{.7cm} $\text{III}_c$   & $0\leq c \leq h^{2,1}-2$        & $\text{rank}(N,N^2,N^3)=(4+c,2,0)$                                                 \\ \hline
\rule[-.25cm]{.0cm}{.7cm} $\text{IV}_d$    & $1\leq d \leq h^{2,1}$         & $\text{rank}(N,N^2,N^3)=(2+d,2,1)$                                                 \\ \hline
\end{tabular}
\renewcommand{\arraystretch}{1}
\caption{Classification of singularity types according to the properties of $N$ and $\eta$. For $\mathrm{III}_c$ and $\mathrm{IV}_d$ singularities we do not need to know the signature of $\eta N$ to tell the types apart from each other.}
\label{table:singularitytypes}
\end{table}

Considering a limit involving the moduli $t^1,\ldots, t^k$, the singularity type associated with this limit can be determined from the log-monodromy matrices $N_1,\ldots, N_k$ and the symplectic pairing matrix $\eta$. One can take any linear combination of these nilpotent matrices with positive coefficients, which we denote by $N=c_1 N_1+\ldots+c_k N_k$ with $c_i>0$, and classify the limit based on the properties of $N$ and $\eta$ as given in table \ref{table:singularitytypes}. The resulting singularity type does not depend on the choice of positive coefficients $c_i$, hence we take simply $N=N_{(k)}=N_1+ \ldots + N_k$. One finds one of the $4h^{2,1}$ different types of singularities, denoted by
\begin{equation}
\mathrm{I}_a,\ \ \mathrm{II}_b ,\ \ \mathrm{III}_c ,\ \ \mathrm{IV}_d ,\ \
\end{equation}
where the range for the subindices $a,b,c,d$ is listed in table \ref{table:singularitytypes}.

Besides the properties of the matrices $N$ and $\eta$, the main types $\mathrm{I}$, $\mathrm{II}$, $\mathrm{III}$ and $\mathrm{IV}$ can also be characterized by the asymptotic behavior given by the nilpotent orbit approximation \eqref{eq:NilpotentOrbit}. Namely, the singularity type fixes the number of nilpotent matrices that can be applied on the form $a_{0}$ before it vanishes. Considering a limit involving moduli $t^1,\ldots,t^k$, one finds that
\begin{equation}\label{eq:didef}
N_{(k)}^{d_k} a_0 \neq 0\, ,
\end{equation}
where $d_k=0,1,2,3$ correspond to the four main types $\mathrm{I}$, $\mathrm{II}$, $\mathrm{III}$ and $\mathrm{IV}$  respectively. Based on this estimate for the holomorphic three-form, it can be argued that all $\mathrm{I}_a$ singularities lie at finite distance, whereas infinite distance points are necessarily on type $\mathrm{II}_b$, $\mathrm{III}_c$ and $\mathrm{IV}_d$ singularities \cite{Wang1}. The subscript on the type 
does not play a role in this discussion and is not fixed by \eqref{eq:didef}.

For one-modulus limits a single singularity type suffices to characterize the limit, but when moduli scale at different rates asymptotic Hodge theory provides us with a more refined structure. We can think of a limit with $v^1 \gg v^2 \gg \ldots \gg v^n$ as an ordered limit, where we first take $v^1 \to \infty$, thereafter $v^2 \to \infty$, up to $v^n \to \infty$. One can then determine the singularity types associated with each of the matrices $N_{(i)}$. The picture that emerges is a singularity type that enhances as we send additional moduli to their limit, resulting in an \textit{enhancement chain}
\begin{equation}\label{eq:enhancementchain}
\text{I}_0\xrightarrow{\ t^{1} \rightarrow i\infty\ }\  {{\sf Type\ A}_{(1)}}\ \xrightarrow{\ t^{2} \rightarrow i\infty\ }\  {\sf Type\ A}_{(2)} \ \xrightarrow{\ t^{3} \rightarrow i\infty\ }\  
\ldots \ \xrightarrow{\ t^{n} \rightarrow i\infty\ }\  {\sf Type\ A}_{(n)} \, ,
\end{equation}
where $\mathrm{A}_{(i)}$ denotes the singularity type associated with sending $t^1,...,t^i \to i \infty$, i.e.~one of the types $\mathrm{I}_a$, $\mathrm{II}_b$, $\mathrm{III}_c$, $\mathrm{IV}_d$. The steps that occur in these enhancement chains can be constrained \cite{Kerr2017}, and not all enhancements are 
actually possible. Most important for our purposes is the fact that the Latin number labelling the main types can only stay equal or increase. In terms
of the $d_i$ introduced in \eqref{eq:didef} we see that in an enhancement chain \eqref{eq:enhancementchain} we have 
\beq \label{d-bound}
   0 \leq d_1 \leq ... \leq d_n \leq 3\ .  
\eeq
Clearly, this implies that for limits involving many moduli it will often be the case that $d_i = d_{i+1}$.

For the purposes of this work we do not need to know the subindices $a,b,c,d$ that appear in the enhancement chain. We are only interested in vectors that descend from $a_0$ by application of nilpotent matrices $N_i$, and hence we only need to know the integers $d_i$ given in \eqref{eq:didef}. We therefore use the following replacements for segments of the enhancement chain
\begin{equation}
\begin{aligned}
\mathrm{I}  &\equiv  \mathrm{I}_{a_1} \to \ldots \to \mathrm{I}_{a_p}\, , \\
\mathrm{II}  &\equiv  \mathrm{II}_{b_1} \to \ldots \to \mathrm{II}_{b_q}\, ,\\
\mathrm{III}  &\equiv  \mathrm{III}_{c_1} \to \ldots \to \mathrm{III}_{c_r}\, ,\\
\mathrm{IV}  &\equiv  \mathrm{IV}_{d_1} \to \ldots \to \mathrm{IV}_{d_s}\, . \\
\end{aligned}
\end{equation}
Note that enhancement chains always start from $\mathrm{I}_0$, so when using the shorthand notation we always start from the segment $\mathrm{I}$ with the subscript $a_1=0$. Going through all possible enhancement chains by either including or excluding the segments $\mathrm{II}$, $\mathrm{III}$ and $\mathrm{IV}$, we find that we have to consider eight different kinds of enhancement chains in total.

\subsection{A special sl(2)-split three-form basis and strict asymptotic behavior}
The powerful machinery of asymptotic Hodge theory does not stop with the nilpotent orbit formulation. There are two more key structures that can be associated to a given $n$-parameter limit; (1) a set of $n$ commuting $sl(2,\mathbb{R})$-algebras, and (2) a unique Hodge decomposition of $H^{3}(Y_{3},\mathbb{C})$ in terms of $H^{p,q}_{\infty}$. These structures allow us for instance to pick a special basis of three-forms that decomposes $H^3(Y_3,\mathbb{R})$ into a direct sum of subspaces, where three-forms grouped in the same subspace are characterized by similar asymptotic behavior as we approach the degeneration loci in $\mathcal{M}^{cs}(Y_3)$.  In the following we will only give a brief outline of the underlying formalism
\cite{CKS}, see for example \cite{Grimm:2018cpv} for a more detailed exposition and an explicitly computed example.

In order to apply these techniques, we first have to divide the moduli space around the singularity into sectors. These sectors are dubbed \textit{growth sectors}, since the growth estimates provided by asymptotic Hodge theory apply for limits that lie within these sectors. One such growth sector is given by\footnote{As an aside, let us already mention that we will need to divide these growth sectors into even more refined subsectors in section \ref{ssec:bounds}, when we study the charge-to-mass spectrum of electric BPS states. Namely, the physical charge of an electric BPS state should become small (or finite) asymptotically, which requires us to treat the relative scaling of the moduli in more detail.}
\begin{equation}\label{eq:growthsector}
\cR_{12\cdots n} = \big\{t^i = b^i+iv^i\,\Big|\, v^1\geq  v^2 \, \ldots \geq  v^n > 1, \, | b^i | <1 \big\}\, .
\end{equation}
The other sectors can be obtained by permuting the moduli $t^i$. The collection of all such sets $\cR_{i_1...i_n}$ covers the asymptotic 
region. Larger values in the directions $b^i$ can be obtained by using the transformations $b^i \rightarrow b^i +1$ with an action given in \eqref{monodromy_v}.     
Note that picking a growth sector fixes a particular ordering for the moduli $t^i$, since each coordinate $v^i$ is bounded from below by the next coordinate $v^{i+1}$. 
Hence, in every growth sector there is a sequential limit, where we first send $t^1 \to i \infty$, then $t^2 \to i \infty$, up to $t^n \to i \infty$. Enhancement chains as described in \eqref{eq:enhancementchain} therefore naturally characterize limits taken within a growth sector. 

Given a growth sector, we can now associate various structures to its asymptotic boundary. The first structure we discuss makes use of $sl(2,\mathbb{R})$-algebras. Taking the log-monodromy matrices $N_i$ and the filtration $F^p_0$ as input data, together with the ordering specified by the growth sector, it was shown in \cite{CKS} that one can construct
\begin{equation}
\text{$n$ commuting $sl(2,\mathbb{R})$-triples:}\qquad (N_i^{-}, N_i^{+}, Y_i) \, , \quad i=1,\ldots,n\, .
\end{equation}
These $sl(2,\mathbb{R})$-triples satisfy the standard commutation relations 
\begin{equation}
[Y_i, N_i ^{\pm}]= \pm 2 N_i^{\pm}\, , \quad [N_i^+,N_i^-]=Y_i\, .
\end{equation}
The procedure to obtain these $sl(2,\mathbb{R})$-triples is rather non-trivial. We therefore refer the reader to \cite{Grimm:2018cpv} for a detailed review on their construction, where the authors also worked out an explicit example. In this work we simply assume the required steps have already been performed, and that the resulting $sl(2,\mathbb{R})$-triples are handed to us.

We can now use these $sl(2,\mathbb{R})$-algebras to decompose $H^3(Y_3,\mathbb{R})$ into eigenspaces of their weight operators $Y_{i}$. This decomposition is independent of the moduli $t^i$ but can vary with changes in the spectator moduli $\zeta^{k}$. It can be written as
\begin{equation}
H^3(Y_3,\mathbb{R}) = \bigoplus_{\Bell \in \cE} V_{\Bell}\, , \qquad \Bell = (\ell_1, \dots , \ell_n)\, , \label{eq:Sl2Decomp} 
\end{equation}
where the integers $\ell_i \in \{0, \dots , 6 \}$ denote the eigenvalues of $Y_{(i)}=Y_1+\dots+Y_i $, i.e.
\begin{equation}
v_{\Bell} \in V_{\Bell}:\qquad Y_{(i)} v_{\Bell} = (\ell_i-3) v_{\Bell}\, .\label{eq:LevelOp}
\end{equation}
In this decomposition we use $\cE$ to denote the set of all labels $\Bell$ indicating non-empty spaces $V_{\Bell}$. The values that these integers $\ell_i$ can take depend on the details of the singularity under consideration. For instance the range for the integers $\ell_i$ is determined by the singularity type associated with $N_{(i)}$. For type $\mathrm{I}_a$ and $\mathrm{II}_b$ singularities we find as range $\ell_i=2,\ldots,4$, for $\mathrm{III}_c$ singularities $\ell_i=1,\ldots,5$, and for $\mathrm{IV}_d$ singularities $\ell_i = 0,\ldots,6$.

For later reference, let us record a few useful relations that can be obtained for these eigenspaces $V_{\Bell}$. For example, one can use the commutation relations of the $sl(2,\mathbb{R})$-triples to show that
\begin{equation}\label{eq:lowering}
N^-_i V_{\Bell} \subseteq V_{\Bell'} \text{ with }\Bell'=(\ell_1,\ldots,\ell_{i-1},\ell_i-2,\ldots,\ell_n-2)\, .
\end{equation}
We can also write down orthogonality conditions between the subspaces $V_{\Bell}$ with respect to the symplectic pairing. For two elements $w_{\Bell}\in V_{\Bell}$ and $w_{\mathbf{r}} \in V_{\mathbf{r}}$ we find that
\begin{equation}\label{eq:orthogonality}
\langle w_{\Bell} , w_{\mathbf{r}} \rangle = 0\, , \quad \text{unless $\ell_i+r_i=6$ for all $i$}\, ,
\end{equation}
which can be shown by using that $\langle \cdot, Y_i \cdot \rangle  = - \langle Y_i \cdot, \cdot \rangle$. 

The other key structure that arises at the boundary is provided by a boundary Hodge decomposition of the space of three-forms $H^3(Y_3,\mathbb{C})$, given by
\begin{equation}\label{eq:boundaryHodge}
H^3(Y_3,\mathbb{C}) = \bigoplus H^{p,3-p}_\infty\, ,
\end{equation}
where $\overline{H^{p,q}_\infty}=H^{q,p}_\infty$. This decomposition is independent of the moduli $t^i$ sent to their limit, but still varies with changes in the moduli $\zeta^k$ that are kept finite, similar to the decomposition \eqref{eq:Sl2Decomp} following from the $sl(2,\mathbb{R})$-algebras. There exists a boundary Weil operator $C_\infty$ associated with this Hodge decomposition, whose dependence on the moduli $\zeta^k$ we suppress. It acts on individual elements $w^{p,q} \in H^{p,q}_\infty$ as
\begin{equation}\label{eq:Cinftyaction}
C_{\infty} w^{p,q} = i^{p-q} w^{p,q}\, .
\end{equation}
We can relate the subspaces $H^{p,q}_\infty$ to the filtration $F^p_0$ that appeared in the nilpotent orbit \eqref{eq:NilpotentOrbit}. First we construct a new filtration $\tilde{F}^p_0$ via two matrices $\zeta',\delta$, given by 
\begin{equation}\label{eq:Ftilde}
\tilde{F}^p_0 = e^{\zeta'}e^{i\delta}F^p_0\, .
\end{equation} 
These matrices $\zeta',\delta$ play an important role in the construction of the $sl(2,\mathbb{R})$-triples, so we refer again to \cite{Grimm:2018cpv} for their precise form. Both the filtration $F^p_0$ and the filtration $\tilde{F}^p_0$ do not define a pure Hodge structure via intersections of the sorts of \eqref{eq:Hpqdef}. However, by using the lowering operators $N_i^-$ we can construct a new filtration that does define a pure Hodge structure. This filtration can be written as
\begin{equation}\label{eq:boundaryfiltration}
F^p_\infty = e^{iN^-_{(n)}} \tilde{F}^p_0\, .
\end{equation}
The boundary Hodge structure $H^{p,q}_\infty$ is then obtained via
\begin{equation}
H^{p,q}_\infty = F^p_\infty \cap \bar{F}^q_\infty \, .
\end{equation}
We can use the Weil operator $C_\infty$ associated with this boundary Hodge decomposition to define an inner product on $H^3(Y_3,\mathbb{C})$. This product satisfies the orthogonality condition
\begin{equation}\label{eq:Cinftyorthogonality}
\langle C_{\infty} w_{\Bell},  w_{\Bell'} \rangle = 0\, , \quad \text{unless $\Bell = \Bell'$}\, .
\end{equation}
Note that this orthogonality condition tells us that $C_{\infty}$ maps $V_{\Bell}$ to $V_{\mathbf{6}-\Bell}$, as can be seen from \eqref{eq:orthogonality}.

Let us now elaborate on the role of the three-form $a_0$ in these structures. As a representative of $F^3_0$, we can use the matrices $\zeta',\delta$ to rotate $a_0$ to
\begin{equation}
\tilde{a}_0 = e^{\zeta'}e^{i\delta} a_0\, .
\end{equation}
This three-form $\tilde{a}_0$ has a well-defined location in one of the eigenspaces $V_{\Bell}$, to be precise
\begin{equation}\label{eq:a0position}
\Re \tilde{a}_0\, , \, \Im \tilde{a}_0 \ \in V_{\mathbf{3}+\mathbf{d}}\, , 
\end{equation}
where $ \mathbf{d}=(d_1, \dots , d_n)$, with the $d_i$ defined in \eqref{eq:didef}. Following \eqref{eq:boundaryfiltration} we can apply lowering operators $N_i^-$ to construct another three-form out of $a_0$. This three-form can be placed in one of the subspaces $H^{p,q}_\infty$ of the boundary Hodge decomposition. It is given by
\begin{equation}
\Omega_\infty \equiv e^{iN_{(n)}^-} \tilde{a}_0 \in H^{3,0}_\infty\, .
\end{equation}
From \eqref{eq:Cinftyaction} we then know that $C_\infty \Omega_\infty = -i\Omega_\infty$, and we can use this relation to fix the action of $C_{\infty}$ on $\tilde{a}_0$ and its descendants $N_i^- \tilde{a}_0$, $N_i^- N_j^- \tilde{a}_0$, $N_i^- N_j^- N_k^- \tilde{a}_0$. Its action is compactly summarized by the identity\footnote{This identity is derived by expanding the exponential in $e^{iN_{(n)}^-} \tilde{a}_0$. The weights of the various terms with respect to $Y_{(i)}$ follow from the weights of $\tilde{a}_0$ given in \eqref{eq:a0position}, together with how the $N_i^-$ lower the weights as described in \eqref{eq:lowering}. By using that the Weil operator acts as $C_{\infty} V_{\Bell} \subseteq V_{\mathbf{6}-\Bell}$ one can then match terms with the same weights, resulting in the given identity.} 
\begin{equation}\label{eq:Cinftyidentity}
 \boxed{\quad \rule[-.4cm]{0cm}{1.1cm}  
 C_{\infty}\,\prod_{i=1}^n \frac{i^{k_i}}{k_i!}(N_i^-)^{k_i} \, \tilde{a}_0= -i \prod_{i=1}^n \frac{i^{d_i-d_{i-1}-k_i}}{(d_i-d_{i-1}-k_i)!} (N_i^-)^{d_i-d_{i-1}-k_i}\, \tilde{ a}_0\, , \quad}
\end{equation}
This identity will be crucial in determining charge-to-mass ratios in section \ref{sec:generalanalysis}. Let us also point out its similarity with the relation $\ast J^k/k! = J^{n-k}/(n-k)!$, as can be used for the K\"ahler form $J$ of a Calabi-Yau $n$-fold. 

Having covered the boundary structure, we now move slightly away from the singular loci in order to discuss the dependence on the coordinates $v^i$. In section \ref{ssec:nilpotentorbits} we already discussed some techniques to control this moduli dependence, when we introduced the nilpotent orbit approximation which allowed us to drop corrections in $e^{2\pi i t^i}$. This region corresponds to taking $v^{i}\gg 1$, which we refer to as the \textit{asymptotic regime}. One gains even more control if one considers the sector
\beq \label{strictgrowthsector}
   \cR^{\rm strict}_{12\cdots n} = \Big\{t^i = b^i+iv^i\,\Big|\,\frac{v^1}{v^2} > \gamma,\, \ldots , \,\frac{v^{n-1}}{v^n}>\gamma,\, v^n> \gamma; \,|b^i| <1 \Big\}_{\gamma \gg 1}\, ,
\eeq
which we call the \textit{strict asymptotic regime} in the growth sector \eqref{eq:growthsector}. 
In this strict asymptotic regime one is then allowed to drop subleading polynomial terms in $v^{i+1}/v^{i}$ and $1/v^{n} $ as well. The mathematical structure that captures the moduli dependence in this regime is provided by the so-called sl(2)-orbit, which we will discuss in the remainder or this section. A summary of the different regimes is given in table \ref{table:regimes}.

\begin{table}[htb]
\centering
\renewcommand{\arraystretch}{1.3}
\begin{tabular}{|c||c|c|c|}
\hline
regime & asymptotic & strict asymptotic & boundary \\
\hline \hline
validity & $e^{2\pi i t^i} \ll 1$ & $v^{1} \gg ... \gg v^n \gg 1$ &  $t^i = i \infty$ \\ \hline
filtration & $F^p_{\rm nil}$ & $F^p_{\rm sl(2)}$ & $F^p_{\infty}$ \\
\hline
\end{tabular}
\renewcommand{\arraystretch}{1}
\caption{Summary of the different regimes. We indicated the corrections that can be dropped in each of these regimes, as well as what limiting structure ($F^p_{\rm nil}$, $F^p_{\rm sl(2)}$ or $F^p_{\infty}$) should be used to describe these regimes.}\label{table:regimes} 
\end{table}

Before turning to the sl(2)-orbit approximation, let us stress that for both approximations one should be careful with the order of taking derivatives and dropping the corrections. Recall from the discussion below~\eqref{eq:NilpDescend} that the nilpotent orbit approximation for just the holomorphic $(3,0)$-form $\Omega$ in terms of $a_{0}$ did not necessarily suffice to study the whole space $H^{3}(Y_{3},\mathbb{C})$. Similarly its sl(2)-orbit approximation in terms of $\tilde{a}_{0}$ does not necessarily provide the complete picture. Rather one should take the limit to the strict asymptotic regime for each of the subspaces $F^{p}_{\rm nil}$, similar to how we moved from $F^{p}$ to $F^{p}_{\rm nil}$ before. 

Let us begin by introducing the sl(2)-orbit itself. It can be obtained from the filtration $\tilde{F}^{p}_{0}$ defined in \eqref{eq:Ftilde} via
\begin{equation}
F^{p}_{\rm sl(2)}(v, \zeta) = e^{iv^i N^{-}_{i}} \tilde{F}^{p}_{0}(\zeta^{k})  \, .
\end{equation}
The sl(2)-orbit theorem then states that the spaces $F^{p}_{\rm sl(2)}$ approximate the nilpotent orbit $F^{p}_{\rm nil}$ \eqref{eq:NilpotentOrbit} in the strict asymptotic regime where $\gamma \gg 1$. We can make this statement more precise by introducing the operator\footnote{As an example, application of $e(v)$ on a three-form in one of the eigenspaces $V_{\Bell}$ simply multiplies it with $(\frac{v^{1}}{v^{2}})^{\frac{\ell_1-3}{2}} \cdots (\frac{v^{n-1}}{v^{n}})^{\frac{\ell_{n-1}-3}{2}} (v^{n})^{\frac{\ell_{n}-3}{2}}$.} 
\begin{align}\label{eq:evdef}
e(v)= \exp \Big( \frac{1}{2} \sum_{j=1}^{n-1}  \log(\frac{v^j}{v^{j+1}})Y_{(j)} + \frac{1}{2} \log(v^n )Y_{(n)} \Big)\, .
\end{align}
We can use $e(v)$ to rewrite the sl(2)-orbit $F^{p}_{\rm sl(2)}$ in terms of the boundary filtration $F^{p}_{\infty}$ as
\begin{equation}
F^{p}_{\rm sl(2)}(v, \zeta) = e^{{-1}}(v) F^{p}_{\infty}(\zeta^{k})\, .
\end{equation}
The approximation of the nilpotent orbit by the sl(2)-orbit can then be seen by applying $e(v)$ on the spaces $F^{p}_{\rm nil}$ and taking the limit $ \gamma \gg 1$ in \eqref{strictgrowthsector}. Namely, it can be shown that
\begin{equation}
\lim_{\gamma \to \infty} e(v) F^{p}_{\rm nil}= F^{p}_{\infty}\, .
\end{equation}
where we require the axions $b^{i}$ to remain bounded as in \eqref{eq:growthsector}, \eqref{strictgrowthsector} above. 

Note that the above discussion concerned spaces of three-forms, hence we were free to apply rescalings when application of $e(v)$ yielded an overall factor that vanishes or diverges asymptotically. If one is interested in a specific three-form, one has to account for this rescaling by hand. For example, the sl(2)-orbit approximation of the holomorphic $(3,0)$-form $\Omega$ is given by
\begin{align}\label{eq:sl2orbit3form}
\Omega_{\rm sl(2)} = \Big(\frac{v^{1}}{v^{2}}\Big)^{\frac{d_1}{2}} \cdots \Big(\frac{v^{n-1}}{v^{n}}\Big)^{\frac{d_{n-1}}{2}} (v^{n})^{\frac{d_{n}}{2}} e^{-1}(v)  \Omega_{\infty}\, .
\end{align}
An important corollary of the sl(2)-orbit theorem is that it tells us how the Hodge norm \eqref{Hodge-product} behaves asymptotically. It is at this stage that the decomposition of $H^3(Y_3,\mathbb{R})$ into eigenspaces $V_{\Bell}$ becomes useful. Namely, for a vector $w_{\Bell}\in V_{\Bell}$ it can be shown that the strict asymptotic behavior of its Hodge norm is given by
\begin{equation}\label{eq:growth}
\| w_{\Bell} \|^{2} = \langle w_{\ell} , C_{\rm sl(2)} w_{\ell} \rangle  + \cO\Big(\frac{v^{i+1}}{v^i}\Big) \, ,
\end{equation}
where we have introduced the sl(2) hodge norm 
\begin{align}
\langle w_{\ell} , C_{\rm sl(2)} w_{\ell} \rangle  = (v^1)^{\ell_1-3}(v^2)^{\ell_2-\ell_1}\cdots (v^n)^{\ell_n-\ell_{n-1}} \langle w_{\Bell}, C_{\infty} w_{\Bell} \rangle .
\end{align}
We introduced the so-called sl(2) Weil operator $C_{\rm sl(2)}$ captures the coordinate dependence of the strict asymptotic Hodge norm. This is related to its counterpart at the boundary by 
\begin{align}
C_{\rm sl(2)}= e^{-1}(v) C_{\infty} e(v) \,. \label{eq:Csl2ToCInf}
\end{align}
Note that the orthogonality property given in \eqref{eq:Cinftyorthogonality} tells us that the Hodge norm decomposes into blocks in the strict asymptotic regime.

As an application, one can use these techniques to determine for instance the leading order behavior of the K\"ahler potential \eqref{eq:kahlerpotential} in the strict asymptotic regime. We find that
\begin{equation}\label{eq:Kahlerpotasymp}
K_{\rm sl(2)}=-\log i \langle \Omega_{\infty} , \bar{\Omega}_{\infty} \rangle \, (v^1)^{d_1} (v^2)^{d_2-d_1} \cdots (v^n)^{d_n-d_{n-1}} \, ,
\end{equation}
where the integers $d_{i}$ are defined by \eqref{eq:didef}. We will need to know the precise form of this coefficient $\langle \Omega_{\infty} , \bar{\Omega}_{\infty} \rangle$ at a later stage to determine the charge-to-mass ratios of particular BPS states, so let us record that
\begin{equation}\label{eq:Kinf}
\langle \Omega_{\infty} , \bar{\Omega}_{\infty} \rangle = 2^{d_n} i^{d_n}\langle  (N_1^{-})^{d_1} (N_2^{-})^{d_2-d_1} \cdots (N_n^{-})^{d_n-d_{n-1}} \tilde{a}_0 \, , \ \bar{\tilde{a}}_{0}  
\rangle\, . 
\end{equation}
It is tempting to now try to compute the K\"ahler metric by taking derivatives of \eqref{eq:Kahlerpotasymp}. However, as we stressed before, one should be careful with interchanging the order of taking derivatives and taking limits. In fact, the K\"ahler metric provides us with a good example why this order matters. Namely, \eqref{eq:Kahlerpotasymp} does not depend on the coordinate $v^{i}$ whenever $d_{i}-d_{i-1}=0$. The integers $d_{i}$ are bounded by $0 \leq d_{i} \leq d_{i+1} \leq 3$ as already noted in \eqref{d-bound}, so this already happens if one considers limits where four or more moduli are scaled at different rates. In particular, note that $d_{i}-d_{i-1}=0$ is therefore not only specific to finite distance singularities, but can also occur for infinite distance singularities such as the large complex structure point we discussed in section \ref{ssec:LCS}. Thus simply taking \eqref{eq:Kahlerpotasymp} can result in a degenerate K\"ahler metric, and one should include corrections when necessary in order to resolve this issue.

\section{Asymptotic analysis of the charge-to-mass spectrum}\label{sec:generalanalysis}
Here we study the charge-to-mass spectrum of BPS states for \textit{any} limit in complex structure moduli space, both at finite and infinite distance. In order to perform this analysis we apply the machinery introduced in the previous section. First we derive a formula for the charge-to-mass ratio of sl(2)-elementary BPS states. 
We then apply this formula to give general bounds on the electric charge-to-mass spectrum. These bounds are obtained by computing the radii of the ellipsoid that is spanned by the charge-to-mass vectors of electric BPS states. The values found for these radii are listed in table \ref{table:WGCradii}. To illustrate these results, we conclude by considering some examples where we demonstrate how to use this formula for charge-to-mass ratios.

\subsection{Formula for asymptotic charge-to-mass ratios}
In this work we want to put bounds on the charge-to-mass spectrum of BPS states. In general one finds that the charge-to-mass ratio of a state depends in a non-trivial manner on the complex structure moduli. This behavior simplifies when we move towards the boundary of moduli space, where we can give a precise description of how the charges and masses of BPS states scale in the moduli via the techniques introduced in section \ref{sec:asympHodgetheory}. The aim of this paper is therefore to study BPS states in these limits in moduli space, and obtain asymptotic bounds on their charge-to-mass ratios in this way.

For a BPS state with a generic set of charges, it is however still a rather complicated problem to give its asymptotic charge-to-mass ratio. In order to simplify this problem, we turn to the sl$(2)^n$-splitting \eqref{eq:Sl2Decomp} that plays a central role in asymptotic Hodge theory. This splitting decomposes the charge space $H^3(Y_3,\mathbb{R})$ into irreducible sl$(2)^n$ representations, where the weights $\ell_i$ ($i=1,\ldots,n$) of a charge fix the scaling in the moduli via equations such as \eqref{eq:growth}. We restrict our attention for now to states that can be specified by a single set of weights $\ell_i$, which were referred to as  single-charge states in \cite{Gendler:2020dfp}, but we will adopt the name \textit{sl(2)-elementary} states. Let us denote the set of charges for these states by
\begin{equation}
\cQ_{\rm sl(2)} = \{ q \in H^3(Y_3,\mathbb{R}) \, | \ q \in V_{\Bell} \text{ for some } \Bell \}\, .  \label{eq:SetElementary}
\end{equation}
The space $\cQ_{\rm sl(2)} $ is a union of vector spaces if we consider the charges to be continuous. We note that this 
split can also be performed over the rational numbers to accommodate quantized charges, but we will not address this issue any further in the following. 
At first, the restriction to this particular set of states limits the generality of our results. However, let us point out that a formula for the charge-to-mass ratio of these sl$(2)$-elementary  states suffices to obtain the 
asymptotic shape of the charge-to-mass spectrum of all electric BPS states, as we will see in section \ref{ssec:bounds}.

Let us examine the asymptotic behavior of the charge-to-mass ratio for a candidate 
sl$(2)$-elementary  BPS state $q_{\Bell} \in V_{\Bell}$ piece by piece. By using the growth theorem \eqref{eq:growth} we find that the physical charge \eqref{eq:chargehodgenorm} of this state asymptotes in the strict asymptotic regime \eqref{strictgrowthsector} to
\begin{equation}
\label{eq:piece1}
Q^2 = -\frac{1}{2} (v^1)^{\ell_1-3} (v^2)^{\ell_2-\ell_1} \cdots (v^n)^{\ell_n - \ell_{n-1}} \langle q_{\Bell},\, C_{\infty} q_{\Bell} \rangle + \cO\Big(\frac{v^{i+1}}{v^i}\Big) \, ,
\end{equation}
where we remind the reader that $C_{\infty}$ is the Weil operator associated with the boundary. The mass of a BPS state $M(q_{\Bell})$, given in \eqref{eq:centralcharge}, consists of two factors. The first one involves the K\"ahler potential, and by using \eqref{eq:Kahlerpotasymp} we find that its leading term is given by
\begin{equation}
\label{eq:piece2}
e^{-K} = (v^1)^{d_1} (v^2)^{d_2-d_1}\cdots (v^n)^{d_n-d_{n-1}}\ i \langle \Omega_{\infty} , \bar{\Omega}_{\infty} \rangle + \cO\Big(\frac{v^{i+1}}{v^i}\Big) \, .  
\end{equation}
The second factor asymptotes to
\begin{equation}
\label{eq:piece3}
| \langle q_{\Bell}, \Omega \rangle |^2 = (v^1)^{\ell_1+d_1-3} (v^2)^{\ell_2+d_2-\ell_1-d_1} \cdots (v^n)^{\ell_n+d_n-\ell_{n-1}-d_{n-1}}  \ |\langle q_{\Bell},  \Omega_\infty \rangle |^2 + \cO\Big(\frac{v^{i+1}}{v^i}\Big) \, .
\end{equation}
This follows by using the sl$(2)$-orbit approximation for the $(3,0)$-form $\Omega$ given in \eqref{eq:sl2orbit3form} 
which is valid in the strict asymptotic regime \eqref{strictgrowthsector}. The operator $e^{-1}(v)$ defined in \eqref{eq:evdef} can be moved to the other side via $\langle e(v) \cdot, \cdot \rangle = \langle \cdot, e^{-1}(v) \cdot \rangle$, after which it can be applied on the charge $q_{\Bell}$ to obtain part of the parametrical scaling.

Now we can put the pieces of the charge-to-mass ratio back together. When we compare their scalings in the moduli, we find that the factors of $v^i$ cancel out precisely. However, this relies crucially on the coefficients of these leading terms being non-zero. The coefficients of $|Q|^2$ and $e^{-K}$ are indeed non-zero, since both can be interpreted as a vector norm computed with the metric $\langle \cdot, C_{\infty} \bar{\cdot} \rangle$, where we note that $C_\infty \Omega_\infty = -i\Omega_\infty$. However, the coefficient in \eqref{eq:piece3} is trickier, and we require that
\begin{equation}
\label{eq:graviphotoncoupling}
\langle q_{\Bell} , \ \Omega_\infty  \rangle \neq 0\, .
\end{equation}
This quantity has the natural interpretation as the asymptotic coupling of the state to the graviphoton. We can see this by looking at the scaling of the charge-to-mass ratio for states for which \eqref{eq:graviphotoncoupling} vanishes. Namely, when this product is zero, a term subleading to \eqref{eq:piece3} sets the asymptotic behavior of the mass $M(q_{\Bell})$. Previously the scaling of the different pieces of the charge-to-mass ratio precisely matched, so now $|Q|$ grows parametrically compared to $M(q_{\Bell})$. This means that the charge-to-mass ratio of such states must diverge along the limit, which leads us to consider \eqref{eq:graviphotoncoupling} as the asymptotic coupling to the graviphoton.

For sl$(2)$-elementary states with a non-vanishing coupling to the graviphoton, we find that the charge-to-mass ratio is given by
\begin{equation}
\label{eq:centralchargeasymptotically2}
\bigg( \frac{Q}{M}\bigg)^2 \bigg|_{q_{\Bell}} =  \frac{\langle q_{\Bell}, C_{\infty} q_{\Bell} \rangle \ i \langle \Omega_{\infty} , \bar{\Omega}_{\infty} \rangle }{2 | \langle q_{\Bell}, \Omega_{\infty} \rangle |^2} + \cO\Big(\frac{v^{i+1}}{v^i}\Big)  \, .
\end{equation}
To compute this ratio, one first has to identify the sl$(2)$-elementary  states that have a non-vanishing coupling to the graviphoton, i.e.~charges satisfying \eqref{eq:graviphotoncoupling}. However, this condition does not fix a unique set of charges. One is free to add any charges with a vanishing coupling to charges with a non-vanishing coupling to the asymptotic graviphoton, so we have to specify how we pick these charges. A natural choice is to consider charges that sit in the same irreducible sl$(2)^n$ representation as the asymptotic graviphoton. These charges can be obtained from $\tilde{a}_0$ by applying lowering operators $N_i^-$. We write this set of charges as
\begin{equation}\label{eq:gravitystates}
\cQ_{\rm G} = \{ q \in \cQ_{\rm sl(2)} \, | \ q = (N_1^-)^{k_1}  \cdots (N_n^-)^{k_n} \, (a \Re \tilde{ a}_0 + b  \Im \tilde{a}_0) \text{ with $a,b \in \mathbb{R}$}   \} \, .
\end{equation}
For the remaining charges we define the set
\begin{equation}\label{eq:fieldstates}
\cQ_{\rm F} = \{ q \in \cQ_{\rm sl(2)} \, | \ \langle q , \, \Omega_{\infty} \rangle = 0   \} \, .
\end{equation}
Together $\cQ_{\rm G}$ and $\cQ_{\rm F} $ provide us with a complete basis for the charges of BPS states
and have been discussed in \cite{Grimm:2018ohb,Grimm:2018cpv} in the context of the distance conjecture. One can then apply identity \eqref{eq:Cinftyaction} to compute the charge-to-mass ratio for the states in $\cQ_{\rm G}$, the details of which have been moved to appendix \ref{app:charge-to-mass}. In the end, one finds that the charge-to-mass ratio of an sl(2)-elementary state with non-vanishing coupling to the graviphoton is given by the formula
\begin{equation}\label{eq:chargetomass}
\boxed{\quad \rule[-.5cm]{.0cm}{1.2cm} \lim_{\gamma \to \infty} \bigg( \frac{Q}{M} \bigg)^{-2} \bigg|_{q_{\rm G}}= 2^{1-d_n}  \prod_{i=1}^n {\Delta d_i\choose{(\Delta d_i - \Delta \ell_i)/2}} \times \begin{cases}
1 \text{ for $d_n = 3$}\, ,\\
\frac{1}{2} \text{ for $d_n \neq 3$}\, ,
\end{cases}\, }
\end{equation}
where $\gamma$ denotes the constant involved in the definition of the strict asymptotic regimes \eqref{strictgrowthsector} and we used the abbreviations $\Delta d_i = d_i - d_{i-1}$ and $\Delta \ell_i = \ell_i - \ell_{i-1}$. We stress that sending $\gamma \rightarrow \infty$ can 
also be viewed as performing a consecutive limit sending $v^1 \rightarrow \infty$, then $v^2 \rightarrow \infty$, up to $v^n \rightarrow \infty$.
A different order of limits requires to consider another choice of sector  \eqref{strictgrowthsector} and will, in general, change the 
integers appearing in \eqref{eq:chargetomass}. The formula \eqref{eq:chargetomass} admits a straightforward generalization 
for any Calabi-Yau $D$-fold as we show in appendix \eqref{eq:Cinftyaction}. Explicitly, we find  
\beq
\lim_{\gamma \to \infty} \bigg( \frac{Q}{M} \bigg)^{-2} \bigg|_{q_{\rm G}}= 2^{1-d_n}  \prod_{i=1}^n {\Delta d_i\choose{(\Delta d_i - \Delta \ell_i)/2}} \times \begin{cases}
1 \text{ for $d_n = D$}\, ,\\
\frac{1}{2} \text{ for $d_n \neq D$}\, ,
\end{cases}\, \label{eq:GeneralChargetomass}
\eeq
which trivially agrees with \eqref{eq:chargetomass} when setting $D=3$. While we will not use this formula in this generality any further, it is nice to see 
that the same general pattern arises in any dimension. 

Before we continue, let us briefly summarize our findings. We studied the asymptotic behavior of the charge-to-mass ratio for sl(2)-elementary BPS states. We found that this behavior depends crucially on whether the charges of these states couple to the asymptotic graviphoton via \eqref{eq:graviphotoncoupling} or not. When this coupling vanishes the charge-to-mass ratio diverges, whereas if this coupling is non-vanishing the charge-to-mass ratio stays finite and is given by \eqref{eq:chargetomass}. This formula expresses the charge-to-mass ratio purely in terms of the discrete data $d_i,\ell_i$ that characterizes the limit and the choice of sl(2)-elementary state. In particular, note that these charge-to-mass ratios are independent of the spectator moduli $\zeta^k$ that are not taken to a limit and therefore remain constant to leading order, up to suppressed corrections in $v^{i+1}/v^{i}$ and $1/v^{n}$.

Let us now take a closer look at these charge-to-mass ratios given in \eqref{eq:chargetomass}. First of all, it is interesting to point out that this formula even applies for limits with $\Delta d_i = 0$, since it involves binomial coefficients. This is the upshot of working with the boundary Hodge structure via identities such as \eqref{eq:Cinftyidentity}, instead of an asymptotic approximation for the K\"ahler metric that follows from \eqref{eq:Kahlerpotasymp}. Secondly, notice that the charge-to-mass ratios are symmetric under
\begin{equation}\label{eq:electromagneticduality}
\Delta \ell_i \to -\Delta \ell_i:\quad \frac{Q}{M} \to \frac{Q}{M}\, .
\end{equation}
This symmetry has a natural interpretation from a physics perspective, since it tells us that dual electric and magnetic states have the same charge-to-mass ratio. Namely, recall from the orthogonality condition \eqref{eq:orthogonality} that dual electric and magnetic charges are related by $\ell_i \to 6 - \ell_i$, which is equivalent to sending $\Delta \ell_i \to -\Delta \ell_i$.

The formula we presented in \eqref{eq:chargetomass} is only applicable for the charge-to-mass ratios of sl(2)-elementary states, but one might wonder if it can be extended to apply for BPS states with generic charges. Ideally one could simply identify its elementary charge with the largest parametrical growth according to \eqref{eq:growth}, and argue that this elementary charge fixes its charge-to-mass ratio. However, when looking more carefully at the spaces $V_{\Bell}$ in which these elementary charges reside, one realizes that things can become more complicated. The first issue arises when the parametrical growth associated with two (or more) of these spaces via \eqref{eq:growth} is the same for a given path. In that case both charges contribute to the charge-to-mass ratio of the state, such that one ends up with some combination between their charge-to-mass ratios. Another issue arises when we try to add one of the charges in \eqref{eq:fieldstates} that does not couple to the asymptotic graviphoton to a charge in \eqref{eq:gravitystates} that does have a non-vanishing coupling. Assuming that both charges lie in the same eigenspace $V_{\Bell}$, we find that this added charge does contribute asymptotically to the physical charge of the state but not its mass, so the charge-to-mass ratio changes. It would be interesting to see what the generalized formula for the charge-to-mass ratio that resolves these issues looks like, but this lies beyond the scope of this work. We will, however, argue in the next subsection that our results for sl(2)-elementary states allows us to make statements about the asymptotic shape of the  general charge-to-mass spectrum of electric BPS states.

\subsection{Asymptotic shape of the electric charge-to-mass spectrum}\label{ssec:bounds}
Now that we have derived a formula for the charge-to-mass ratio of sl(2)-elementary BPS states with \eqref{eq:chargetomass}, we can put it to use to determine more properties of the charge-to-mass spectrum of electric BPS states. An elegant way to do so was given in \cite{Gendler:2020dfp}, where it was shown that the charge-to-mass vectors of electric BPS states lie on an ellipsoid with at most two nondegenerate directions. We will determine the asymptotic shape of this ellipsoid, by deriving the asymptotic values for its radii. It turns out that these radii can then be determined from the charge-to-mass ratios of electric sl(2)-elementary states via \eqref{eq:asymptoticradii}. Besides specifying a structure for the electric charge-to-mass spectrum, this also provides the limiting value for the smallest radius as lower bound on the asymptotic charge-to-mass ratio of any electric BPS state. 

In order to study the charge-to-mass spectrum of electric BPS states, we first have to establish how we can identify electric charges. When we look at the prepotential formulation of 4d $\mathcal{N}=2$ supergravities, a natural choice is to pick the charges $q_I$ that couple to the periods $X^I$ in \eqref{eq:centralcharge}. However, this method is not suitable for our purposes. First of all, the general techniques that we borrow from asymptotic Hodge theory simply do not make use of a prepotential. Secondly, one wants the physical charge \eqref{eq:charge} of an electric BPS state to be small in order to provide a weakly-coupled description for the $U(1)$ gauge fields. For instance, if we recall our analysis of the large complex structure point in section \ref{ssec:LCS}, we found that some of the charges $q_I$ had to be replaced by $p^I$ as electric charges when considering limits outside the sector \eqref{eq:pathcondition}. This teaches us that we should study the asymptotic behavior of the physical charge \eqref{eq:charge} carefully in order to identify the electric charges correctly. As a first step let us therefore take  sl(2)-elementary states as basis for the electric charges, since the parametrical behavior of their physical charges is described by \eqref{eq:growth}.  A complication that can then arise is that the physical charge of a BPS state does not diverge or vanish asymptotically, but stays finite instead. In that case one can use that sl(2)-elementary states with finite physical charge come in pairs that are each others electro-magnetic dual, as can be seen by using \eqref{eq:orthogonality} and \eqref{eq:growth}. This allows us to pick the electric charge out of each pair by hand, which in particular means that our choice of electric charges is not necessarily unique.

The task that remains is then to fix a sector in complex structure moduli space such that we know precisely what sl(2)-elementary charges are electric and magnetic. We find that we can specify these sectors simply by imposing constraints on the scalings of the moduli. To begin with we limit ourselves to considering strict asymptotic regimes, which already restricts the saxions $v^i$ via the constraints given in \eqref{eq:growthsector} with $\gamma \gg 1$. Subsequently we want electric states to have an asymptotically vanishing physical charge, which leads to additional conditions such as $(v^{1})^2\gg v^2$ by imposing \eqref{eq:growth} to decrease for a given set of $\ell_i$. In practice the allowed values for $\ell_i$ are fixed by the type of singularity under consideration, so one can systematically determine all possible subsectors. While we do not outline a procedure to construct these subsectors here, let us refer to section \ref{ssec:II-III-IV} where we show how this works in an example, and to appendix \ref{app:radii} where we perform the general analysis.  

To be more precise, let us summarize the above conditions that specify the choice of electric charges in terms of an equation. We define the set of elementary electric charges in a given sector by 
\begin{align}\label{eq:defelectric}
\cQ_{\text{el}}= \{ q, q' \in \cQ_{\text{sl(2)}} \, \, | \, \, \|q \|^2 , \|q'\|^2 < \infty \text{ and } \langle q , q' \rangle =0 \}\, ,
\end{align}
where the mutual non-locality condition makes sure that among the electric charges we picked, none of the ones with finite physical charge are dual to each other. As stated above, there can of course be more than one possibility to make this choice, so this definition of $\cQ_{\text{el}}$ does not define a unique set. Having defined our space of elementary electric charges, we can define the dual magnetic charges by application of $C_{\infty}$ as follows
\begin{align}
\cQ_{\text{mag}}= C_{\infty} \cQ_{\text{el}} \,. \label{eq:EMduality}
\end{align}
Note that this choice of magnetic charges ensures that products between electric and magnetic charges computed with the asymptotic Hodge norm $\langle \cdot, C_{\infty} \cdot \rangle $ vanish, as can be shown by using that $C_{\infty}^{2}=-1$. In other words, the gauge kinetic functions $\Re \cN_{IJ}$ that describe the coupling between electric and magnetic charges via \eqref{def-cM} vanish in the strict asymptotic regime. This follows from the asymptotic behavior of the Hodge norm \eqref{eq:growth}, and by expressing the Hodge norm in terms of the gauge kinetic functions via \eqref{eq:normtogkfunctions}.

In order to compute the radii of the electric charge-to-mass spectrum we now want to make a particular choice of symplectic 
basis $(\tilde \alpha_I, \tilde \beta^J)$, $I=1,...,h^{2,1}+1$. We first pick linearly independent $(\tilde \alpha_I,\tilde \beta^J)$ 
such that 
\beq
  \tilde \alpha_I \in \cQ_{\rm el}\ , \qquad \tilde \beta^I \in \cQ_{\rm mag}\ . 
\eeq 
Crucially, we make sure that these elements satisfy a number of further conditions that 
will be useful below.  
As a start we pick a basis that preserves the splitting in terms of the sets $\mathcal{Q}_{\rm G}$ and $\mathcal{Q}_{\rm F}$, the reasons for which are twofold. On the one hand, this splits the charges based on whether they couple to the asymptotic graviphoton or not, which provides us with a precise description of their charge-to-mass ratios via expressions such as \eqref{eq:chargetomass}. On the other hand, it proves to be useful to pick a basis that diagonalizes the gauge kinetic functions $\Im \cN_{IJ}$ in the strict asymptotic regime.  The advantage of splitting our basis elements in $\mathcal{Q}_{\rm G}$ and $\mathcal{Q}_{\rm F}$ is then that mixed terms between these subsets vanish in this setting. This follows from expressing the gauge kinetic functions in terms of the Hodge norm via \eqref{eq:normtogkfunctions}, which in turn can be described by the boundary Hodge norm $\langle \cdot, C_{\infty} \cdot \rangle$ via the approximation \eqref{eq:growth}.\footnote{It can then be argued that mixed terms vanish from the fact that $C_{\infty}$ maps $\mathcal{Q}_{\rm G}$ back into $\mathcal{Q}_{\rm G}$ as follows from \eqref{eq:Cinftyidentity}, and that charges in $\mathcal{Q}_{\rm F}$ have a vanishing symplectic product with elements of $\mathcal{Q}_{\rm G}$ by construction.} 

Let us now construct a particular basis for the charges in $\cQ_{\rm el}  \cap \mathcal{Q}_{\rm G}$ in detail. 
This basis will make up parts of the elements $\tilde \alpha_I$. 
From the expressions for the radii given in \eqref{eq:radiiperiod} we know that the coupling of these charges to the real and imaginary parts of the holomorphic $(3,0)$-form $\Omega$ plays an important role. This motivatives us to define our basis elements via
\begin{align}
\mathcal{Q}_{\Re}&= \{ q \in \cQ_{\rm el} \cap \cQ_{\rm G} \, \,  | \, \,  \text{linearly independent and }  \langle q , \Im (\Omega_{\infty}) \rangle =0 \}\, ,  \nonumber \\
\mathcal{Q}_{\Im}&= \{ q \in \cQ_{\rm el} \cap \cQ_{\rm G}  \, \,  | \, \,  \text{linearly independent and }  \langle q , \Re (\Omega_{\infty}) \rangle =0 \} \, . \label{eq:ReImBasisSets}
\end{align}
It can be argued that this choice of basis provides us with a diagonalization for the boundary Hodge norm $\langle \cdot, C_{\infty} \cdot \rangle$.\footnote{To be precise, this follows from the action of $C_{\infty}$ on elements of $Q_{\rm G}$ as given by \eqref{eq:Cinftyidentity}, where one also needs to use that $C_{\infty}$ is a real map together with the orthogonality conditions \eqref{eq:orthogonality} and polarization conditions \eqref{eq:polarization1} and \eqref{eq:polarization2}.} In principle one can then complete the basis $\tilde \alpha_I$ for $\mathcal{Q}_{\rm el}$ by picking linearly independent elements of $\cQ_{\rm el} \cap \cQ_{\rm F}$ that also diagonalize $\langle \cdot, C_{\infty} \cdot \rangle$, but for our purposes we do not need to derive more explicit expressions for these charges.

The above choice of basis $\tilde \alpha_I$ for the electric charges allows us to diagonalize the gauge kinetic functions $\Im \cN_{ IJ}$ in the strict asymptotic regime, since they can be expressed in terms of the Hodge norm via \eqref{eq:normtogkfunctions}. The appropriate quantity to describe the Hodge norm in the strict asymptotic regime is the sl(2)-norm $\langle \cdot, C_{\rm sl(2)} \cdot \rangle $ introduced in \eqref{eq:growth}, so let us write the gauge kinetic functions as
\begin{align}
\langle \tilde \alpha_I , C_{\rm sl(2)} \tilde \alpha_I \rangle = (\Im \tilde{\cN}_{II})^{-1} \, , \label{eq:Csl2DiagonalElements}
\end{align}
where the matrix $ \tilde{\cN}_{IJ}$ is the gauge coupling function associated to the sl(2)-orbit.
 
We now express the radii \eqref{eq:radiigeneral} in terms of charge-to-mass ratios of sl(2)-elementary electric charges in the strict asymptotic regime (see table \ref{table:regimes} for a reminder of this notion). As the derivation is analogous for both radii we will only be explicit for $\gamma_1$. We start by rewriting the expression for $\gamma_1^{-2}$ in the sl(2)-basis outlined above as
\begin{equation}
\begin{aligned}
\gamma_1^{-2} &=-2 e^{K_{\rm sl(2)}} \Im \tilde{\cN}_{II} \Re \tilde{X}^I \Re \tilde{X}^I + \cO\Big(\frac{v^{i+1}}{v^i}\Big)  \, ,
\end{aligned}
\end{equation}
where the $\tilde{X}^I$ are obtained by expanding $\Omega_{\rm sl(2)}$ along $\tilde \alpha_I$, and the quantities $K_{\rm sl(2)}, \tilde{\cN}_{II}$
are the $\cN=2$ data associated to the sl(2)-orbit. We can now manipulate this strict asymptotic expression using the techniques introduced in section \ref{sec:asympHodgetheory}, which gives
\begin{align}
-2 e^{K_{\rm sl(2)}} \Im \tilde{\cN}_{II} \Re \tilde{X}^I \Re \tilde{X}^I  &=- 2 e^{K_{\rm sl(2)}} \sum_{I} \frac{\langle \tilde \alpha_I , \Re \Omega_{\rm sl(2)} \rangle^2 }{\langle \tilde \alpha_I , C_{\rm sl(2)} \tilde \alpha_I \rangle} \nonumber \\ 
&= - 2 e^{K_{\rm sl(2)}} \sum_{I} \frac{\langle e(v) \tilde \alpha_I , e(v)\Re \Omega_{\rm sl(2)} \rangle^2 }{ \langle e(v) \tilde \alpha_I , C_{\infty} e(v) \tilde \alpha_I \rangle} \nonumber \\
&=-2  \sum_{I} \frac{\langle \tilde \alpha_I , \Re \Omega_{\infty} \rangle^2 }{\langle \Omega_{\infty} , \bar{\Omega}_{\infty} \rangle \langle \tilde \alpha_I , C_{\infty} \tilde \alpha_I \rangle}  \nonumber \\
&=  -2   \sum_{  \tilde \alpha_I \in \mathcal{Q}_{\Re} } \frac{\langle \tilde \alpha_I , \Omega_{\infty} \rangle^2 }{\langle \Omega_{\infty} , \bar{\Omega}_{\infty} \rangle \langle \tilde \alpha_I , C_{\infty} \tilde \alpha_I \rangle} \, . \label{eq:radDerivation}
\end{align}
In the first step, we used \eqref{eq:Csl2DiagonalElements} and reformulated things in the language of forms. In the second step, we inserted the identity in the form of $e^{-1}(v) e(v)$ into the symplectic products and used that $\langle e^{-1}(v) \cdot , \cdot \rangle= \langle \cdot , e(v) \cdot \rangle$. For the third equality, we used \eqref{eq:sl2orbit3form}-\eqref{eq:Kinf} which makes clear that all the parametrical scaling cancels out. In the last step, we used the defining property of the set $\cQ_{\Re}$. We want to emphasize that the basis charges sitting in $\cQ_{\rm el} \cap \cQ_{F}$ can be safely ignored here as they give a vanishing contribution to the sum. We recognize that \eqref{eq:radDerivation} is a sum over the inversed of strict asymptotic charge-to-mass ratios \eqref{eq:centralchargeasymptotically2}. As the strict asymptotic expression for $\gamma_2^{-2}$ can be rewritten in a similar manner, we directly state the result from \cite{Gendler:2020dfp}
\begin{align} \label{eq:asymptoticradii} 
\gamma_1^{-2}= \sum_{\tilde \alpha_I \in \cQ_{\Re}} \bigg( \frac{Q}{M} \bigg)^{-2} \bigg|_{q=\tilde \alpha_I} \!\!\!\!+\cO\Big(\frac{v^{i+1}}{v^i}\Big)\, , \quad \quad \gamma_2^{-2}=\sum_{\tilde \alpha_I \in \cQ_{\Im}} \bigg( \frac{Q}{M} \bigg)^{-2} \bigg|_{q=\tilde \alpha_I} \!\!\!\!+\cO\Big(\frac{v^{i+1}}{v^i}\Big)\,.
\end{align}
We can provide a quick check of our formula for the charge-to-mass ratios of sl(2)-elementary states \eqref{eq:chargetomass} by verifying the $\mathcal{N}=2$ constraint \eqref{gamma_constr} on the radii. To derive this relation, we will make use of a well known identity for binomial coefficients, which in our specific setup reads
\begin{equation}
\sum_{\Delta \ell_i}  {{\Delta d_i}\choose{\frac{\Delta d_i-\Delta \ell_i}{2}}} = 2^{\Delta d_i}\, .
\end{equation}
The sum $\gamma_1^{-2}+ \gamma_2^{-2}$ amounts to adding up the inverse squares of charge-to-mass ratios for all electric states, as can be seen from \eqref{eq:asymptoticradii}. Since we found that the charge-to-mass ratios are the same for dual electric and magnetic charges, one can just as well sum over all charges and compensate by dividing by two. Then one finds that
\begin{equation}
\begin{aligned}
\gamma_1^{-2}+ \gamma_2^{-2} &= \frac{1}{2} \sum_{\Bell} \bigg(\frac{Q}{M}\bigg)^{-2} \times \begin{cases}
1 \text{ for $d_n = 3$}\, ,\\
2 \text{ for $d_n \neq 3$}\, ,
\end{cases} \\
&= 2^{-d_n} \prod_{i=1}^n \sum_{\Delta\ell_i}  {{\Delta d_i}\choose{\frac{\Delta d_i-\Delta \ell_i}{2}}}  \\
&= 2^{-d_n} \prod_{i=1}^n 2^{\Delta d_i} = 1\, ,
\end{aligned}
\end{equation}
Here the extra factor of two in the first line for the case that $d_n \neq 3$ follows from the fact that each sl(2)-level site is populated by two states. Namely, one has both states coming from applying lowering operators $N_i^-$ on $\Re \tilde{a}_0$ and on $\Im \tilde{a}_0$, whereas for $d_n=3$ they only come from the real three-form $\tilde{a}_0$. In the next line this factor of two cancels against the factor of two that has to be included in the expression for the charge-to-mass ratio in \eqref{eq:chargetomass}.

This relation already gives us some insight into the bounds for the charge-to-mass ratio. As mentioned before, the smallest radius of the ellipsoid serves as lower asymptotic bound on the charge-to-mass spectrum via
\begin{equation}\label{eq:radiibound}
 \frac{Q}{M} \bigg|_{\rm asym} \gtrsim \min(\gamma_1,\gamma_2)\, .
\end{equation}
Let us briefly explain what we mean by this asymptotic bound. Firstly, we can consider the bound after taking the asymptotic limit, i.e.~to the boundary of the moduli space, by taking the limit $\gamma \rightarrow \infty$ in \eqref{strictgrowthsector} or sending consecutively $v^1,...,v^n \rightarrow \infty$.  
In this limit \eqref{eq:radiibound} turns into a proper inequality and we can replace $\gtrsim$ with $\geq$. However, as soon as we go away from the boundary, there are corrections to the ratio as seen in \eqref{eq:asymptoticradii}. These are suppressed in the strict asymptotic regime \eqref{strictgrowthsector}, but we did not infer any information about the signs of these corrections. Therefore, also any bound on the general expression for 
$\frac{Q}{M}$ for an electric  state can have already in the strict asymptotic regime small 
corrections that become increasingly irrelevant near the boundary. In the following 
we will compute the radii $\gamma_1,\gamma_2$ for all possible asymptotic limits. 
Before doing so let us briefly note that \eqref{gamma_constr} implies that the radii are bounded from below by $\gamma_{1,2} \geq 1$. This tells us that the charge-to-mass ratio of any electric BPS state is bounded by
\begin{equation}\label{eq:generallowerbound}
 \frac{Q}{M}  \geq 1\, .
\end{equation}
This nicely agrees with our knowledge from 4d $\mathcal{N}=2$ supergravities, since \eqref{eq:N=2identity} predicts the same lower bound.

\begin{table}[]
\centering
\begin{tabular}{|c|c|c|c|}
\hline
enhancement chain & subsector & $\gamma_1^{-2}$ & $\gamma_2^{-2}$ \\
\hline \hline
$\mathrm{I}$ & & 1 & 0 \\ \hline
$\mathrm{I} \to \mathrm{II}$ & & 1/2 & 1/2 \\ \hline
$\mathrm{I} \to \mathrm{III}$ & & 3/4 & 1/4 \\ \hline
$\mathrm{I} \to \mathrm{II} \to \mathrm{III}$ & & 1/2 & 1/2  \\ \hline
$\mathrm{I} \to \mathrm{IV}$ & & 3/4 & 1/4 \\ \hline
$\mathrm{I} \to \mathrm{II} \to \mathrm{IV}$ & \begin{minipage}{2.5cm}\vspace{0.1cm}\centering
$v^{\mathrm{II}} \gg (v^\mathrm{IV})^2 $ \\ 
$v^{\mathrm{II}} \ll (v^\mathrm{IV})^2 $ \vspace{0.1cm}
\end{minipage} & \begin{minipage}{1cm}\centering
1/2 \\
3/4
\end{minipage} & \begin{minipage}{1cm}\centering
1/2 \\
1/4
\end{minipage} \\ \hline
$\mathrm{I} \to \mathrm{III} \to \mathrm{IV}$ & & 3/4 & 1/4  \\ \hline
$\mathrm{I} \to \mathrm{II} \to \mathrm{III} \to \mathrm{IV}$ &\begin{minipage}{2.5cm}\vspace{0.1cm}\centering
$v^{\mathrm{II}} \gg v^\mathrm{III} v^\mathrm{IV} $ \\
$v^{\mathrm{II}} \ll v^\mathrm{III} v^\mathrm{IV} $ \vspace{0.1cm}
\end{minipage} & \begin{minipage}{1cm}\centering
1/2 \\
3/4
\end{minipage} & \begin{minipage}{1cm}\centering
1/2 \\
1/4
\end{minipage} \\ \hline
\end{tabular}
\caption{Asymptotic values for the radii of the ellipsoid that forms the charge-to-mass spectrum of electric BPS states. These values both depend on the enhancement chain that characterizes the limit, and on the subsector of the strict asymptotic regime that is being considered \eqref{strictgrowthsector}. For the latter we indicated the constraints on scaling of the saxions when relevant, where $v^{\mathrm{A}}$ corresponds to the saxion that sources an increase to roman numeral $\mathrm{A}$ for the enhancement chain.}\label{table:WGCradii} 
\end{table}

Let us now make the bound \eqref{eq:radiibound} precise by explicitly computing the radii. In order 
to do that we 
 apply \eqref{eq:asymptoticradii} to determine the radii from the charge-to-mass ratios of sl(2)-elementary states. We can then turn to our formula for charge-to-mass ratios given in \eqref{eq:chargetomass}, and go through all possible limits by considering all possible enhancement chains. These results are summarized in table \ref{table:WGCradii}, and the details are included in appendix \ref{app:radii}. We found only three different sets of values for the radii $\gamma_1,\gamma_2$, namely
\begin{equation}
(\gamma_1^{-2},\gamma_2^{-2})=\quad (1,0),\quad (\frac{3}{4},\frac{1}{4}),\quad (\frac{1}{2},\frac{1}{2})\, .
\end{equation}
Let us briefly elaborate on the sector-dependence of the results given in table \ref{table:WGCradii}. The underlying reason is that depending on in what subsector of the strict asymptotic regime we are, we pick a different set of charges for our electric states. To be more precise, the constraints given in \ref{table:WGCradii} ensure that for each eigenspace $V_{\Bell}$ we can make a definite statement about whether its asymptotic Hodge norm given in \eqref{eq:growth} diverges, stays finite or vanishes asymptotically. We need this information because we require the physical charge of our electric states to be bounded as described by \eqref{eq:defelectric}.  In fact, the limtis towards large complex structure that we ignored in section \ref{ssec:LCS} are precisely those that lie in the sectors $v^{\mathrm{II}} \gg (v^\mathrm{IV})^2 $ or $v^{\mathrm{II}} \gg v^\mathrm{III} v^\mathrm{IV} $. In other words, we deduce from table \ref{table:WGCradii} that the radii are given by $\gamma_{1}=\gamma_{2}=\sqrt{2}$ when we take a limit outside of the sector \eqref{eq:pathcondition}.

Having determined the radii for all possible limits, let us first look at the differences between finite and infinite distance limits. Finite distance limits only involve $\mathrm{I}_a$ singularities, whereas infinite distance limits include one of the other types of singularities, i.e.~$\mathrm{II}_b$, $\mathrm{III}_c$ or $\mathrm{IV}_d$. Then we observe from table \ref{table:WGCradii} that one of the radii always diverges for finite distance singularities, resulting in an ellipse that degenerates into two lines, separated from each other by a distance of 2. On the other hand, for infinite distance limits both radii remain finite, and we find either an ellipse with radii $\gamma_1=2/\sqrt{3}$ and $\gamma_2=2$, or a circle with radius $\gamma_1=\gamma_2=\sqrt{2}$. 

We can now use these values for the radii to bound the charge-to-mass ratio of electric BPS states based on the singularity under consideration. For finite distance limits we find that the lower bound for charge-to-mass ratios given in \eqref{eq:generallowerbound} can be saturated, since the smallest radius is given by $\gamma_1=1$, so we do not obtain a new bound. For infinite distance limits we do find new bounds, and depending on the limit one obtains $Q/M \geq 2/\sqrt{3}$ or $Q/M \geq \sqrt{2}$ as bound for the charge-to-mass ratio. In either case, the charge-to-mass ratio is bounded from below by
\begin{equation}
 \frac{Q}{M}  \geq \frac{2}{\sqrt{3}}\, .
\end{equation} 
Before we move on to the examples, it is interesting to point out that the state with minimal charge-to-mass ratio need not be a sl(2)-elementary state. Namely, when the sum over charge-to-mass ratios for one of the radii in \eqref{eq:asymptoticradii} runs over only one state, then the charge-to-mass ratio of this state is equal to the radius. But once the sum runs over multiple states, we find that for none of these states the charge-to-mass ratio can be equal to the radius. In fact, since there are $h^{2,1}+1$ electric charges, one finds for $h^{2,1}>1$ that for at least one of the radii multiple charges should contribute, so the state corresponding to this radius cannot be sl(2)-elementary. As an example we want to point out the LCS point discussed in section \ref{ssec:LCS}. There the smallest radius $\gamma_2$ in equation \eqref{eq:radii} is obtained by summing over multiple periods $\Im(\mathbf{\Pi}_E)$, and thus the state with minimal charge-to-mass must be realized as a linear combination of sl(2)-elementary states. It is quite remarkable that the charge-to-mass ratios of sl(2)-elementary states fix this minimal charge-to-mass ratio via \eqref{eq:asymptoticradii}, even though the formula for the charge-to-mass ratio \eqref{eq:chargetomass} only applies for sl(2)-elementary states.

\subsection{Examples}\label{ssec:formalexamples}  
Here we consider a few examples to demonstrate how asymptotic Hodge theory can be used to describe the charge-to-mass spectrum of BPS states in practice. We focus on sl(2)-elementary states that couple to the asymptotic graviphoton as defined by \eqref{eq:gravitystates}. We discuss the properties of these states in detail, such as how their charge-to-mass ratios can be obtained from formula \eqref{eq:chargetomass}, and how one can compute the radii of the electric charge-to-mass spectrum from this data. The first example puts the conifold point discussed in section \ref{ssec:conifold} into a more general perspective. The other three examples highlight different aspects of infinite distance limits, and in particular shed some light on the differences between our formula \eqref{eq:chargetomass} and the result of \cite{Gendler:2020dfp}. Let us also point out that each of these infinite distance limits can be realized as a different path to the same singularity in moduli space, namely a particular large complex structure point with $h^{2,1}=3$.\footnote{ \label{footnote:CICY} To be precise, in \cite{Grimm:2019bey} it was found that these limits could be realized at the large volume point (mirror of the large complex structure point) of CICY 7875 and Kreuzer-Skarke polytopes 103 and 111 (their fine regular star triangulations are unique). The respective databases were originally constructed in \cite{Candelas:1987kf} and \cite{Kreuzer:2000xy}.}
 
\subsubsection{One-modulus limit: $\mathrm{I}_1$ singularity}
As our first example we consider a finite distance limit, namely a one-modulus $\mathrm{I}_1$ singularity. Note in particular that the conifold point discussed in section \ref{ssec:conifold} is of this type, as can be inferred from the monodromy data \eqref{eq:conifoldmonodromy} by cross referencing with table \ref{table:singularitytypes}. For simplicity we set $h^{2,1}=1$, although in principle one could easily include spectator moduli that are not sent to a limit in this discussion. We find that many of the observations made for the conifold point carry over to this more general setting and, in fact, turn out to be characteristics of finite distance limits as discussed in the previous subsection.

Let us begin by determining the sl(2)-elementary states that couple to the asymptotic graviphoton. We know from \eqref{eq:gravitystates} that these states can be obtained from $\tilde{a}_{0}$ by application of the lowering operator $N^{-}=N$, which coincides with the log monodromy matrix $N$ when there is just a single modulus taken to the boundary. The discrete data that characterizes the $\mathrm{I}_{1}$ singularity is given by $d=0$, which tells us that the lowering operator $N$ annihilates $\tilde{a}_{0}$, i.e.~$N\tilde{a}_{0} = 0$. Hence there are only two elementary charges that couple to the asymptotic graviphoton, $\Re \tilde{a}_{0}$ and $\Im \tilde{a}_{0}$. Their properties have been summarized in table \ref{table:Iexample}. In principle there are two more elementary charges to consider which do not couple to the asymptotic graviphoton, but their charge-to-mass ratios are divergent. These states correspond to exponentially suppressed terms in the saxion $v$, and for the conifold point in section \ref{ssec:conifold} we found that they coupled to a exponentially suppressed term in the expansion of the holomorphic $(3,0)$-form.

\begin{table}[htb]\centering
\renewcommand{\arraystretch}{1.3}
\begin{tabular}{|c|c|c|c|c|c|}
\hline
charges & sl(2)-level & scaling  & $Q/M$ & period \\ \hline
$\Re \tilde{a_{0}}$ & $3$ & const. & 1 & imaginary \\ \hline
$\Im \tilde{a}_{0}$ & $3$& const.  & 1 & real\\ \hline
\end{tabular}
\renewcommand{\arraystretch}{1}
\caption{Properties of the charges that couple to the asymptotic graviphoton.}
\label{table:Iexample}
\end{table}

Let us briefly go over the properties of these sl(2)-elementary states $\Re \tilde{a}_{0}$ and $\Im \tilde{a}_{0}$. Their eigenvalues under application of the weight operator $Y$ of the $sl(2,\mathbb{R})$-triple follow simply from $\ell=3+d=3$ as described by \eqref{eq:a0position}. By using \eqref{eq:piece1} we find that their physical charges scale asymptotically as a constant. Furthermore, we can plug the discrete data $(d_{1},\ell_{1})=(0,3)$ that characterizes the singularity and the charges into formula \eqref{eq:chargetomass} for the charge-to-mass ratios, and we obtain
\begin{equation}
\bigg(\frac{Q}{M}\bigg)^{-2} \bigg|_{\Re \tilde{a}_{0}} = \bigg(\frac{Q}{M}\bigg)^{-2}\bigg|_{\Im\tilde{a}_0} = {0 \choose{0}} = 1\, ,
\end{equation}
where we used that $d_0=0$ and $\ell_0=3$. 

We also need to determine whether the charges $\Re \tilde{a}_{0}$ and $\Im \tilde{a}_{0}$ couple to the real or imaginary part of $\Omega_{\infty}$ in order to compute the radii of the ellipsoid from \eqref{eq:asymptoticradii}. This can be inferred from the constraint
\begin{equation}
i\langle \tilde{a}_{0}, \bar{\tilde{a}}_{0} \rangle > 0\, ,
\end{equation}
which tells us that $\Re \tilde{a}_{0}$ couples to $\Im \tilde{a}_{0}$ through the symplectic product and vice versa. This constraint follows from the so-called polarization conditions given in \eqref{eq:polarization1} and \eqref{eq:polarization2}. Although we did not introduce these conditions in our review in section \ref{sec:asympHodgetheory}, note that the given constraint can also be derived by using that the boundary Hodge norm $\langle \cdot, C_{\infty}  \bar{\cdot} \rangle$ is non-degenerate together with $\tilde{a}_{0}\in H^{3,0}_{\infty}$. We can use this constraint to read off to what part of $\Omega_{\infty}$ these charges couple when we compute the mass of a BPS state, cf.~\eqref{eq:piece3}. We find that  $\Re \tilde{a}_{0}$ couples to $\Im \Omega_{\infty}$, whereas $\Im \tilde{a}_{0}$ couples to $\Re \Omega_{\infty}$.

Next we want to compute the radii of the ellipsoid that forms the charge-to-mass spectrum of electric BPS states. Following \eqref{eq:asymptoticradii}, we can sum over the inverse squares of the charge-to-mass ratios of electric sl(2)-elementary states. We pick $\Re \tilde{a}_{0}$ as electric charge, and $\Im \tilde{a}_{0}$ is its dual magnetic charge.\footnote{ 
Taking instead some linear combination of $\Re \tilde{a}_{0}$ and $\Im \tilde{a}_{0}$ is also possible, but this can be accounted for by rotating the holomorphic $(3,0)$-form by an overall phase as discussed for instance in section \ref{ssec:conifold}.} In principle there is one more electric charge that should be considered, but we can ignore this sl(2)-elementary state since its charge-to-mass ratio diverges as already mentioned above. From the couplings of the charges to the periods given in table \ref{table:Iexample} we know that $\Re \tilde{a}_{0}$ belongs to the set $\mathcal{Q}_{\rm Im}$ given in \eqref{eq:ReImBasisSets}, whereas the set $\mathcal{Q}_{\rm Re}$ is empty. Thus we obtain as radii
\begin{equation}
\gamma_1^{-2}=0\, , \qquad \gamma_2^{-2} =  \bigg(\frac{Q}{M}\bigg)^{-2} \bigg|_{\Re \tilde{a}_{0}} = 1\, .
\end{equation}
To put these results into perspective, let us compare with the conifold point considered in section \ref{ssec:conifold}. We again found two sl(2)-elementary states with a charge-to-mass ratio equal to one, which means that the usual lower bound on the charge-to-mass ratio given in \eqref{eq:generallowerbound} can be saturated.  These states are each others electromagnetic duals, so only one of them can contribute to the radii of the electric charge-to-mass spectrum. Moreover there are no other sl(2)-elementary states with a finite charge-to-mass ratio, so for one of the radii we found $\gamma_1^{-2}=0$. Therefore the ellipsoid again degenerates into two lines at the finite distance singularity, separated from each other by a distance of two. These observations turn out to be generic features of finite distance singularities, since they do not only arise for the one-modulus finite distance limits considered here, but are found for any finite distance limit as discussed in the previous subsection.

\subsubsection{One-modulus limit: $\mathrm{III}_0$ limit}
\label{ssec:IIIexample}
For our next example we turn to a one-modulus $ \mathrm{III}_0$ singularity. We choose this example because our formula \eqref{eq:chargetomass} predicts charge-to-mass ratios that are different from the results of \cite{Gendler:2020dfp} for this singularity. In general $\mathrm{III}_{c}$ singularities can only be realized in moduli spaces with dimension $h^{2,1}\geq 2+c$ with $c \geq 0$. Here we take $h^{2,1}=3$, because then this example and the next two can then all be realized as limits in the same Calabi-Yau threefold as steps in an enhancement chain leading up to a large complex structure point (cf.~footnote \ref{footnote:CICY}). For the example studied here it means there are two moduli that are kept finite, whereas one saxion, which we shall denote by $t$, is sent to the boundary.

First we determine the sl(2)-elementary states that couple to the asymptotic graviphoton. We know from \eqref{eq:gravitystates} that these states can be obtained from $\tilde{a}_{0}$ by application of the lowering operator $N$. As already stated in the previous example, the lowering operator $N^{-}$ coincides with the log monodromy matrix $N$ when we are dealing with only a single modulus taken to the limit . The discrete data that characterizes the $\mathrm{III}_{0}$ singularity is given by $d=2$. This tells us that we can apply the lowering operator $N$ twice on $\tilde{a}_{0}$ before it vanishes. Furthermore, the charges obtained from $\Re \tilde{a}_{0}$ and $\Im \tilde{a}_{0}$ are linearly independent for singularities with $d \neq 3$, which results in six charges for $\mathcal{Q}_{\rm G}$ in total. Their properties have been summarized in table \ref{table:IIIexample}. In principle there are two more elementary charges to consider for $\mathcal{Q}_{\rm F}$ which do not couple to the asymptotic graviphoton, but their charge-to-mass ratios are divergent. As a sidenote, when this limit is realized at the large complex structure point one finds that these charges couple to polynomially suppressed terms in $1/t$ of the holomorphic $(3,0)$-form that are dropped in the sl(2)-orbit approximation. 

\begin{table}[htb]\centering
\renewcommand{\arraystretch}{1.3}
\begin{tabular}{|c|c|c|c|c|c|}
\hline
charges & sl(2)-level & scaling & $Q/M$ & electric/magnetic & period \\ \hline
$\Re \tilde{a}_{0}$ & $5$ & $t^{2}$ &2 & magnetic & imaginary \\ \hline
$\Im \tilde{a}_{0}$ & $5$& $t^{2}$ & 2 & magnetic & real\\ \hline
$N \Re \tilde{a}_{0}$ & $3$ & const. &  $\sqrt{2}$ & electric & real \\ \hline
$N \Im \tilde{a}_{0}$ & $3$& const. &  $\sqrt{2}$ & magnetic & imaginary \\ \hline
$N^{2} \Re \tilde{a}_{0}$ &  $1$ & $\frac{1}{t^{2}}$ & 2& electric & imaginary\\ \hline
$N^2 \Im \tilde{a}_{0}$ & $1$& $\frac{1}{t^{2}}$ & 2 & electric & real \\ \hline
\end{tabular}
\renewcommand{\arraystretch}{1}
\caption{Properties of the charges that couple to the asymptotic graviphoton. The parametrical scaling of the physical charges of $N \Re \tilde{a}_{0}$ and $N \Im \tilde{a}_{0}$ allows us to pick the electric charge by hand, and we chose $N \Re \tilde{a}_{0}$ as electric charge and $N \Im \tilde{a}_{0}$ as dual magnetic charge.}
\label{table:IIIexample}
\end{table}

Now let us briefly go over the properties of the states in $\mathcal{Q}_{\rm G}$. The eigenvalues under application of the weight operator $Y$ of the $sl(2,\mathbb{R})$-triple $(N,N^{+},Y)$ follow by looking at the eigenvalue of $\tilde{a}_{0}$ as follows from \eqref{eq:a0position}, together with how the $N$ lowers the eigenvalues according to \eqref{eq:lowering}. One finds that $N$ lowers the eigenvalue of $\tilde{a}_{0}$ by two, so we obtain $\ell=5,3,1$.  By using \eqref{eq:piece1} we can then infer the parametrical behavior of the physical charge for each of these states, and pick the electric charges based on whether the physical charge vanishes asymptotically.\footnote{Note that the physical charge of $N\Re \tilde{a}_{0}$ and $N\Im \tilde{a}_{0}$ is asymptotically constant, so we have to designate the electric and magnetic charge by hand. We pick $N \Re \tilde{a}_{0}$ as electric charge, but similar to the $\mathrm{I}_{1}$ singularity a different choice of electric charge can be accounted for by rotating $\Omega$ by a phase.} Furthermore, we can plug the discrete data $(d_{1},\ell_{1})=(2,5),(2,3),(2,1)$ into formula \eqref{eq:chargetomass} and obtain as charge-to-mass ratios
\begin{equation}
\begin{aligned}
\bigg( \frac{Q}{M} \bigg)^{-2} \bigg|_{(d_{1},\ell_{1})=(2,5)} &=  \frac{1}{4} {2 \choose 0} = \frac{1}{4}\, , \\
 \bigg( \frac{Q}{M} \bigg)^{-2} \bigg|_{(d_{1},\ell_{1})=(2,3)}&= \frac{1}{4} {2 \choose 1} =\frac{1}{2}\, , \\
 \quad \bigg( \frac{Q}{M} \bigg)^{-2} \bigg|_{(d_{1},\ell_{1})=(2,1)} &=  \frac{1}{4} {2 \choose 2} = \frac{1}{4}\, , 
\end{aligned}
\end{equation} 
where we used that $(d_{0},\ell_{0})=(0,3)$.

In order to determine the radii of the electric charge-to-mass spectrum, we have to determine whether a charge couples to the real or imaginary part of $\Omega_{\infty}=e^{iN}\tilde{a}_{0}$. This can be inferred from the constraint
\begin{equation}
-i\langle \tilde{a}_{0}, N^{2} \bar{\tilde{a}}_{0} \rangle > 0\, .
\end{equation}
This constraint can be obtained for instance from the fact that $\langle \cdot, C_{\infty} \bar{\cdot} \rangle$ defines a non-degenerate norm, where one needs to use that $e^{iN}\tilde{a}_{0} \in H^{3,0}_{\infty}$ together with the orthogonality conditions \eqref{eq:orthogonality}. It tells us that charges obtained from $\Re \tilde{a}_{0}$ couple to charges obtained from $\Im \tilde{a}_{0}$. It then follows that $\Re \tilde{a}_{0}$, $N \Im \tilde{a}_{0}$ and $N^{2} \Re \tilde{a}_{0}$ couple to the imaginary part of $\Omega_{\infty}$, whereas $\Im \tilde{a}_{0}$, $N \Re \tilde{a}_{0}$ and $N^{2} \Im \tilde{a}_{0}$ couple to the real part of $\Omega_{\infty}$.

Having gathered the necessary information about sl(2)-elementary states in $\mathcal{Q}_{G}$, we now compute the radii for the ellipsoid that forms the electric charge-to-mass spectrum. Plugging the charge-to-mass ratios of the electric sl(2)-elementary states into \eqref{eq:asymptoticradii}, we find that
\begin{equation}
\begin{aligned}
\gamma_1^{-2}&=\bigg(\frac{Q}{M}\bigg)^{-2}\bigg|_{N^{2} \Re \tilde{a}_{0}} =   \frac{1}{4} \, ,\\
\gamma_2^{-2}&=\bigg(\frac{Q}{M}\bigg)^{-2}\bigg|_{N^{2} \Im \tilde{a}_{0}} +\bigg(\frac{Q}{M}\bigg)^{-2}\bigg|_{N \Re \tilde{a}_{0}}  = \frac{1}{4}+ \frac{1}{2}=\frac{3}{4}\, .
\end{aligned}
\end{equation}
Before we move on to the next example, let us compare our results to what formula \eqref{eq:ratioGV} would have predicted for the charge-to-mass ratios. If we insert the discrete data $d_{1},\ell_{1}$ into this expression for the charge-to-mass ratios, we find that the sl(2)-elementary states $N \Re \tilde{a}_0, N \Im \tilde{a}_0$ would have charge-to-mass ratio equal to 1 instead of $\sqrt{2}$, and the states $\Re \tilde{a}_{0}, \Im \tilde{a}_{0}, N^{2}  \Re \tilde{a}_0, N^2  \Im \tilde{a}_0$ have charge-to-mass ratios equal to $\sqrt{3}$ instead of 2. In fact, if one compares \eqref{eq:chargetomass} and the formula derived in \cite{Gendler:2020dfp} for the asymptotic charge-to-mass ratio, 
\begin{equation}\label{eq:ratioGV}
\lim_{\gamma \to \infty} \bigg(\frac{Q}{M}\bigg)^{2} \bigg|_{q_\ell} = 1+\sum_i \frac{(\ell_i - \ell_{i-1})^2}{d_i-d_{i-1}}\, ,
\end{equation}
one finds that this mismatch shows up for any limit whose enhancement chain ends with $d_{n}=2$, and the square of the charge-to-mass ratio predicted by \eqref{eq:ratioGV} is one lower than what we obtain from \eqref{eq:chargetomass} for these cases. We can figure out the underlying reason for this difference by studying limits whose enhancement chain does involve $\mathrm{III}_{c}$ singularities, but also enhances further to a $\mathrm{IV}_{d}$ singularity.

\subsubsection{Two-moduli limit: $ \mathrm{III}_0 \to \mathrm{IV}_3$} 
To explain why \eqref{eq:ratioGV} predicts different charge-to-mass ratios for limits that end with $d_{n}=2$ compared to our result \eqref{eq:chargetomass}, we study how the $\mathrm{III}_{0}$ singularity considered before enhances to a $\mathrm{IV}_{3}$ singularity. The underlying reason for this mismatch is that terms related to spectator moduli are ignored in \eqref{eq:ratioGV}. By studying limits where these moduli are also sent to the boundary, we can uncover how these terms contribute to the charge-to-mass ratios. We denote the additional saxion that sources this enhancement step by $u$, such that the limit $t\gg u\gg 1$ can be summarized by the enhancement chain
\begin{equation}
\text{I}_0 \xrightarrow{\ t \rightarrow \infty\ }\   \mathrm{III}_0 \ \xrightarrow{\ u \rightarrow \infty\ }\  \mathrm{IV}_3\, .\end{equation}
The discrete data that characterizes this limit is given by $(d_{1},d_{2})=(2,3)$.

We begin by determining the states that couple to the asymptotic graviphoton. These states can be obtained from $\tilde{a}_{0}$ by application of the lowering operators as described by \eqref{eq:gravitystates}, where we denote the lowering operator associated with $t$ by $N_{\rm III}^{-}$ and with $u$ by $N_{\rm IV}^{-}$. From the discrete data $d_{i}$ we can infer that $N_{\mathrm{III}}^{-}$ can be applied twice on $\tilde{a}_{0}$, whereas $N_{\mathrm{IV}}^{-}$ can only be applied $d_{2}-d_{1}=1$ time. In total this leads to six different elementary charges for $\mathcal{Q}_{\rm G}$. Their properties have been summarized in table \ref{table:III-IVexample}. Similar to the $\mathrm{III}_{0}$ limit there are two more elementary charges, but these will not be considered since they lie in $\mathcal{Q}_{\rm F}$ and therefore have divergent charge-to-mass ratios. 

\begin{table}[htb]\centering
\renewcommand{\arraystretch}{1.3}
\begin{tabular}{|c|c|c|c|c|c|}
\hline
charges & sl(2)-levels & scaling & $Q/M$ & electric/magnetic & period \\ \hline
$\tilde{a}_{0}$ & $(5,6)$ & $t^{2}u$  & 2 & magnetic & imaginary \\ \hline
$N_{\rm III} \tilde{a}_{0}$ & $(3,4)$ &$u$ & $\sqrt{2}$ & magnetic & real\\ \hline
$ N_{\rm IV} \tilde{a}_{0}$ & $(5,4)$&  $\frac{t^{2}}{u}$ & 2 &magnetic & real\\ \hline
$N_{\rm III}^{2} \tilde{a}_{0}$ & $(1,2)$& $\frac{u}{t^{2}}$ &  2 &electric & imaginary\\ \hline
$N_{\rm III} N_{\rm IV} \tilde{a}_{0}$ &  $(3,2)$ & $\frac{1}{u}$ & $\sqrt{2}$ & electric & imaginary\\ \hline
$N_{\rm III}^2 N_{\rm IV} \tilde{a}_{0}$ & $(1,0)$ & $\frac{1}{t^{2}u}$ & 2 & electric & real \\ \hline
\end{tabular}
\renewcommand{\arraystretch}{1}
\caption{Properties of the charges that couple to the asymptotic graviphoton.}
\label{table:III-IVexample}
\end{table}

Let us briefly go over the properties of the states in $\mathcal{Q}_{\rm G}$. The eigenvalues under application of the weight operator $Y_{i}$ of the $sl(2,\mathbb{R})$-triple can be determined by looking at the eigenvalues of $\tilde{a}_{0}$ as follows from \eqref{eq:a0position}, together with how the $N_{i}^{-}$ lower the eigenvalues according to \eqref{eq:lowering}. The eigenvalues of $\tilde{a}_{0}$ are $(\ell_{1},\ell_{2})=(3+d_{1},3+d_{2})=(5,6)$. By acting with $N_{\mathrm{III}}^{-}$ we find that both values are lowered by two, whereas acting with $N_{\rm IV}^{-}$ only lowers the second value by two. The scaling of the physical charge in the saxions can then be obtained by simply plugging these eigenvalues into \eqref{eq:piece1}. Each of these physical charges either diverges or vanishes asymptotically for $t \gg u \gg 1$, so there are no ambiguities in distinguishing the electric and magnetic charges for this limit. We can also use the eigenvalues $\ell_{i}$ of these sl(2)-elementary states to compute their charge-to-mass ratios via \eqref{eq:chargetomass}, and we find that
\begin{equation}
\begin{aligned}
\bigg( \frac{Q}{M} \bigg)^{-2} \bigg|_{(\ell_{1},\ell_{2})=(5,6)} &=  \frac{1}{4} {2 \choose 0} {1 \choose 0}= \frac{1}{4}\, , \qquad \bigg( \frac{Q}{M} \bigg)^{-2} \bigg|_{(\ell_{1},\ell_{2})=(1,0)} &=  \frac{1}{4} {2 \choose 2} {1 \choose 1}= \frac{1}{4}\, , \\
\bigg( \frac{Q}{M} \bigg)^{-2} \bigg|_{(\ell_{1},\ell_{2})=(3,4)} &=  \frac{1}{4} {2 \choose 1} {1 \choose 0}= \frac{1}{2}\, , \qquad 
\bigg( \frac{Q}{M} \bigg)^{-2} \bigg|_{(\ell_{1},\ell_{2})=(3,2)} &=  \frac{1}{4} {2 \choose 1} {1 \choose 1}= \frac{1}{2}\, , \\
\bigg( \frac{Q}{M} \bigg)^{-2} \bigg|_{(\ell_{1},\ell_{2})=(5,4)} &=  \frac{1}{4} {2 \choose 0} {1 \choose 1}= \frac{1}{4}\, , \qquad \bigg( \frac{Q}{M} \bigg)^{-2} \bigg|_{(\ell_{1},\ell_{2})=(1,2)} &=  \frac{1}{4} {2 \choose 2} {1 \choose 0}= \frac{1}{4}\, , \\
\end{aligned}
\end{equation} 
where we used that $(d_{0},\ell_{0})=(0,3)$. 

Before we compute the radii of the electric charge-to-mass spectrum, we have to determine how the sl(2)-elementary states in $\mathcal{Q}_{\rm G}$ couple to $\Omega_{\infty}= e^{i(N_{\rm III}^{-}+N_{\rm IV}^{-})} \tilde{a}_{0}$. This coupling can be inferred from the constraint
\begin{equation}
\langle \tilde{a}_{0}, (N_{\rm III}^{-})^{2} N_{\rm IV}^{-} \bar{\tilde{a}}_{0} \rangle > 0\, .
\end{equation}
This constraint can be derived for instance by using that the metric $\langle \cdot, C_{\infty} \bar{\cdot} \rangle$ is non-degenerate together with $\Omega_{\infty} \in H^{3,0}_{\infty}$, keeping in mind that the orthogonality condition \eqref{eq:orthogonality} has to be satisfied. It tells us that charges obtained by an odd number of applications of lowering operators $N_{i}^{-}$ couple to $\Im \Omega_{\infty}$, whereas for an even number of lowering operators they couple to $\Re \Omega_{\infty}$.

Having gathered the necessary information about the sl(2)-elementary states that couple to the asymptotic graviphoton, let us now determine the radii of the ellipsoid that forms the charge-to-mass spectrum of electric BPS states. By using \eqref{eq:asymptoticradii} we find as radii
\begin{equation}
\begin{aligned}
\gamma_1^{-2}&=\bigg(\frac{Q}{M}\bigg)^{-2}\bigg|_{N^{-}_{\mathrm{III}} (N^{-}_{\mathrm{IV}})^{2}  \tilde{a}_{0}} =   \frac{1}{4} \, ,\\
\gamma_2^{-2}&=\bigg(\frac{Q}{M}\bigg)^{-2}\bigg|_{(N_{\mathrm{III}}^{-})^{2}  \tilde{a}_{0}} +\bigg(\frac{Q}{M}\bigg)^{-2}\bigg|_{N_{\mathrm{III}}^{-}N_{\mathrm{IV}}^{-} \tilde{a}_{0}}  = \frac{1}{4}+ \frac{1}{2}=\frac{3}{4}\, .
\end{aligned}
\end{equation}
To conclude, we explain how this example teaches us why \eqref{eq:ratioGV} predicts lower charge-to-mass ratios for limits with $d_{n}=2$ compared to \eqref{eq:chargetomass}. The underlying reason is that the inverse K\"ahler metric appears in expressions such as \eqref{eq:N=2identityrewritten} for the physical charge, and in general it is a non-trivial task to describe this quantity. It is only in asymptotic regimes that its form can be made more precise, where one can use \eqref{eq:Kahlerpotasymp} as approximation for the K\"ahler potential. However, it is then still hard to deal with components of the K\"ahler metric related to spectator moduli, but one can gain some intuition for these terms by sending the remaining moduli to the boundary as well. For this example we notice that the contribution of the spectator moduli adds $(\ell_2-\ell_1)^2/(d_2-d_1)$ to the square of the charge-to-mass ratio, which is always equal to one for the sl(2)-elementary states in $\mathcal{Q}_{G}$. This is exactly the difference that our expression for the charge-to-mass ratios \eqref{eq:chargetomass} predicted in section \ref{ssec:IIIexample}. We circumvented these issues with the K\"ahler metric because we computed the charge-to-mass ratios by working with the boundary Weil operator $C_{\infty}$ via identities such as \eqref{eq:Cinftyidentity}. In particular, we did not need to drop any terms related to the spectator moduli with this approach, so we can be certain that our formula \eqref{eq:chargetomass} determines the charge-to-mass ratios correctly. In this context it is interesting to recall that $\mathrm{III}_{c}$ limits can only occur in moduli spaces of dimension $h^{2,1}\geq 2+c$ with $c \geq 0$, so for a one-modulus $\mathrm{III}_{0}$ limit there must be at least one spectator modulus present. One might now wonder if sending this spectator modulus to the boundary could result also in a $\mathrm{III}_{c}$ singularity instead of a $\mathrm{IV}_{d}$ singularity. The interplay between formulas \eqref{eq:chargetomass} and \eqref{eq:ratioGV} for the charge-to-mass ratios leads us to speculate that $\mathrm{III}_{c}$ cannot be realized as $h^{2,1}$-parameter limits, since the mismatch between the values for the charge-to-mass ratios is accounted for by terms related to spectator moduli. From the results of \cite{Kerr2017}, it is known that this statement is true for $h^{2,1}=2$. It would be interesting to see whether one can also show this rigorously for $h^{2,1} > 2$, but for now we leave this task for future work.

\subsubsection{Three-moduli limit: $\mathrm{II}_2 \to \mathrm{III}_0 \to \mathrm{IV}_3$}\label{ssec:II-III-IV}
For our last example we consider a three-moduli limit characterized by the enhancement chain $\mathrm{II}_2 \to \mathrm{III}_0 \to \mathrm{IV}_3$. The purpose of this example is to demonstrate how one should divide growth sectors into smaller subsectors in order to identify the electric charges of BPS states unambiguously. In addition to the saxions $t,u$ that we used in the previous two examples for the $\mathrm{III}_0$ and $\mathrm{IV}_3$ singularities, we introduce another saxion $s$ that corresponds to the $\mathrm{II}_{2}$ singularity. The limit $s\gg t\gg u \gg 1$ can then be summarized by
\begin{equation}
\text{I}_0\xrightarrow{\ s \rightarrow \infty\ }\  \mathrm{II}_{2}\ \xrightarrow{\ t \rightarrow \infty\ }\   \mathrm{III}_0 \ \xrightarrow{\ u \rightarrow \infty\ }\  \mathrm{IV}_3\, .
\end{equation}
The discrete data that characterizes this limit is given by $(d_1,d_2,d_3)=(1,2,3)$.

Let us first identify the elementary charges that couple to the asymptotic graviphoton. As described by \eqref{eq:gravitystates} we can obtain these states from $\tilde{a}_{0}$ by applying lowering operators, where we denote the new lowering operator associated with $s$ by $N^{-}_{\rm II}$.\footnote{The lowering operators $N^{-}_{\rm III}$ and $N^{-}_{\rm IV}$ are not precisely the same matrices as in the previous example, since the procedure to construct them out of the log-monodromy matrices $N_{t}$ and $N_{u}$ changes when the saxion $s$ is also involved in the limit. For instance, one finds now that $(N^{-}_{\rm III})^{2}=0$ whereas this was not the case before. We refer again to \cite{Grimm:2018cpv} for a detailed review on the procedure to construct the $sl(2,\mathbb{R})$-triples.} From the discrete data $d_{i}$ we can infer that each lowering operator $N_{\rm II}^-, N_{\rm III}^-, N_{\rm IV}^-$ can be applied once on $\tilde{a}_0$, resulting in eight different elementary charges for $\mathcal{Q}_{\rm G}$ in total. Their properties have been summarized in table \ref{table:II-III-IVexample}. Note that these elementary charges provide us with a complete basis, so the set $\mathcal{Q}_F$ consisting of sl(2)-elementary states with divergent charge-to-mass ratios is empty.

\begin{table}[htb]\centering
\renewcommand{\arraystretch}{1.3}
\begin{tabular}{|c|c|c|c|c|c|}
\hline
charges & sl(2)-levels & scaling & $Q/M$ & electric/magnetic & period \\ \hline
$\tilde{a}_{0}$ & $(4,5,6)$ & $stu$  & 2 & magnetic & imaginary \\ \hline
$N_{\rm II} \tilde{a}_{0}$ & $(2,3,4)$ &$\frac{tu}{s}$ & 2 & sector-dep. & real\\ \hline
$N_{\rm III} \tilde{a}_{0}$ & $(4,3,4)$  &$\frac{su}{t}$ & 2 &magnetic & real\\ \hline
$ N_{\rm IV} \tilde{a}_{0}$ & $(4,5,4)$&  $\frac{st}{u}$ & 2 &magnetic & real\\ \hline
$N_{\rm II} N_{\rm III} \tilde{a}_{0}$ &  $(2,1,2)$ & $\frac{u}{st}$ & 2 & electric & imaginary\\ \hline
$N_{\rm II} N_{\rm IV} \tilde{a}_{0}$ & $(2,3,2)$&  $\frac{t}{su}$  &  2 &electric & imaginary\\ \hline
$N_{\rm III} N_{\rm IV} \tilde{a}_{0}$ & $(4,3,2)$& $\frac{s}{ut}$ & 2 & sector-dep. & imaginary\\ \hline
$N_{\rm II} N_{\rm III} N_{\rm IV}\tilde{a}_{0}$ & $(2,1,0)$ & $\frac{1}{stu}$ & 2 & electric & real \\ \hline
\end{tabular}
\renewcommand{\arraystretch}{1}
\caption{Properties of the charges that couple to the asymptotic graviphoton. The distinction between $N_{\rm II} \tilde{a}_{0}$ and $N_{\rm III} N_{\rm IV} \tilde{a}_{0}$ as electric or magnetic charge depends on the subsector of the growth sector that is being considered.}
\label{table:II-III-IVexample}
\end{table}

Next let us briefly go over the properties of these states. Eigenvalues under application of the weight operators $Y_i$ of the $sl(2,\mathbb{R})$-triple can be determined by looking at the eigenvalues of $\tilde{a}_{0}$ as follows from \eqref{eq:a0position}, together with how the $N_{i}^{-}$ lower the eigenvalues according to \eqref{eq:lowering}. The eigenvalues of $\tilde{a}_{0}$ are $\Bell=(3+d_{1},3+d_{2},3+d_{3})=(4,5,6)$. By acting with $N_{\mathrm{II}}^{-}$ we find that all values are lowered by two, acting with $N_{\mathrm{III}}^-$ lowers the last two values by two and acting with $N_{\mathrm{IV}}^-$ only lowers the last value by two. The scaling of the physical charge in the saxions can be obtained by simply plugging these eigenvalues into \eqref{eq:piece1}. We can also use these eigenvalues to compute the charge-to-mass ratios of these states, and by applying \eqref{eq:chargetomass} we obtain
\begin{equation}
\begin{aligned}
\bigg( \frac{Q}{M} \bigg)^{-2} \bigg|_{\Bell=(4,5,6)} &=  \frac{1}{4} {1 \choose 0} {1 \choose 0} {1 \choose 0}= \frac{1}{4}\, , \quad \bigg( \frac{Q}{M} \bigg)^{-2} \bigg|_{\Bell=(2,1,0)} &=  \frac{1}{4} {1 \choose 1} {1 \choose 1} {1 \choose 1}= \frac{1}{4}\, , \\
\bigg( \frac{Q}{M} \bigg)^{-2} \bigg|_{\Bell=(2,3,4)} &=  \frac{1}{4} {1 \choose 1} {1 \choose 0} {1 \choose 0}= \frac{1}{4}\, , \quad \bigg( \frac{Q}{M} \bigg)^{-2} \bigg|_{\Bell=(4,3,2)} &=  \frac{1}{4} {1 \choose 0} {1 \choose 1} {1 \choose 1}= \frac{1}{4}\, , \\
\bigg( \frac{Q}{M} \bigg)^{-2} \bigg|_{\Bell=(4,3,4)} &=  \frac{1}{4} {1 \choose 0} {1 \choose 1} {1 \choose 0}= \frac{1}{4}\, , \quad \bigg( \frac{Q}{M} \bigg)^{-2} \bigg|_{\Bell=(2,3,2)} &=  \frac{1}{4} {1 \choose 1} {1 \choose 0} {1 \choose 1}= \frac{1}{4}\, , \\
\bigg( \frac{Q}{M} \bigg)^{-2} \bigg|_{\Bell=(4,5,4)} &=  \frac{1}{4} {1 \choose 1} {1 \choose 1} {1 \choose 0}= \frac{1}{4}\, , \quad \bigg( \frac{Q}{M} \bigg)^{-2} \bigg|_{\Bell=(2,1,2)} &=  \frac{1}{4} {1 \choose 0} {1 \choose 0} {1 \choose 1}= \frac{1}{4}\, , \\
\end{aligned}
\end{equation} 
where we used that $(d_{0},\ell_{0})=(0,3)$. 

In order to compute the radii of the electric charge-to-mass spectrum we need to know how the charges in $\mathcal{Q}_{\rm G}$ couple to $\Omega_{\infty}= e^{i(N_{\rm II}^{-}+N_{\rm III}^{-}+N_{\rm IV}^{-})} \tilde{a}_{0}$. The relevant constraint for this example is given by
\begin{equation}
-\langle \tilde{a}_{0}, N_{\rm II}^{-}N_{\rm III}^{-} N_{\rm IV}^{-} \bar{\tilde{a}}_{0} \rangle > 0\, .
\end{equation}
This constraint can be derived for instance by using that the metric $\langle \cdot, C_{\infty} \bar{\cdot} \rangle$ is non-degenerate together with $\Omega_{\infty} \in H^{3,0}_{\infty}$, keeping in mind that the orthogonality condition \eqref{eq:orthogonality} has to be satisfied. It tells us that charges obtained by an odd number of applications of lowering operators $N_{i}^{-}$ couple to the imaginary part of $\Omega_{\infty}$, whereas an even number of lowering operators corresponds to the real part of $\Omega_{\infty}$.

The last property of the sl(2)-elementary states that we need to discuss is whether they are electric or magnetic. Recall that electric BPS states are characterized by a physical charge that vanishes asymptotically. Looking at table \ref{table:II-III-IVexample}, we find that the physical charges of $N^{-}_{\rm II} N^{-}_{\rm III} \tilde{a}_{0},$ $N^{-}_{\rm II} N^{-}_{\rm IV} \tilde{a}_{0}$ and $N^{-}_{\rm II} N^{-}_{\rm III} N^{-}_{\rm IV}\tilde{a}_{0}$ always go to zero. However, whether the physical charge of $N^{-}_{\rm II} \tilde{a}_{0}$ or $N^{-}_{\rm III} N_{\rm IV} \tilde{a}_{0}$ vanishes depends on what region of the growth sector we consider.  It is at this stage that we have to subdivide the growth sector $s\gg t\gg u \gg 1$ into smaller subsectors.  When we take $s \gg t u$ the physical charge of $N^{-}_{\rm II} \tilde{a}_{0}$ vanishes asymptotically, whereas when we take $s \ll t u$ this happens for $N^{-}_{\rm III} N^{-}_{\rm IV} \tilde{a}_{0}$. Imposing either of these constraints on the scaling of the saxions ensures that we can make a definite statement about whether the physical charge diverges or vanishes asymptotically. Let us stress that introducing these subsectors is only necessary in order to identify the electric charges, and in particular it is not a requirement that follows from asymptotic Hodge theory. Below we go through the computation of the radii for each of these subsectors.

\textbf{Subsector 1:} $s\ll tu$. In this regime the set of elementary charges for electric BPS states is given by $N^{-}_{\rm II} N^{-}_{\rm III} \tilde{a}_{0}$, $N^{-}_{\rm II} N^{-}_{\rm IV} \tilde{a}_{0}$, $N^{-}_{\rm III} N_{\rm IV} \tilde{a}_{0}$ and $N^{-}_{\rm II} N^{-}_{\rm III} N^{-}_{\rm IV}\tilde{a}_{0}$. We can use the information gathered about these states in table \ref{table:II-III-IVexample} and apply \eqref{eq:asymptoticradii} to compute the radii
\begin{equation}
\begin{aligned}
\gamma_1^{-2}&=\bigg(\frac{Q}{M}\bigg)^{-2}\bigg|_{N^{-}_{\mathrm{II}} N^{-}_{\mathrm{III}} N^{-}_{\mathrm{IV}}  \tilde{a}_{0}} =   \frac{1}{4} \, ,\\
\gamma_2^{-2}&=\bigg(\frac{Q}{M}\bigg)^{-2}\bigg|_{N^{-}_{\mathrm{II}} N^{-}_{\mathrm{III}} \tilde{a}_{0}} +\bigg(\frac{Q}{M}\bigg)^{-2}\bigg|_{N^{-}_{\mathrm{II}} N^{-}_{\mathrm{IV}} \tilde{a}_{0}} +  \bigg(\frac{Q}{M}\bigg)^{-2}\bigg|_{N^{-}_{\mathrm{III}} N^{-}_{\mathrm{IV}} \tilde{a}_{0}}  = \frac{1}{4}+ \frac{1}{4}+\frac{1}{4}=\frac{3}{4}\, .
\end{aligned}
\end{equation}
\textbf{Subsector 2:} $s\gg tu$. In this regime the set of elementary charges for electric BPS states is given by $N^{-}_{\rm II} \tilde{a}_{0}$, $N^{-}_{\rm II} N^{-}_{\rm III} \tilde{a}_{0},$ $N^{-}_{\rm II} N^{-}_{\rm IV} \tilde{a}_{0}$ and $N^{-}_{\rm II} N^{-}_{\rm III} N^{-}_{\rm IV}\tilde{a}_{0}$. We can use the information gathered about these states in table \ref{table:II-III-IVexample} and apply \eqref{eq:asymptoticradii} to compute the radii
\begin{equation}
\begin{aligned}
\gamma_1^{-2}&=\bigg(\frac{Q}{M}\bigg)^{-2}\bigg|_{N^{-}_{\mathrm{II}} N^{-}_{\mathrm{III}} N^{-}_{\mathrm{IV}}  \tilde{a}_{0}} +   \bigg(\frac{Q}{M}\bigg)^{-2}\bigg|_{N^{-}_{\mathrm{II}} \tilde{a}_{0}} =   \frac{1}{4}+\frac{1}{4} = \frac{1}{2} \, ,\\
\gamma_2^{-2}&=\bigg(\frac{Q}{M}\bigg)^{-2}\bigg|_{N^{-}_{\mathrm{II}} N^{-}_{\mathrm{III}} \tilde{a}_{0}} +\bigg(\frac{Q}{M}\bigg)^{-2}\bigg|_{N^{-}_{\mathrm{II}} N^{-}_{\mathrm{IV}} \tilde{a}_{0}}   = \frac{1}{4}+ \frac{1}{4}=\frac{1}{2}\, .
\end{aligned}
\end{equation}
Thus we find that the asymptotic radii of the electric charge-to-mass spectrum differ depending on the subsector we consider. This difference comes about purely by considering other states to be electric when we move between subsectors, since the asymptotic charge-to-mass ratios of states in $\mathcal{Q}_{\rm G}$ do not change.

\section{Remarks on other swampland conjectures}
\label{sec:remarks}
In this section we discuss connections between the order-one coefficients in various swampland conjectures. In the previous section we derived a formula for the charge-to-mass ratio of sl(2)-elementary BPS states that applies to any limit in complex structure moduli space $\cM^{\rm cs}(Y_3)$, which provides us with an order-one coefficient for the Weak Gravity Conjecture in the strict asymptotic regime. First, we point out that these charge-to-mass ratios also appear in the order-one coefficient in the asymptotic de Sitter conjecture for a particular class of flux potentials. Then we review the connection between the Weak Gravity Conjecture and the Swampland Distance Conjecture, and comment on the order-one coefficient that we obtain for the Swampland Distance Conjecture via this connection.

\subsection{Bounds for the de Sitter conjecture}\label{ssec:dSbounds}
We first study the order-one coefficient in the asymptotic de Sitter conjecture \cite{Obied:2018sgi}. This conjecture states that the gradient of a scalar potential $V$ in a theory coupled to gravity must obey in the asymptotic regime of field space the bound 
\begin{equation}\label{eq:dSconjecture}
\frac{|\nabla V|}{V} \geq c\, ,
\end{equation}
for some positive constant $c$ of order-one in Planck units. In its refined version it also states that an alternative to the bound \eqref{eq:dSconjecture} is the existence of an unstable direction. 
For the class of flux potentials we consider here we only study the bound \eqref{eq:dSconjecture} in the asymptotic regime, keeping in 
mind that there are unfixed directions. 

The expression for the flux potentials for Type IIB orientifold compactification on Calabi-Yau threefolds $Y_3$ can be expressed in terms of the superpotential \cite{Gukov:1999ya}
\begin{align}
W(\tau,t) = \langle F_3-\tau H_3, \Omega(t) \rangle \,, \label{eq:Superpotential}
\end{align}
which depends on the complex structure moduli $t^i$ and the axio-dilaton $\tau$ but is independent of the K\"ahler moduli at tree-level. The K\"ahler potential is given by
\begin{equation}
K(\tau,\bar{\tau},t,\bar{t}, T,\bar T)= -\log i(\bar{\tau}-\tau)+K^{cs}(t,\bar{t})-  2\log \cV(T, \bar{T}) \, , 
\end{equation}
where $\cV$ denotes the volume of $Y_3$ in ten-dimensional Einstein frame, which depends non-trivially on 
the K\"ahler structure moduli denoted by $T_\alpha$, see e.g.~\cite{Grimm:2004uq} for details. 
It is well-known \cite{Giddings:2001yu} that the K\"ahler moduli are not stabilized by the listed tree-level $\cN=1$ data and so quantum corrections to the superpotential $W$ or K\"ahler potential need to be included when one is trying to construct vacua. In terms of the superpotential, the tree-level effective potential takes the form
\begin{equation} 
V=e^{K}K^{I\bar{J}}D_I W D_{\bar{J}} \bar{W}\, , \label{eq:IIBPotential}
\end{equation}
where the sum over $I$ runs only over the complex structure moduli $t^i$ and the axiodilaton $\tau$.  

An alternative formulation for the scalar potential \eqref{eq:IIBPotential} that is more suited for our purposes is given by
\begin{equation}  \label{eq:potentialhodgestar}
V = \cV^{-2} \left( \frac{1}{4}e^{\phi} \langle F_3, \ast F_3 \rangle + \frac{1}{4}  e^{-\phi} \langle H_3, \ast H_3 \rangle - \frac{1}{2} \langle F_3, H_3 \rangle  \right) \,.
\end{equation}
Our goal now is to establish a link between the order-one coefficients that appear in the Weak Gravity Conjecture and the one appearing in the de Sitter conjecture. For this we will rewrite a specific class of flux potentials, i.e.~arising from either picking just $F_3$ or just $H_3$ flux, in terms of charge-to-mass ratios of some BPS state\footnote{Let us note that we use the term BPS state here very loosely, since for our argument to work we do not need to make sure that the charge lattice site we pick is actually populated by a physical BPS state. In fact, it would be more appropriate to view these potentials as sourced by domain walls, as recently considered in \cite{Lanza:2020qmt}.}. This will allow us to use our asymptotic expression for the  charge-to-mass ratio \eqref{eq:chargetomass} in order to evaluate the order-one constant $c$ from \eqref{eq:dSconjecture} numerically at the boundary of complex structure moduli space.

The calculation for both flux choices is similar and only differs slightly in the dilaton factor. So we will only be explicit for the case where $F_3=q$ and $H_3=0$. By using the expressions for the charge \eqref{eq:charge} and mass \eqref{eq:centralcharge} for a BPS state that would be associated with this charge $q$, we can suggestively rewrite the potential \eqref{eq:potentialhodgestar} as 
\begin{equation}
V = \frac{1}{2 \cV^2} \bigg( \frac{Q}{M}\bigg)^2 e^{\phi} M^2\, .
\end{equation}
We are now restricting to fluxes for which the above `charge-to-mass' ratio approaches a constant value along the limit. As explained in section \ref{sec:generalanalysis}, this can be realized by requiring the charge $q$ to belong to $\cQ_G$ defined in \eqref{eq:gravitystates}. In 
the following we will assume that
\begin{equation} \label{QMassumption}
\bigg| \nabla  \frac{Q}{M} \bigg|^2_{q \in \cQ_G} = 2 K^{i \bar{j}} \partial_i  \frac{Q}{M} \partial_{\bar{j}} \frac{Q}{M} \to 0\, ,
\end{equation}
along the limit. It should be noted, however, that we inferred this condition from studying a number of examples, and did not yet manage to show it rigorously in the framework of asymptotic Hodge theory. From there, we can see that the ratio that is of interest in the de Sitter conjecture reduces to
\begin{equation}
\frac{\big| \nabla  V \big|^2}{V^2}  = \frac{2K^{A \bar{B}} \partial_A V \partial_{\bar{B}}V }{V^2} =  \frac{2K^{A \bar{B}} \partial_A ( \cV^{-2} e^{\phi} M^2 ) \partial_{\bar{B}} (\cV^{-2}e^{\phi} M^2 ) }{\cV^{-4} e^{2\phi} M^4}\, ,
\end{equation}
with the indices $A,\bar{B}$ running over all the moduli, i.e.~the complex structure moduli, the K\"ahler structure moduli and the axio-dilaton. The Cauchy-Schwarz inequality tells us that we do not have to consider the mixed term between $\partial_I (Q/M)$ and $\partial_I( \cV e^{\phi} M^2)$. Furthermore, by making use of the identity \eqref{eq:N=2identityrewritten} we then find that asymptotically
\begin{equation}
\lim_{\lambda \to \infty} \frac{\big| \nabla  V \big|^2}{V^2} \bigg|_{F_3 \in \cQ_G} = 2\bigg[ \bigg( \frac{Q}{M} \bigg)^2-1\bigg]+ 2 + 6\, . \label{eq:AsyRatioH}
\end{equation}
where the last two terms represent the positive contribution that arise from including the axio-dilaton and K\"ahler moduli respectively. By making instead the flux choice $F_3=0$ and $H_3=q $, we obtain an identical relation for potentials, given by\footnote{In that case the potential is given by $V=\frac{1}{2} \cV^{-2}(\frac{Q}{M})^2e^{ -\phi}M^2$ instead, and the gradient reduces to $2K^{A\bar{B}} \partial_A V \partial_{\bar{B}} V =  2\cV^{4} e^{2\phi}\big(\frac{Q}{M}\big)^4 K^{A \bar{B}} \partial_A (\cV^{-2}e^{-\phi} M^2 ) \partial_{\bar{B}} (\cV^{-2}e^{-\phi} M^2 )$. }
\begin{equation}
\lim_{\lambda \to \infty} \frac{\big| \nabla  V \big|^2}{V^2} \bigg|_{H_3 \in \cQ_G}  = 2\bigg[\bigg( \frac{Q}{M} \bigg)^2-1\bigg]+2+6\, . \label{eq:AsyRatioF}
\end{equation}
where the individual contributions from the axio-dilation and the K\"ahler moduli are again given separately in the second and third terms respectively. We see that \eqref{eq:AsyRatioH} produces the same values as \eqref{eq:AsyRatioF}, so in order to give a bound it does not matter which of the two scenarios we choose. Furthermore, we neglect the contributions from the K\"ahler moduli, as their stabilization would anyway require including quantum corrections for them. Depending on the type of singularity, we found different lowest values for the charge-to-mass ratios. To keep things compact, we will only distinguish between the finite and infinite distance singularities, which gives 
\begin{equation}
\frac{\big| \nabla  V \big|}{V} \bigg|_{\rm asym}   \gtrsim   \begin{cases} \, \sqrt{2} &\text{ finite distance}  \\ 2 \sqrt{2/3}   &\text{ infinite distance} \end{cases}\, ,
\end{equation}
where the bound is to be understood as explained below \eqref{eq:radiibound}. A more refined analysis can of course be performed for the infinite distance case by considering the different enhancement chains. The bounds we obtain here coincide with bounds that were found recently in \cite{Andriot:2020lea,Lanza:2020qmt}, and also with previously established no-go theorems \cite{Blaback:2010sj, Andriot:2016xvq}. Ignoring contributions coming from the axio-dilaton, note that we recover the recently proposed Trans-Planckian Censorship Conjecture bound \cite{Bedroya:2019snp}, i.e.~$c \geq \sqrt{2/3}$. 

In our analysis, we also neglected D7-brane moduli which would also give a contribution to the superpotential \eqref{eq:Superpotential} and enter in the K\"ahler potential at the next to leading order in the string coupling. A systematic way to include these moduli would be to look at F-theory flux vacua where they become together with the axio-dilaton part of the complex structure moduli of the relevant Calabi-Yau fourfold \cite{Sen:1996vd}. In fact, such setups were already studied within the framework of asymptotic Hodge theory in \cite{Grimm:2019ixq}, and it would be interesting to revisit these flux potentials in the future. 

\subsection{Comments on the Swampland Distance Conjecture}
We next turn to the order-one coefficient of the Swampland Distance Conjecture \cite{Ooguri:2006in,Klaewer:2016kiy}. It states that when approaching infinite distance loci in field space an infinite tower of states should become exponentially light in the field distance. For two points $P,Q$ in field space, this means that the masses of these states behave asymptotically as
\begin{equation}
M \sim M_0 \, e^{-\lambda d(Q,P)}\, ,
\end{equation}
where $d(Q,P)$ denotes the geodesic distance between these points, and $\lambda$ is the relevant order-one coefficient. This conjecture motivated detailed studies of moduli spaces in string compactifications, where evidence towards it was provided by identifying the towers of states that become light in these asymptotic regimes \cite{Grimm:2018ohb,Blumenhagen:2018nts,Lee:2018urn,Grimm:2018cpv,Corvilain:2018lgw,Lee:2018spm,Lee:2019tst,Font:2019cxq,Marchesano:2019ifh,Lee:2019xtm,Grimm:2019wtx,Erkinger:2019umg,Lee:2019wij,Baume:2019sry,Lanza:2020qmt,Klaewer:2020lfg}. 

For our purposes it is important to point out the towers of wrapped D3-brane states constructed in \cite{Grimm:2018ohb,Grimm:2018cpv}, since these form the infinite towers of states that become massless at infinite distance loci in complex structure moduli space for Type IIB Calabi-Yau compactifications. This construction starts from a particular sl(2)-elementary state that belongs to $\mathcal{Q}_{\rm G}$, i.e.~it couples to the asymptotic graviphoton. This state becomes light close to the singular loci, and the infinite tower of states is generated by acting with monodromy transformations on this `seed charge'. In studying the bounds put by the Swampland Distance Conjecture it then suffices to consider the mass of this sl(2)-elementary state, since it sets the parametrical behavior for the masses of all states in this infinite tower. 

Our goal is now to relate the order-one coefficient we computed for the Weak Gravity Conjecture to its counterpart for the Swampland Distance Conjecture. The connection between these conjectures has already been studied before, and how to relate their order-one coefficients was spelled out in \cite{Lee:2018spm,Gendler:2020dfp}. Following \cite{Gendler:2020dfp}, we can express $\lambda$ in terms of the gradient of the mass of the sl(2)-elementary state as
\begin{equation}\label{eq:lambda}
\lambda = 2 \Big| K^{ij}\frac{\partial_i M}{M} u_j \Big|\, ,
\end{equation}
where $u_i$ denotes the unit vector that points along the geodesic. By making use of \eqref{eq:N=2identityrewritten} one can then bound the coefficient $\lambda$ via a Cauchy-Schwarz inequality. By picking a geodesic with $u_i = \partial_i \log M$ one can saturate this bound which yields
\begin{equation}\label{eq:SDCbound}
\lambda^2 = 2  \Big| K^{ij}\frac{\partial_i M \partial_j M}{M^2} \Big| = \frac{1}{2} \bigg( \Big(\frac{Q}{M}\Big)^2-1\bigg)\, .
\end{equation}
One can then try to study the coefficient $\lambda$ of the Swampland Distance Conjecture either directly from \eqref{eq:lambda}, or indirectly via the bound given in \eqref{eq:SDCbound}. For the former approach one needs to have control over the asymptotic behavior of the inverse K\"ahler metric $K^{ij}$, which is achieved to some extend by approximations such as \eqref{eq:Kahlerpotasymp}. However, as mentioned before this approximation does not necessarily provide the complete picture of the K\"ahler metric, and in particular it can lead to a mismatch for charge-to-mass ratios as discussed for an example in section \ref{ssec:IIIexample}. We therefore take for the latter approach, and provide an upper bound for $\lambda$ via the charge-to-mass ratio of the sl(2)-elementary state. By using \eqref{eq:chargetomass} we obtain the bound
\begin{equation}
\lambda^2 =  \begin{cases} 2^{d_n-2}  \prod_{i=1}^n \frac{1}{ {\Delta d_i \choose{ (\Delta d_i - \ell_i)/2}}} -\frac{1}{2}
 \text{ for $d_n = 3$}\, ,\\
2^{d_n-1} \prod_{i=1}^n \frac{1}{ {\Delta d_i \choose{ (\Delta d_i - \ell_i)/2}}} -\frac{1}{2} \text{ for $d_n \neq 3$}\, .
\end{cases}
\end{equation}
In comparison to \cite{Gendler:2020dfp} this extends the bounds obtained for $\lambda$ to limits characterized by discrete data with $d_i=d_{i-1}$ for some $i$. Overall we find that the lowest value attained by $\lambda$ is still given by
\begin{equation}
\lambda \geq \frac{1}{\sqrt{6}}\, .
\end{equation}
Another way to obtain the order-one coefficient of the Swampland Distance Conjecture has been noted in \cite{Andriot:2020lea}, where it was conjectured that it can be related to the order-one coefficient of the de Sitter Conjecture via $\lambda = c/2$. In our setting this relation holds true when contributions from the K\"ahler moduli and axio-dilaton to the gradient of the Type IIB flux potential are ignored, cf.~\eqref{eq:AsyRatioH} and \eqref{eq:AsyRatioF}. We only considered infinite distance limits that involved complex structure moduli for the SDC, so it would be interesting to see if limits that also involve the axio-dilaton and/or K\"ahler moduli lead to a matching value for $\lambda$.\footnote{Work in this direction has already been performed in \cite{Font:2019cxq}, where they studied the Swampland Distance Conjecture in the mirror Type IIA setup and found tensionless branes when the dilaton was also sent to a limit. Moreover in \cite{Lanza:2020qmt} connections between swampland conjectures were studied by looking at such extended objects, and it would be interesting to see if the approach taken in our work for computing order-one constants leads to new insights into this matter. }

\section{Conclusions}\label{sec:conclusions}
In this paper we have studied the asymptotic charge-to-mass spectrum of BPS states in 4d $\mathcal{N}=2$ supergravity theories. Specifically we focused on Calabi--Yau threefold compactifications of Type IIB string theory, where these BPS states arise from D3-branes wrapped on three-cycles. Both the physical charges and the masses of such states vary with changes in the complex structure moduli. Using powerful tools from asymptotic Hodge theory we can make their leading behavior explicit when moving towards the boundary of the moduli space. This description relies on the universal structure that emerges at every such limit and can be formulated without referring to specific examples. We used this structure to derive a general formula \eqref{eq:chargetomass} for the charge-to-mass ratios of a particular set of states, which we called sl(2)-elementary, at strict asymptotic regimes in complex structure moduli space. Given this formula we were then able to obtain numerical bounds for the Weak Gravity Conjecture, and also indirectly for the asymptotic de Sitter Conjecture and Swampland Distance Conjecture.

For computing the charge-to-mass ratios of these sl(2)-elementary states two structures played a key role: First, a set of $n$ commuting $sl(2,\mathbb{R})$-algebras, and second a unique Hodge decomposition at the boundary. The former was used to describe the parametrical behavior of physical quantities, and the latter specifies the leading coefficients that appear with these scalings. In particular it turned out to be important to pay special attention to the coupling of these sl(2)-elementary states to the asymptotic graviphoton. For a vanishing coupling we found that their charge-to-mass ratio diverges asymptotically, whereas for states with a non-vanishing coupling the charge-to-mass ratio remains finite and can be given by \eqref{eq:chargetomass} in terms of the discrete data associated with the state and the boundary. Let us note that in \cite{Gendler:2020dfp} such a formula was already derived for a certain class of infinite distance limits. Our results are general and thus cover any limit in complex structure moduli space, both at finite and infinite distance. Moreover, we learned that 
 for certain limits our formula contains additional terms for the charge-to-mass ratio.

In order to put general bounds on the charge-to-mass spectrum of electric BPS states, it proved useful to first investigate how the electric charge-to-mass spectrum looks like for a generic 4d $\mathcal{N}=2$ supergravity. In \cite{Gendler:2020dfp} it was already shown that the charge-to-mass vectors of these states must lie on an ellipsoid with two non-degenerate directions, whose radii $\gamma_i$ can be computed from the supergravity data.
These radii satisfy the relation $\gamma^{-2}_1 + \gamma_2^{-2} = 1$ in any $\cN=2$ supergravity. Considering strict asymptotic regimes in complex structure moduli space, we recalled that the asymptotic values for these radii can be determined from the charge-to-mass ratio of sl(2)-elementary states
and hence our formula \eqref{eq:chargetomass} can be applied. As a consistency check, we showed that the above $\cN=2$ relation 
for the radii is always satisfied. Moreover, we were able to compute the radii $\gamma_i$ for all possible limits, the results of which are summarized in table \ref{table:WGCradii}. We found only three possible asymptotic shapes for the electric charge-to-mass spectrum; for finite distance limits the ellipsoid degenerates into two separate sheets ($\gamma_1^{-2}=1$, $\gamma_2^{-2}=0$), and for infinite distance limits it either forms a circle with radius $\gamma_1=\gamma_2 = \sqrt{2}$ or an ellipse with radii $\gamma_1 = 2$ and $\gamma_2= 2/\sqrt{3}$. We have then discussed how these 
asymptotic radii lead to concrete bounds for the charge-to-mass ratio for electric BPS states in the strict asymptotic regime. 
Let us stress that we have entirely focused on BPS states in this analysis. It is not fully settled if such states suffice to verify the Weak Gravity Conjecture, or if one needs non-BPS states to satisfy its convex hull condition. While it was confirmed in \cite{Gendler:2020dfp} on an example basis
that BPS states suffice to enclose the black hole extremality region, it is an important open problem to show this generally.

We then used these results to obtain numerical bounds for several swampland conjectures. For the Weak Gravity Conjecture these ellipsoids not only constrain the form of the black hole extremality region, but the smallest radius also serves as a lower bound on the charge-to-mass ratio of electric states, resulting in $Q/M \geq \sqrt{3}/2$ for infinite distance limits. For the de Sitter Conjecture we studied flux potentials arising from turning on only R-R flux $F_3$ or only NS-NS flux $H_3$. The respective flux was chosen to correspond to a sl(2)-elementary charge, and it was found that the gradient of the associated potentials can be expressed in terms of the corresponding `charge-to-mass ratio', thereby providing the asymptotic lower bound of $c \geq  \sqrt{2}$ or $c\geq 2\sqrt{2/3}$ on the coefficient of the asymptotic dS conjecture for finite or infinite distance singularities respectively. We noted that getting bounds this low required us to neglect the contribution of an overall K\"ahler moduli dependent volume factor to the gradient, and upon even further neglecting the contributions from the axio-dilaton we could also recover the recently proposed Trans-Planckian Censorship Conjecture bound \cite{Bedroya:2019snp}. Finally, using a connection with the Swampland Distance Conjecture we found $\lambda \geq 1/\sqrt{6}$ matching previous literature \cite{Grimm:2018ohb,Andriot:2020lea,Gendler:2020dfp}.

There are several interesting directions for future research. One natural thing to do is the extension of our analysis by going further away from the boundary into the bulk of the moduli space and thus leaving in a first step the strict asymptotic regime. In that case our formula \eqref{eq:chargetomass} is subject to polynomial correction terms that become increasingly important in the process. In a future project \cite{toappear} we will attempt to systematically analyze these corrections and in particular check whether they come with a positive or a negative sign. This would indicate if and how far our asymptotic lower bound \eqref{eq:radiibound} extends away from the boundary. 
Another promising direction is the extension of our analysis to Calabi-Yau fourfolds in the framework of F-theory along the lines of \cite{Grimm:2019ixq}. In this paper we have already provided in \eqref{eq:GeneralChargetomass} the generalization of our formula for the charge-to-mass ratio to Calabi-Yau manifolds of generic dimension $D$. Of course, this formula does not always have the interpretation as the charge-to-mass ratio of a certain BPS state, but it can nevertheless be used to study the asymptotic behavior of flux potentials similarly to section \ref{sec:remarks}. A direct benefit being that the axio-dilaton and D7-brane moduli in the IIB orientifold compacitifcation are part of the complex structure moduli space of the Calabi-Yau fourfold in F-theory \cite{Sen:1996vd}, thus allowing for a more efficient study in terms of asymptotic Hodge theory.

\subsection*{Acknowledgments}
It is a pleasure to thank Chris Couzens, Stefano Lanza, Chongchuo Li, Miguel Montero and Irene Valenzuela for useful discussions.  This research is partly supported by the Dutch Research Council (NWO) via a Start-Up grant and a Vici grant.

\appendix

\section{Derivation of the formula for charge-to-mass ratios}
\label{app:charge-to-mass}
In this appendix we derive formula \eqref{eq:chargetomass} for the charge-to-mass ratio of sl(2)-elementary BPS states that couple asymptotically to the graviphoton. For the sake of generality we perform these computations for Calabi-Yau manifolds of arbitrary complex dimension $D$. In this work we consider $D=3$, but it turns out to be fairly simple to compute quantities such as \eqref{eq:centralchargeasymptotically2} for generic dimension $D$. We make a separation of cases based on the integer $d_{n}$ that characterizes the limit, since the value that this integer takes matters for the construction of our charges. Namely, when a limit ends with $d_{n}=D$ one can take $\tilde{a}_{0}$ to be real, whereas for $d_{n } \neq D$ we have that $\Re \tilde{a}_0$ and $\Im \tilde{a}_0$ are linearly independent. This split between real and imaginary parts extends to 3-forms obtained from $\tilde{a}_{0}$ by application of lowering operators $N_{i}^{-}$. Recalling the definition of charges that couple to the asymptotic graviphoton from \eqref{eq:gravitystates}, we notice that we have to `double' the amount of charges we consider for $\mathcal{Q}_{\rm G}$ when $d_{n} \neq D$.

Before we make this separation of cases, let us make some general comments about computing the charge-to-mass ratios first. To begin we recall expression \eqref{eq:centralchargeasymptotically2} for the charge-to-mass ratio, which reads
\begin{equation}\label{eq:startingpoint}
\bigg( \frac{Q}{M}\bigg)^2 \bigg|_{q} =  \frac{\langle q, C_{\infty} q \rangle \ i^{D} \langle  \bar{\Omega}_\infty \, , \ \Omega_\infty \rangle }{2 | \langle q,\ \Omega_\infty \rangle |^2} \, ,
\end{equation}  
where we replaced the factor of $i^{3}$ by $i^{D}$. This expression serves as our starting point for computing the charge-to-mass ratios. Without knowledge of the form of the charges, we can already write out the second factor of the numerator as (see also \eqref{eq:Kinf})
\begin{equation}\label{eq:1}
 \langle \Omega_\infty , \ \bar{\Omega}_\infty \rangle = (-2i)^{d_n} \langle \tilde{a}_0 , \ \prod_i \frac{(N_i^-)^{d_i-d_{i-1}}}{(d_i-d_{i-1})!} \bar{\tilde a}_0 \rangle\, ,
\end{equation}
where we expanded the exponentials in $\Omega_\infty = e^{iN_{(n)}^-} \tilde{a}_0$ and its conjugate into lowering operators $N_i^-$. We also used the relation $\langle \cdot, N_i^- \cdot \rangle = -\langle N_i^- \cdot, \cdot \rangle$, and the powers of each $N_i^-$ are fixed by the orthogonality condition \eqref{eq:orthogonality}.

Then remain the other two factors that appear in the charge-to-mass ratio, both of which involve the charge $q$. For convenience in notation we write the charges as 
\begin{equation}
q^r_{\mathbf{k}} = (N_1^-)^{k_1} \ldots (N_n^-)^{k_n} \Re  \tilde{a}_0\, , \qquad q^i_{\mathbf{k}} = (N_1^-)^{k_1} \ldots (N_n^-)^{k_n} \Im  \tilde{a}_0\, .
\end{equation}
The integers $\mathbf{k}=(k_1,\ldots,k_n)$ that label the charges should not be confused with the eigenvalues of the level operators $Y_i$, which follow from $\ell_i-\ell_{i-1} = d_i-d_{i-1}-2k_i$. In the case that $\tilde{a}_0$ is real only the charges $q^r_{\mathbf{k}}$ matter, which in turn allows us to ignore the superscript. 

In order to evaluate the remaining two factors in the charge-to-mass ratio, let us introduce some relations relevant for the above charges. The boundary structure provides us with positivity conditions on products between certain three-forms, known as polarization conditions. The condition of interest to us involves the three-form $\tilde{a}_0$, and is given by
\begin{equation}
\label{eq:polarization1}
-(-i)^{D+d_n}\langle \tilde{a}_0, \ (N_1^-)^{d_1} (N_2^-)^{d_2-d_1} \cdots (N_n^-)^{d_n-d_{n-1}}  \bar{\tilde{a}}_0 \rangle > 0\, .
\end{equation}
In the case that $d_n \neq D$ this positivity condition can be supplemented by the vanishing constraint
\begin{equation}
\label{eq:polarization2}
\langle \tilde{a}_0, \ (N_1^-)^{d_1} (N_2^-)^{d_2-d_1} \cdots (N_n^-)^{d_n-d_{n-1}}  \tilde{a}_0 \rangle = 0\, .
\end{equation}
Finally, we can write \eqref{eq:Cinftyidentity} for a Calabi-Yau manifold of dimension $D$ as
\begin{equation}\label{eq:CinftyidentityD}
 C_{\infty}\,\prod_{i=1}^n \frac{i^{k_i}}{k_i!}(N_i^-)^{k_i} \, \tilde{a}_0= i^D \prod_{i=1}^n \frac{i^{d_i-d_{i-1}-k_i}}{(d_i-d_{i-1}-k_i)!} (N_i^-)^{d_i-d_{i-1}-k_i}\, \tilde{ a}_0\, , 
\end{equation}
Together these relations suffice to evaluate products between charges of BPS states constructed out of $\Re \tilde{a}_0$, $\Im \tilde{a}_0$ and its descendants. In the following two subsections we now write out the remaining two factors of the charge-to-mass ratios in \eqref{eq:startingpoint} for the cases $d_n=D$ and $d_n \neq D$.

\subsection{Limits with $d_{n} = D$}
Let us first consider the case where the limit is characterized by an integer $d_{n}=D$. For a Calabi-Yau threefold this corresponds to an enhancement chain that ends with a $\mathrm{IV}$ singularity. In this case $\mathbf{\tilde a}_0$ is real, so we only need to consider the charges $\mathbf{q}^r_{\mathbf{k}}$, and therefore we drop the superscript label $r$ in this subsection.

Let us begin with the factor appearing in the denominator in \eqref{eq:startingpoint}. By expanding $e^{i N^-_{(n)}}$ in terms of lowering operators $N_i^-$ we find that
\begin{equation}\label{eq:2D}
|\langle q_{\mathbf{k}},\ e^{i N^-_{(n)} } \tilde{a}_0 \rangle| = \prod_i \frac{1}{(d_i-d_{i-1}-k_i)!} \ |\langle \tilde{a}_0, \ \prod_i (N_i^-)^{d_i-d_{i-1} } \tilde{a}_0 \rangle| \, ,
\end{equation}
where we used that we needed $d_i-d_{i-1}-k_i$ factors of $N_i^-$ in this expansion to satisfy the orthogonality condition \eqref{eq:orthogonality}. Then remains the first factor in the numerator of \eqref{eq:startingpoint}. By using \eqref{eq:CinftyidentityD} for the action of $C_{\infty}$ on the charges, it reduces to
\begin{equation}\label{eq:3D}
\langle q_{\mathbf{k}} , \ C_{\infty} q_{\mathbf{k}} \rangle  =- \prod_i \frac{k_i!}{(d_i-d_{i-1}-k_i)!} \ |\langle \tilde{ a}_0, \ \prod_i (N_i^-)^{d_i-d_{i-1} } \tilde{a}_0 \rangle|\, ,
\end{equation}
Putting all factors together (\eqref{eq:1}, \eqref{eq:2D} and \eqref{eq:3D}), we find the charge-to-mass ratio to be 
\begin{equation}\label{eq:D}
\bigg( \frac{Q}{M} \bigg)^2 = 2^{d_n-1} \prod_i \frac{ (d_i-d_{i-1}-k_i)!k_i!}{(d_i-d_{i-1})!} \, .
\end{equation}

\subsection{Limits with $d_{n} \neq D$}
Now we consider the case where the the limit is characterized by an integer $d_{n} \neq D$. For threefolds this corresponds to an enhancement chain that ends with $\mathrm{I}_a$, $\mathrm{II}_b$ or $\mathrm{III}_c$. Here the computations become slightly more involved, but in the end the factors only differ by some factors of two compared to the previous subsection.

First let us exploit the polarization conditions \eqref{eq:polarization1} and \eqref{eq:polarization2} to write down conditions for products involving the vectors $\Re \tilde{a}_0$ and $\Im \tilde{a}_0$. These identities will be useful for computing the charge-to-mass ratio. Depending on the choice of $d_n$, these identities look different. In the case that $D+d_n$ is odd we find from $\langle v, w \rangle = (-1)^D \langle w, v \rangle$  and $\langle v, N_i^- w \rangle = - \langle N_i^- v, w \rangle$ that
\begin{equation}
\langle \Re \tilde{a}_0, \ \prod_i (N_i^-)^{d_i-d_{i-1} } \Re \tilde{a}_0 \rangle =0\, , \qquad \langle \Im \tilde{a}_0, \ \prod_i (N_i^-)^{d_i-d_{i-1} } \Im \tilde{a}_0 \rangle =0\, ,
\end{equation}
whilst from \eqref{eq:polarization1} we know that
\begin{equation}\label{eq:Dodd}
- i^{D+d_n+1}\langle \Re \tilde{a}_0, \ \prod_i (N_i^-)^{d_i-d_{i-1} } \Im \tilde{a}_0 \rangle >0\, .
\end{equation}
On the other hand when $D+d_n$ is even we find that
\begin{equation}
\langle \Re \tilde{a}_0, \ \prod_i (N_i^-)^{d_i-d_{i-1} } \Im \tilde{a}_0 \rangle = 0\, .
\end{equation}
which follows as a non-trivial constraint from the vanishing of the imaginary part of \eqref{eq:polarization1}. Meanwhile by combining the polarization conditions \eqref{eq:polarization1} and \eqref{eq:polarization2} we find for $D+d_n$ even that
\begin{equation}\label{eq:Deven}
(-i)^{D+d_n}\langle \Re \tilde{a}_0, \ \prod_i (N_i^-)^{d_i-d_{i-1} } \Re \tilde{a}_0 \rangle = (-i)^{D+d_n} \langle \Im \tilde{a}_0, \ \prod_i (N_i^-)^{d_i-d_{i-1} } \Im \tilde{a}_0 \rangle <0\, ,
\end{equation}
where \eqref{eq:polarization2} implied that the products involving $\Re \tilde{a}_0$ and $\Im \tilde{a}_0$ are equal to one another. For threefolds the case $D+d_n$ odd corresponds to a  $\mathrm{I}_a$ or $\mathrm{III}_c$ singularity, while the case $D+d_n$ even corresponds to a $\mathrm{II}_b$ singularity. Keeping these identities in mind, we now write out the remaining two factors of the charge-to-mass ratio below.

We begin with the factor that appears in the denominator of the charge-to-mass ratio in \eqref{eq:startingpoint}. By expanding $e^{i N^-_{(n)}}$ in terms of lowering operators $N_i^-$ we find that
\begin{equation}\label{eq:2!D}
|\langle q^{r,i}_{\mathbf{k}},\ e^{i N^-_{(n)} } \tilde{a}_0 \rangle| = \frac{1}{2} \prod_i \frac{1}{(d_i-d_{i-1}-k_i)!} \ |\langle \tilde{a}_0, \ \prod_i (N_i^-)^{d_i-d_{i-1} }  \bar{\tilde{a}}_0 \rangle |\, ,
\end{equation}
where we divide by two in comparison to \eqref{eq:2D}. This division by two was necessary because when writing out the right-hand side, either the product in \eqref{eq:Dodd} appears twice if $D+d_n$ is odd, or both products in \eqref{eq:Deven} appear when $D+d_n$ is even, whereas these products appear only once on the left-hand side.

Then remains the first factor that appears in the numerator in \eqref{eq:startingpoint}, and we find that
\begin{equation}\label{eq:3!D}
\langle q^{r,i}_{\mathbf{k}} , \ C_{\infty} q^{r,i}_{\mathbf{k}} \rangle  = \frac{1}{2} \prod_i \frac{k_i!}{(d_i-d_{i-1}-k_i)!} \ |\langle \tilde{a}_0, \ \prod_i (N_i^-)^{d_i-d_{i-1} }  \bar{\tilde{a}}_0 \rangle|\, ,
\end{equation}
where we take either both charges with superscript $r$ or both with superscript $i$. In deriving this expression we made use of \eqref{eq:CinftyidentityD} and that $C_{\infty}$ is a real map. Furthermore we divided by two in comparison to \eqref{eq:3D} for reasons similar to \eqref{eq:2!D}.

If we now put all the different factors together (\eqref{eq:1}, \eqref{eq:2!D} and \eqref{eq:3!D}), we find for the charge-to-mass ratio
\begin{equation}
\bigg( \frac{Q}{M} \bigg)^2 = 2^{d_n} \prod_i \frac{ (d_i-d_{i-1}-k_i)!k_i!}{(d_i-d_{i-1})!}\, .
\end{equation}
Note that in the end we picked up an additional factor of two compared to the case $d_n = D$ in \eqref{eq:D}.

\section{Computation of radii of the ellipsoid}
\label{app:radii}
In this appendix we determine the radii of the electric charge-to-mass spectrum for limits in complex structure moduli space $\cM^{\rm cs}(Y_3)$. We go through all possible enhancement chains that classify these limits, and compute the radii from the charge-to-mass ratios of sl(2)-elementary states via \eqref{eq:asymptoticradii}. The states relevant for this computation are the ones that couple to the asymptotic graviphoton as described by \eqref{eq:gravitystates}, since their charge-to-mass ratio stays finite. The discrete data $d_i$ associated with the enhancement chain suffices to characterize this subset of sl(2)-elementary states. This means that the relevant information about the enhancement chain is captured by just the presence or absence of the segments  $ \mathrm{II}$, $\mathrm{III}$ and $\mathrm{IV}$, so the problem reduces to considering eight different kinds of enhancement chains in total.  

Before we go through each of these kinds of enhancement chains, let us briefly summarize how one can obtain the relevant properties of the sl(2)-elementary states under consideration. The first thing we need to know are their eigenvalues under the weight operators $Y_{i}$ of the $sl(2,\mathbb{R})$-algebras. These follow from the level of $\tilde{a}_{0}$ as indicated by \eqref{eq:a0position}, together with how the lowering operators $N_{i}^{-}$ lower these levels according to \eqref{eq:lowering}. The charge-to-mass ratios of these states can then be obtained simply by evaluating \eqref{eq:chargetomass} for their discrete data. We next need to determine whether charges are electric or magnetic. This distinction is based upon whether the physical charge of these states diverges or vanishes asymptotically, which in turn can be deduced from the sl(2)-data by making use of \eqref{eq:piece1}.\footnote{For some particular elementary charges this method does not lead to a definite statement, e.g.~when the physical charge is finite asymptotically. In these cases we clarify whether elementary charges are electric or magnetic on the spot.} The last thing we need for the computation of the radii is whether charges couple to the real or imaginary part of the asymptotic graviphoton $\Omega_{\infty}$, cf.~\eqref{eq:ReImBasisSets}. This follows from the polarization conditions \eqref{eq:polarization1} and \eqref{eq:polarization2}. To be more precise, one finds for even $d_{n}$ (odd $d_{n}+3$) that charges obtained from $\Re \tilde{a}_{0}$ couple to charges obtained from $\Im \tilde{a}_{0}$, whereas for odd $d_{n}$ (even $d_{n}+3$) charges obtained from $\Re \tilde{a}_{0}$ couple to other charges obtained from $\Re \tilde{a}_{0}$ and similarly for charges obtained from $\Im \tilde{a}_{0}$. Keeping in mind that each application of an $N_{i}^{-}$ on $\tilde{a}_{0}$ comes together with an $i$ for $\Omega_{\infty}$, one can then straightforwardly determine whether a charge couples to the real or imaginary part of the asymptotic graviphoton. Having gathered all this information on the sl(2)-elementary states in $\mathcal{Q}_{\rm G}$, one is then finally ready to compute the radii of the ellipsoid.

\subsection*{Enhancement chain $\mathrm{I}$}
\label{app:Ichain}
Here we consider limits characterized by enhancement chains of the form $\mathrm{I}$, i.e.~it consists only of $\mathrm{I}_{a}$ singularities. This sort of limit is a finite distance limit, and the discrete data of such a limit is given by $d_{1},\ldots,d_{n}=0$. From this discrete data we can infer that all lowering operators $N_{i}^{-}$ annihilate the 3-form $\tilde{a}_{0}$, so we only have to consider the sl(2)-elementary charges $\Re \tilde{a}_{0}$ and $\Im \tilde{a}_{0}$. Their properties have been summarized in table \ref{table:I}. From this information we can straightforwardly compute the radii via \eqref{eq:asymptoticradii} to be
\begin{equation}
\begin{aligned}
\gamma_1^{-2} = 0 \, ,  \qquad \gamma_2^{-2} = \Big( \frac{Q}{M} \Big)^{-2} \Big|_{\Re \tilde{a}_0} = 1\, .
\end{aligned}
\end{equation}

\begin{table}[htb]\centering
\renewcommand{\arraystretch}{1.3}
\begin{tabular}{|c|c|c|c|c|}
\hline
charges & sl(2)-level   & $Q/M$ & electric/magnetic &period \\ \hline
$\Re \tilde{a}_{0}$ & $3$  & 1 & electric & imaginary \\ \hline
$\Im \tilde{a}_{0}$ & $3$   & 1 & magnetic & real\\ \hline
\end{tabular}
\renewcommand{\arraystretch}{1}
\caption{Properties of the charges that couple to the asymptotic graviphoton. The choice of electric and magnetic charge was picked by hand since the physical charges of both states are finite asymptotically.}
\label{table:I}
\end{table}

\subsection*{Enhancement chain $\mathrm{I} \to \mathrm{II}$}
Here we consider limits characterized by enhancement chains of the form $\mathrm{I} \to \mathrm{II}$. The discrete data of such a limit is given by $d_{1},\ldots,d_{k-1} = 0$ and $d_{k},\ldots,d_{n}=1$. The enhancement to a $\mathrm{II}_{b}$ singularity occurs at step $k$, and we denote the lowering operator $N_{k}^{-}$ therefore by $N_{\rm II}$. From the discrete data we can infer that $N_{\rm II}$ can be applied once on $\tilde{a}_{0}$. Moreover, charges obtained from $\Re \tilde{a}_{0}$ and $\Im \tilde{a}_{0}$ are linearly independent, so there are four charges to consider in total. Their properties have been summarized in table \ref{table:II}. From this information we can straightforwardly compute the radii via \eqref{eq:asymptoticradii} to be
\begin{equation}
\begin{aligned}
\gamma_1^{-2} = \Big( \frac{Q}{M} \Big)^{-2} \Big|_{N \Re \tilde{a}_0} = \frac{1}{4}\, ,  \qquad \gamma_2^{-2} = \Big( \frac{Q}{M} \Big)^{-2} \Big|_{N \Im \tilde{a}_0} = \frac{1}{4}\, .
\end{aligned}
\end{equation}

\begin{table}[htb]\centering
\renewcommand{\arraystretch}{1.3}
\begin{tabular}{|c|c|c|c|c|}
\hline
charges & sl(2)-level   & $Q/M$ & electric/magnetic &period \\ \hline
$\Re \tilde{a}_{0}$ & $4$  & $\sqrt{2}$ & magnetic & imaginary \\ \hline
$\Im \tilde{a}_{0}$ & $4$   & $\sqrt{2}$ & magnetic & real\\ \hline
$N_{\rm II}\Re \tilde{a}_{0}$ & $2$  & $\sqrt{2}$ & electric & real \\ \hline
$N_{\rm II}\Im \tilde{a}_{0}$ & $2$   & $\sqrt{2}$ & electric & imaginary \\ \hline
\end{tabular}
\renewcommand{\arraystretch}{1}
\caption{Properties of the charges that couple to the asymptotic graviphoton. }
\label{table:II}
\end{table}

\subsection*{Enhancement chain $\mathrm{I} \to \mathrm{III}$}
Here we consider limits characterized by enhancement chains of the form $\mathrm{I} \to \mathrm{III}$. The discrete data of such a limit is given by $d_{1},\ldots,d_{k-1} = 0$ and $d_{k},\ldots,d_{n}=2$. The enhancement to a $\mathrm{III}_{c}$ singularity occurs at step $k$, and we denote the lowering operator $N_{k}^{-}$ therefore by $N_{\rm III}$. From the discrete data we can infer that $N_{\rm III}$ can be applied twice on $\tilde{a}_{0}$. Moreover, charges obtained from $\Re \tilde{a}_{0}$ and $\Im \tilde{a}_{0}$ are linearly independent, so there are six charges to consider in total. Their properties have been summarized in table \ref{table:III}. From this information we can straightforwardly compute the radii via \eqref{eq:asymptoticradii} to be
\begin{equation}
\begin{aligned}
\gamma_1^{-2} &= \Big( \frac{Q}{M} \Big)^{-2} \Big|_{N \Re \tilde{a}_0} + \Big( \frac{Q}{M} \Big)^{-2} \Big|_{N^2 \Im \tilde{a}_0}   = \frac{1}{2}+\frac{1}{4} = \frac{3}{4}\, ,  \\ \gamma_2^{-2} &= \Big( \frac{Q}{M} \Big)^{-2} \Big|_{N^2 \Re \tilde{a}_0} = \frac{1}{4}\, .
\end{aligned}
\end{equation}

\begin{table}[htb]\centering
\renewcommand{\arraystretch}{1.3}
\begin{tabular}{|c|c|c|c|c|}
\hline
charges & sl(2)-level  & $Q/M$ & electric/magnetic & period \\ \hline
$\Re \tilde{a}_{0}$ & $5$  & 2 & magnetic & imaginary \\ \hline
$\Im \tilde{a}_{0}$ & $5$& 2& magnetic & real\\ \hline
$N \Re \tilde{a}_{0}$ & $3$  &$\sqrt{2}$ & electric & real \\ \hline
$N \Im \tilde{a}_{0}$ & $3$ & $\sqrt{2}$ & magnetic & imaginary \\ \hline
$N^{2} \Re \tilde{a}_{0}$ &  $1$ & 2 & electric & imaginary\\ \hline
$N^2 \Im \tilde{a}_{0}$ & $1$ & 2 & electric & real \\ \hline
\end{tabular}
\renewcommand{\arraystretch}{1}
\caption{Properties of the charges that couple to the asymptotic graviphoton. The parametrical scaling of the physical charges of $N \Re \tilde{a}_{0}$ and $N \Im \tilde{a}_{0}$ allows us to pick the electric charge by hand, and we chose $N \Re \tilde{a}_{0}$ as electric charge and $N \Im \tilde{a}_{0}$ as dual magnetic charge.}
\label{table:III}
\end{table}

\subsection*{Enhancement chain $\mathrm{I} \to \mathrm{II} \to \mathrm{III}$}
Here we consider limits characterized by enhancement chains of the form $\mathrm{I} \to \mathrm{II} \to \mathrm{III}$. The discrete data of such a limit is given by $d_{1},\ldots,d_{k-1} = 0$, $d_{k},\ldots,d_{l-1}=1$ and $d_{l},\ldots,d_{n}=2$. The enhancements to $\mathrm{II}_{b}$ and $\mathrm{III}_{c}$ singularities occur at steps $k$ and $l$ respectively, and we denote the lowering operators $N_{k}^{-}$ and $N_{l}^{-}$ therefore by $N_{\rm II}$ and $N_{\rm III}$. From the discrete data we can infer that $N_{\rm II}$ and $N_{\rm III}$ can both be applied once on $\tilde{a}_{0}$. Moreover, charges obtained from $\Re \tilde{a}_{0}$ and $\Im \tilde{a}_{0}$ are linearly independent, so there are eight charges to consider in total. Their properties have been summarized in table \ref{table:II-III}. From this information we can straightforwardly compute the radii via \eqref{eq:asymptoticradii} to be
\begin{equation}
\begin{aligned}
\gamma_1^{-2} &= \Big( \frac{Q}{M} \Big)^{-2} \Big|_{N_{\rm II} \Re \tilde{a}_{0}} +\Big( \frac{Q}{M} \Big)^{-2} \Big|_{N_{\rm II} N_{\rm III} \Im \tilde{a}_{0}} = \frac{1}{4} + \frac{1}{4} = \frac{1}{2}\, ,  \\
\gamma_1^{-2} &= \Big( \frac{Q}{M} \Big)^{-2} \Big|_{N_{\rm II} \Im \tilde{a}_{0}} +\Big( \frac{Q}{M} \Big)^{-2} \Big|_{N_{\rm II} N_{\rm III} \Re \tilde{a}_{0}} = \frac{1}{4} + \frac{1}{4} = \frac{1}{2}\, ,  \\
\end{aligned}
\end{equation}

\begin{table}[htb]\centering
\renewcommand{\arraystretch}{1.3}
\begin{tabular}{|c|c|c|c|c|}
\hline
charges & sl(2)-level  & $Q/M$ & electric/magnetic & period \\ \hline
$\Re \tilde{a}_{0}$ & $(4,5)$  & 2 & magnetic & imaginary \\ \hline
$\Im \tilde{a}_{0}$ & $(4,5)$ & 2& magnetic & real\\ \hline
$N_{\rm II} \Re \tilde{a}_{0}$ & $(2,3)$  & 2 & electric & real \\ \hline
$N_{\rm II} \Im \tilde{a}_{0}$ & $(2,3)$ & 2 & electric & imaginary \\ \hline
$N_{\rm III} \Re \tilde{a}_{0}$ & $(4,3)$  & 2 & magnetic & real \\ \hline
$N_{\rm III} \Im \tilde{a}_{0}$ & $(4,3)$ & 2 & magnetic & imaginary \\ \hline
$N_{\rm II} N_{\rm III} \Re \tilde{a}_{0}$ &  $(2,1)$ & 2 & electric & imaginary\\ \hline
$N_{\rm II} N_{\rm III} \Im \tilde{a}_0$ & $(2,1)$ & 2 & electric & real \\ \hline
\end{tabular}
\renewcommand{\arraystretch}{1}
\caption{Properties of the charges that couple to the asymptotic graviphoton. }
\label{table:II-III}
\end{table}

\subsection*{Enhancement chain $\mathrm{I} \to \mathrm{IV}$ }
Here we consider limits characterized by enhancement chains of the form $\mathrm{I} \to \mathrm{IV}$. The discrete data of such a limit is given by $d_{1},\ldots,d_{k-1} = 0$ and $d_{k},\ldots,d_{n}=3$. The enhancement to  a $\mathrm{IV}_{d}$ singularity occurs at step $k$, and we denote the lowering operator $N_{k}^{-}$ therefore by $N_{\rm IV}$. From the discrete data we can infer that $N_{\rm IV}$ can be applied three times on $\tilde{a}_{0}$, so there are four charges to consider in total. Their properties have been summarized in table \ref{table:IV}. From this information we can straightforwardly compute the radii via \eqref{eq:asymptoticradii} to be
\begin{equation}
\begin{aligned}
\gamma_1^{-2} = \Big( \frac{Q}{M} \Big)^{-2} \Big|_{N^3 \tilde{a}_0} = \frac{3}{4}\, ,  \qquad \gamma_2^{-2} = \Big( \frac{Q}{M} \Big)^{-2} \Big|_{N^2 \tilde{a}_0} = \frac{1}{4}\, .
\end{aligned}
\end{equation}

\begin{table}[htb]\centering
\renewcommand{\arraystretch}{1.3}
\begin{tabular}{|c|c|c|c|c|}
\hline
charges & sl(2)-levels  & $Q/M$ & electric/magnetic & period \\ \hline
$\tilde{a}_{0}$ & $6$   & 2 & magnetic & imaginary \\ \hline
$N \tilde{a}_{0}$ & $4$  & $2/\sqrt{3}$ & magnetic & real\\ \hline
$ N^2 \tilde{a}_{0}$ & $2$ & $2/\sqrt{3}$ & real & imaginary\\ \hline
$N^{3} \tilde{a}_{0}$ & $0$ &  2 &electric & real\\ \hline
\end{tabular}
\renewcommand{\arraystretch}{1}
\caption{Properties of the charges that couple to the asymptotic graviphoton.}
\label{table:IV}
\end{table}

\subsection*{Enhancement chain $\mathrm{I} \to \mathrm{II} \to \mathrm{IV}$}
Here we consider limits characterized by enhancement chains of the form $\mathrm{I} \to \mathrm{II} \to \mathrm{IV}$. The discrete data of such a limit is given by $d_{1},\ldots,d_{k-1} = 0$, $d_{k},\ldots,d_{l-1}=1$ and $d_{l},\ldots,d_{n}=3$. The enhancements to $\mathrm{II}_{b}$ and $\mathrm{IV}_{d}$ singularities occur at s $k$ and $l$ respectively, and we denote the lowering operators $N_{k}^{-}$ and $N_{l}^{-}$ therefore by $N_{\rm II}$ and $N_{\rm IV}$. From the discrete data we can infer that $N_{\rm II}$ can be applied once on $\tilde{a}_{0}$ and  $N_{\rm IV}$ twice, so there are six charges to consider in total. Their properties have been summarized in table \ref{table:II-IV}. The computation of the radii for these limits depends on the subsector of the growth sector \eqref{strictgrowthsector} we consider, since moving between these sectors changes what charges we consider to be electric. Below we go through both sectors.

\begin{table}[htb]\centering
\renewcommand{\arraystretch}{1.3}
\begin{tabular}{|c|c|c|c|c|}
\hline
charges & sl(2)-levels  & $Q/M$ & electric/magnetic & period \\ \hline
$\tilde{a}_{0}$ & $(4,6)$   & 2 & magnetic & imaginary \\ \hline
$N_{\rm II} \tilde{a}_{0}$ & $(2,4)$  & $2$ & sector-dep. & real\\ \hline
$ N_{\rm IV} \tilde{a}_{0}$ & $(4,4)$ & $\sqrt{2}$ & magnetic & real\\ \hline
$N_{\rm II} N_{\rm IV} \tilde{a}_{0}$ & $(2,2)$ &  $\sqrt{2}$ &electric & imaginary\\ \hline
$ N^2_{\rm IV} \tilde{a}_{0}$ &  $(4,2)$  & 2 & electric & imaginary\\ \hline
$N_{\rm II} N_{\rm IV}^{2}\tilde{a}_{0}$ & $(2,0)$  & 2 & electric & real \\ \hline
\end{tabular}
\renewcommand{\arraystretch}{1}
\caption{Properties of the charges that couple to the asymptotic graviphoton.}
\label{table:II-IV}
\end{table}

\textbf{Subsector 1:} $v^{\rm II} \gg (v^{\rm IV})^2$. In this subsector we find that the charge $N_{\rm IV}^{2} \tilde{a}_{0}$ has a decreasing physical charge and is therefore electric, whereas the dual charge $N_{\rm II} \tilde{a}_{0}$ is magnetic. From the information in table \ref{table:II-IV} we then compute the radii via \eqref{eq:asymptoticradii} to be
\begin{equation}
\begin{aligned}
\gamma_1^{-2} &= \Big( \frac{Q}{M} \Big)^{-2} \Big|_{N_{\rm II} \tilde{a}_{0}} + \Big( \frac{Q}{M} \Big)^{-2} \Big|_{N_{\rm II} N_{\rm IV}^{2}\tilde{a}_{0}} = \frac{1}{4} + \frac{1}{4} =\frac{1}{2}\, ,  \\
 \gamma_2^{-2} &= \Big( \frac{Q}{M} \Big)^{-2} \Big|_{N_{\rm II} N_{\rm IV} \tilde{a}_{0}} = \frac{1}{4}\, .
\end{aligned}
\end{equation}
\textbf{Subsector 2:} $v^{\rm II} \ll (v^{\rm IV})^2$. In this subsector we find that the charge $N_{\rm II} \tilde{a}_{0}$ has a decreasing physical charge instead and is therefore electric, whereas the dual charge $N^{2}_{\rm IV} \tilde{a}_{0}$ is now magnetic. From the information in table \ref{table:II-IV} we then compute the radii via \eqref{eq:asymptoticradii} to be
\begin{equation}
\begin{aligned}
\gamma_1^{-2} &= \Big( \frac{Q}{M} \Big)^{-2} \Big|_{N_{\rm II} N_{\rm IV}^{2}\tilde{a}_{0}} = \frac{1}{4}\, ,  \\
 \gamma_2^{-2} &= \Big( \frac{Q}{M} \Big)^{-2} \Big|_{N_{\rm II} N_{\rm IV} \tilde{a}_{0}}+  \Big( \frac{Q}{M} \Big)^{-2} \Big|_{N^2_{\rm IV} \tilde{a}_{0}} = \frac{1}{4}+\frac{1}{2} = \frac{3}{4}\, .
\end{aligned}
\end{equation}

\subsection*{Enhancement chain $\mathrm{I} \to \mathrm{III} \to \mathrm{IV}$}
Here we consider limits characterized by enhancement chains of the form $\mathrm{I} \to \mathrm{III} \to \mathrm{IV}$. The discrete data of such a limit is given by $d_{1},\ldots,d_{k-1} = 0$, $d_{k},\ldots,d_{l-1}=2$ and $d_{l},\ldots,d_{n}=3$. The enhancements to $\mathrm{III}_{c}$ and $\mathrm{IV}_{d}$ singularities occur at s $k$ and $l$ respectively, and we denote the lowering operators $N_{k}^{-}$ and $N_{l}^{-}$ therefore by $N_{\rm III}$ and $N_{\rm IV}$. From the discrete data we can infer that $N_{\rm III}$ can be applied twice on $\tilde{a}_{0}$ and  $N_{\rm IV}$ once, so there are six charges to consider in total. Their properties have been summarized in table \ref{table:III-IV}. From this information we can straightforwardly compute the radii via \eqref{eq:asymptoticradii} to be
\begin{equation}
\begin{aligned}
\gamma_1^{-2} &= \Big( \frac{Q}{M} \Big)^{-2} \Big|_{N_{\rm III} N_{\rm IV}^{2}\tilde{a}_{0}} = \frac{1}{4}\, ,  \\
 \gamma_2^{-2} &= \Big( \frac{Q}{M} \Big)^{-2} \Big|_{N^2_{\rm III} \tilde{a}_{0}} +\Big( \frac{Q}{M} \Big)^{-2} \Big|_{N_{\rm III} N_{\rm IV} \tilde{a}_{0}}= \frac{1}{4}+\frac{1}{2} = \frac{3}{4}\, .
\end{aligned}
\end{equation}

\begin{table}[htb]\centering
\renewcommand{\arraystretch}{1.3}
\begin{tabular}{|c|c|c|c|c|c|}
\hline
charges & sl(2)-levels & scaling & $Q/M$ & electric/magnetic & period \\ \hline
$\tilde{a}_{0}$ & $(5,6)$ & $t^{2}u$  & 2 & magnetic & imaginary \\ \hline
$N_{\rm III} \tilde{a}_{0}$ & $(3,4)$ &$u$ & $\sqrt{2}$ & magnetic & real\\ \hline
$ N_{\rm IV} \tilde{a}_{0}$ & $(5,4)$&  $\frac{t^{2}}{u}$ & 2 &magnetic & real\\ \hline
$N_{\rm III}^{2} \tilde{a}_{0}$ & $(1,2)$& $\frac{u}{t^{2}}$ &  2 &electric & imaginary\\ \hline
$N_{\rm III} N_{\rm IV} \tilde{a}_{0}$ &  $(3,2)$ & $\frac{1}{u}$ & $\sqrt{2}$ & electric & imaginary\\ \hline
$N^2_{\rm III} N_{\rm IV} \tilde{a}_{0}$ & $(1,0)$ & $\frac{1}{t^{2}u}$ & 2 & electric & real \\ \hline
\end{tabular}
\renewcommand{\arraystretch}{1}
\caption{Properties of the charges that couple to the asymptotic graviphoton.}
\label{table:III-IV}
\end{table}

\subsection*{Enhancement chain $\mathrm{I} \to \mathrm{II} \to \mathrm{III} \to \mathrm{IV}$}
Finally we consider limits characterized by enhancement chains of the form $\mathrm{I} \to \mathrm{II} \to \mathrm{III} \to \mathrm{IV}$. The discrete data of such a limit is given by $d_{1},\ldots,d_{k-1} = 0$, $d_{k},\ldots,d_{l-1}=1$, $d_{l},\ldots,d_{m-1}=2$ and $d_{m},\ldots,d_{n}=3$. The enhancements to $\mathrm{II}_{b}$, $\mathrm{III}_{c}$ and $\mathrm{IV}_{d}$ singularities occur at steps $k$, $l$ and $m$ respectively, and we denote the lowering operators $N_{k}^{-}$, $N_{l}^{-}$ and $N_{m}^{-}$ therefore by $N_{\rm II}$, $N_{\rm III}$ and $N_{\rm IV}$. From the discrete data we can infer that $N_{\rm II}$, $N_{\rm III}$ and $N_{\rm IV}$ each can be applied once on $\tilde{a}_{0}$, so there are six charges to consider in total. Their properties have been summarized in table \ref{table:II-III-IV}. The computation of the radii for these limits depends on the subsector of the growth sector \eqref{strictgrowthsector} we consider, since moving between these sectors changes what charges we consider to be electric. Below we go through both sectors.

\begin{table}[htb]\centering
\renewcommand{\arraystretch}{1.3}
\begin{tabular}{|c|c|c|c|c|}
\hline
charges & sl(2)-levels  & $Q/M$ & electric/magnetic & period \\ \hline
$\tilde{a}_{0}$ & $(4,5,6)$  & 2 & magnetic & imaginary \\ \hline
$N_{\rm II} \tilde{a}_{0}$ & $(2,3,4)$  & 2 & sector-dep. & real\\ \hline
$N_{\rm III} \tilde{a}_{0}$ & $(4,3,4)$   & 2 &magnetic & real\\ \hline
$ N_{\rm IV} \tilde{a}_{0}$ & $(4,5,4)$ & 2 &magnetic & real\\ \hline
$N_{\rm II} N_{\rm III} \tilde{a}_{0}$ &  $(2,1,2)$  & 2 & electric & imaginary\\ \hline
$N_{\rm II} N_{\rm IV} \tilde{a}_{0}$ & $(2,3,2)$  &  2 &electric & imaginary\\ \hline
$N_{\rm III} N_{\rm IV} \tilde{a}_{0}$ & $(4,3,2)$ & 2 & sector-dep. & imaginary\\ \hline
$N_{\rm II} N_{\rm III} N_{\rm IV}\tilde{a}_{0}$ & $(2,1,0)$  & 2 & electric & real \\ \hline
\end{tabular}
\renewcommand{\arraystretch}{1}
\caption{Properties of the charges that couple to the asymptotic graviphoton. The distinction between $N_{\rm II} \tilde{a}_{0}$ and $N_{\rm III} N_{\rm IV} \tilde{a}_{0}$ as electric or magnetic charge depends on the subsector of the growth sector that is being considered.}
\label{table:II-III-IV}
\end{table}

\textbf{Subsector 1:} $v^{\rm II}\ll v^{\rm III} v^{\rm IV}$. In this subsector we find that the charge $N_{\rm III} N_{\rm IV} \tilde{a}_{0}$ has a decreasing physical charge and is therefore electric, whereas the dual charge $N_{\rm II} \tilde{a}_{0}$ is magnetic. From the information in table \ref{table:II-III-IV} we then compute the radii via \eqref{eq:asymptoticradii} to be
\begin{equation}
\begin{aligned}
\gamma_1^{-2}&=\bigg(\frac{Q}{M}\bigg)^{-2}\bigg|_{N^{-}_{\mathrm{II}} N^{-}_{\mathrm{III}} N^{-}_{\mathrm{IV}}  \tilde{a}_{0}} =   \frac{1}{4} \, ,\\
\gamma_2^{-2}&=\bigg(\frac{Q}{M}\bigg)^{-2}\bigg|_{N^{-}_{\mathrm{II}} N^{-}_{\mathrm{III}} \tilde{a}_{0}} +\bigg(\frac{Q}{M}\bigg)^{-2}\bigg|_{N^{-}_{\mathrm{II}} N^{-}_{\mathrm{IV}} \tilde{a}_{0}} +  \bigg(\frac{Q}{M}\bigg)^{-2}\bigg|_{N^{-}_{\mathrm{III}} N^{-}_{\mathrm{IV}} \tilde{a}_{0}}  = \frac{1}{4}+ \frac{1}{4}+\frac{1}{4}=\frac{3}{4}\, .
\end{aligned}
\end{equation}
\textbf{Subsector 2:} $v^{\rm II}\gg v^{\rm III} v^{\rm IV}$. In this subsector we find that the charge $N_{\rm II} \tilde{a}_{0}$ has a decreasing physical charge instead and is therefore electric, whereas the dual charge $N_{\rm III} N_{\rm IV} \tilde{a}_{0}$ is now magnetic. From the information in table \ref{table:II-III-IV} we then compute the radii via \eqref{eq:asymptoticradii} to be
\begin{equation}
\begin{aligned}
\gamma_1^{-2}&=\bigg(\frac{Q}{M}\bigg)^{-2}\bigg|_{N^{-}_{\mathrm{II}} N^{-}_{\mathrm{III}} N^{-}_{\mathrm{IV}}  \tilde{a}_{0}} +   \bigg(\frac{Q}{M}\bigg)^{-2}\bigg|_{N^{-}_{\mathrm{II}} \tilde{a}_{0}} =   \frac{1}{4}+\frac{1}{4} = \frac{1}{2} \, ,\\
\gamma_2^{-2}&=\bigg(\frac{Q}{M}\bigg)^{-2}\bigg|_{N^{-}_{\mathrm{II}} N^{-}_{\mathrm{III}} \tilde{a}_{0}} +\bigg(\frac{Q}{M}\bigg)^{-2}\bigg|_{N^{-}_{\mathrm{II}} N^{-}_{\mathrm{IV}} \tilde{a}_{0}}   = \frac{1}{4}+ \frac{1}{4}=\frac{1}{2}\, .
\end{aligned}
\end{equation}


\bibliographystyle{jhep}
\bibliography{references}

\end{document}